\documentclass[
reprint,
amsmath,amssymb,
aps,
prx,
showpacs,
twocolumn,
longbibliography,
superscriptaddress,
floatfix
]{revtex4-2}
\usepackage{tikz-cd}
\usepackage{bbm}
\usepackage{xcolor}
\usepackage{mathrsfs}
\usepackage[commandnameprefix=always]{changes}
\usepackage{epsfig}

\usepackage{soul,color}
\usepackage{graphicx}
\usepackage{mleftright}
\usepackage{amsfonts}
\usepackage{amsthm}
\usepackage[figuresright]{rotating}
\usepackage{amssymb}
\usepackage{amsmath}
\usepackage{dcolumn}
\usepackage{physics}
\usepackage{float}
\usepackage{bm}
\usepackage{verbatim}
\usepackage{braket}
\usepackage[normalem]{ulem}
\usepackage[ruled,vlined,linesnumbered]{algorithm2e}
\usepackage{setspace}
\usepackage[colorlinks,linkcolor=blue,anchorcolor=blue,citecolor=blue,urlcolor=blue]{hyperref}
\usepackage{stackengine}

\usepackage{tikz}
\usetikzlibrary{quantikz2}

\usepackage{youngtab}

\usepackage{fancyhdr}

\newcommand{\prepare}{\mbox{P{\scriptsize REPARE}}}
\newcommand{\prep}{\mbox{P{\scriptsize REP}}}
\newcommand{\select}{\mbox{S\scriptsize ELECT}}

\usepackage{xcolor}

\definecolor{moss}{RGB}{44,186,0}

\newcommand{\MITChem}{Department of Chemistry, Massachusetts Institute of
Technology, Cambridge, MA 02139, USA}
\newcommand{\MITPh}{Department of Physics, Massachusetts Institute of
Technology, Cambridge, MA 02139, USA}
\newcommand{\MITEECS}{Department of Electrical Engineering and Computer Science, Massachusetts Institute of Technology, Cambridge, MA 02139, USA}
\newcommand{\NTTPHI}{Physics and Informatics Laboratory, NTT Research, Inc., 940 Stewart Dr., Sunnyvale, California, 94085, USA}
\newcommand{\Quantinuum}{Quantinuum, Terrington House, 13-15 Hills Road, Cambridge CB2 1NL, United Kingdom}
\newcommand{\NCSUECE}{Department of Electrical and Computer Engineering,\\ North Carolina State University, Raleigh, NC 27606, USA}
\newcommand{\NCSUCS}{Department of Computer Science, North Carolina State University, Raleigh, NC 27606, USA}
\newcommand{\NCSUPhys}{Department of Physics, North Carolina State University, Raleigh, NC 27606, USA}

\usepackage{orcidlink}

\begin{document}

\newpage

\title{Unification of Finite Symmetries in Simulation of Many-body Systems on Quantum Computers}

\author{Victor M. Bastidas\orcidlink{0000-0001-8091-7860}}
\altaffiliation[]{These authors contributed equally to this work.}
 \affiliation{\MITChem}
 \affiliation{\NTTPHI}

\author{Nathan Fitzpatrick\orcidlink{0000-0001-5819-9129}}
\altaffiliation[]{These authors contributed equally to this work.}
 \affiliation{\Quantinuum}

\author{K. J. Joven\orcidlink{0000-0003-4730-7053}}
 \affiliation{\NTTPHI}

\author{Zane M. Rossi}
 \affiliation{\MITPh}

\author{Shariful Islam\orcidlink{0000-0001-5230-6499}}
 \affiliation{\NCSUPhys}

\author{Troy Van Voorhis}
 \affiliation{\MITChem}
 
\author{Isaac L. Chuang\orcidlink{0000-0001-7296-523X}}
 \affiliation{\MITPh}
 \affiliation{\MITEECS}

\author{Yuan Liu\orcidlink{0000-0003-1468-942X}}\thanks{Corresponding author: q\_yuanliu@ncsu.edu}
 \affiliation{\NCSUECE}
 \affiliation{\NCSUCS}
 \affiliation{\NCSUPhys}

\begin{abstract}
Symmetry is fundamental in the description and simulation of quantum systems. Leveraging symmetries in classical simulations of many-body quantum systems can results in significant overhead due to the exponentially growing size of some symmetry groups as the number of particles increases. Quantum computers hold the promise of achieving exponential speedup in simulating quantum many-body systems; however, a general method for utilizing symmetries in quantum simulations has not yet been established. 
In this work, we present a unified framework for incorporating symmetry group transforms on quantum computers to simulate many-body systems. The core of our approach lies in the development of efficient quantum circuits for symmetry-adapted projection onto irreducible representations of a group or pairs of commuting groups. We provide resource estimations for common groups, including the cyclic and permutation groups. Our algorithms demonstrate the capability to prepare coherent superpositions of symmetry-adapted states and to perform quantum evolution across a wide range of models in condensed matter physics and \textit{ab initio} electronic structure in quantum chemistry.
Specifically, we execute a symmetry-adapted quantum subroutine for small molecules in first-quantization on noisy hardware, and demonstrate the emulation of symmetry-adapted quantum phase estimation for preparing coherent superpositions of quantum states in various irreducible representations of a symmetry group. In addition, we present a discussion of open problems regarding treating symmetries in digital quantum simulations of many-body systems, paving the way for future systematic investigations into leveraging symmetries \emph{quantumly} for practical quantum advantage.
The broad applicability and rigorous resource estimation for symmetry transformations make our framework appealing for achieving provable quantum advantage on fault-tolerant quantum computers, especially for symmetry-related properties.
\end{abstract}

\maketitle

\newpage 
\section{Introduction}

Symmetry is an important concept and it has been widely used in many fields, ranging from unification of fundamental interactions in physics \cite{buras1978aspects,weinberg1980conceptual} and the foundation of complexity theory in computer science \cite{jozsa2001computing,lomont2004hiddensubgroupproblem}, to the description of materials and molecules \cite{girvin2019modern,harris1989symmetry}.

To formally describe symmetries, sophisticated mathematical tools such as group and representation theory have been developed \cite{curtis1966representation,goodman2009symmetry}. 
These tools include the transformation between group elements and their irreducible representations (irreps), i.e., the group Fourier transform, which conveniently captures this relationship~\cite{diaconis1990efficient,maslen1998efficient}. The group Fourier transform has found various applications ranging from engineering to physical sciences \cite{terras1999fourier}. In addition to individual groups, interesting relationships between pairs of groups have been revealed, often called dualities \cite{rowe2012dual,rowe2011simple}, which enable natural and simple decomposition for group actions and their representations. These group duality theorems result in powerful transformations between group elements and their irreps for pairs of commuting groups \cite{howe1989remarks}. One prominent example is the Schur transform, which concerns the permutation and local unitary group pair \cite{howe1995perspectives}.

These mathematical triumphs in group theory are accompanied closely by their powerful applications in the description \cite{buras1978aspects,weinberg1980conceptual} and simulation~\cite{girvin2019modern,harris1989symmetry} of physical systems. For example, successful reduction in computational cost is routinely achieved in state-of-the-art classical simulation of quantum chemistry using point group symmetries \cite{goodlett2024molsym} or periodic systems with translational symmetries \cite{kratzer2019basics}. Other quantum mechanical symmetries, such as permutation of identical particles have also been considered \cite{mcweeny1971methods}.
In many-body systems, the spin of the particles is inherently related to their quantum statistics. In fact, fermions and bosons can be viewed as irreducible representations of the permutation group for half-integer and integer spin, respectively~\cite{Pauli1940}. In the fermionic case,
examples of such symmetry include the construction of many-particle wave functions with a fixed total spin quantum number (or projection of total spin along a fixed axis), such as the spin-flip coupled-cluster theory \cite{krylov2008equation,nooijen1996general} and the celebrated unitary group approach (UGA) \cite{paldus1981unitary,li2014unitary,sonnad2016informal,hinze2012unitary,matsen1986unitary} and its variants \cite{shavitt1981graphical}.  More recently, imposing symmetry constraints of the underlying Hamiltonian in quantum simulations and quantum circuits has also been proposed and realized \cite{kokail2019self,PRXQuantum.5.037001,bauer2023quantum,martinez2016real}. 

However, these efforts on leveraging group transformations are performed with classical resources or human efforts, which can face significant challenges beyond small physical systems. 
One major reason is that the size of underlying group (and the associated transformation) often scales \emph{exponentially or combinatorially} with the number of particles in quantum systems (for example, the symmetric group). This leads to an exponential overhead to execute such transformations on classical computers.

Quantum information science provides novel tools and techniques to realize and make use of these powerful symmetry transformations. Not surprisingly, it has been discovered that \emph{efficient} quantum circuits exist for some group Fourier transforms \cite{hoyer1997efficient,beals1997quantum} and the Schur transform \cite{Bacon2006,Krovi2019efficienthigh}. The group Fourier transform has been combined with controlled group actions to realize the so-called generalized quantum phase estimation algorithm \cite{harrow2005applications,childs2010quantum}. More recently, the Schur transform has been further generalized to include both the local unitary with its dual unitary, where efficient quantum circuits are developed for such mixed Schur transforms  \cite{grinko2023gelfand,nguyen2023mixed,fei2023efficient}. These quantum circuits for symmetry transformations \cite{laborde2024quantum} have been used in quantum information theory for testing states and channel properties \cite{LaBorde2023testingsymmetry,laborde2022quantum,harrow2005applications} and for designing better quantum teleportation protocols \cite{fei2023efficient}.

Despite the progress on leveraging symmetry transformations in quantum information theory, utilization of symmetry transformations in quantum simulation remains rare.
Scattered examples include symmetry-adapted hybrid quantum-classical algorithms to improve their performance \cite{meyer2023exploiting,Lyu2023symmetryenhanced}. Symmetry is also used to protect certain quantum simulation algorithms \cite{minh2021faster,nguyen2022digital,zhao2023making}. A very similar approach to what is presented in this work has been applied to quantum machine learning of point cloud data \cite{heredge2024nonunitaryquantummachinelearning}.

As the description and analysis of many seemingly disparate quantum algorithms is simplified and made more uniform \cite{martyn2021grand,gilyen2019quantum,low2017optimal}, a new opportunity has arisen in standardizing and simplifying the treatment of generic \emph{symmetry-adapted} quantum subroutines. By doing so, one natural question becomes: is it possible to leverage the coherent nature of quantum computers to simultaneously produce a coherent superposition of states in different irreps of a symmetry group? If so, is it possible to achieve significant quantum speedup for a wide class of quantum simulation problems based on efficient quantum subroutines for symmetry transformation?

There are three key challenges for such a unified treatment of symmetry transformations for quantum simulation of physical systems. %For one
First, group theoretical transformations need to be converted into efficient quantum circuits, which are not obvious to achieve. Second, the quantum circuits developed for symmetry transformations often directly operate on qubits. It is not clear how to relate the qubit computational basis to degrees of freedom in physical systems of interest. This depends on how the underlying physical systems are mapped to quantum computers. Third, for different mappings, the intrinsic symmetry of the physical systems may manifest as different symmetries on quantum computers. For example, permutation symmetry of local fermionic operators becomes a non-local operation under the Jordan-Wigner transformation. This creates a gap between pure mathematical group representation theory and realistic many-body physical systems. It makes it even more challenging that this gap intertwines with the quest for \emph{efficient} simulation of many-body systems using quantum algorithms.

In this work, we address these challenges by providing a general framework for leveraging arbitrary symmetry-adapted transformations to perform quantum simulation of many-body systems on quantum computers. As a second major contribution, we articulate open problems for provable quantum advantage in simulating many-body systems by exploiting symmetries. In more detail,
we discuss efficient quantum circuit implementation of symmetry transformations for common groups and pairs of commuting groups relevant for many-body systems. We further demonstrate how such symmetry transformations can be used to transform physical states and operators on quantum computers.

We illustrate the broad applicability of our framework for symmetry-adapted preparation, Hamiltonian simulation and quantum phase estimation. Our methodology can be applied to diverse systems ranging from spin lattices in condensed matter physics to the \emph{ab initio} electronic structure problem in quantum chemistry.
Crucially, for a given many-body system, the symmetry-adapted transformation allows preparation and manipulation of coherent superposition of states within different irreps of a symmetry group on quantum computers. Circuit constructions and resource estimations are provided. Lastly, we validate our symmetry transformation subroutine by preparing a coherent superposition of singlet and triplet states of H$_2$ molecules using a Noisy Intermediate-Scale Quantum (NISQ) device. We also demonstrate that the symmetry transform allows determination of singlet and triplet energies of H$_2$ simultaneously, which offers additional quantum speedups for chemistry simulations.

The rest of the paper is organized as follows. Sec.~\ref{sec:symmetry-group-transform} provides constructions of quantum circuits for group symmetry transforms for single and direct product of commuting groups common in many-body systems. Sec.~\ref{sec:example-application} discusses how to incorporate these symmetry routines in generic quantum algorithms to transform states and operators, with illustrative examples provided. We summarize major open problems in dealing with symmetries in many-body systems on quantum computers in Sec. \ref{sec:open-problems}. Conclusions and outlook are presented in Sec. \ref{sec:conclusion-outlook}.

\section{Quantum Character Transform and Group Symmetry Transformations}
\label{sec:symmetry-group-transform}

Symmetry group transformations and their quantum circuit realizations are discussed in this section. We start by a brief overview of Fourier transform over groups and the generalized quantum phase estimation (GQPE) in Sec. \ref{Newssec:general}. We then discuss the effect of an entangled basis measurement on the group Fourier transform and GPQE in Sec. \ref{ssec:gft2qct}, which serves as a natural transition to a formal discussion of quantum character transforms and their use in constructing projections into particular irreps in Sec. \ref{ssec:symm-projection-from-qct}. Resource estimations for single groups including the cyclic groups, symmetry groups, and general case of finite groups are provided in Sec.~\ref{ssec:single-group}. Direct product of multiple groups are covered in Sec. \ref{ssec:product-groups}.

\subsection{Overview of Group Fourier Transform and Generalized Quantum Phase Estimation}
\label{Newssec:general}

The familiar Fourier transform, which allows us to map functions of time or time series to unveil their frequency spectrum, can be generalized to Fourier transforms associated to finite groups~\cite{moore2006generic}. 

Consider a finite group $G$ of order $|G|$ and a quantum state $\ket{\phi}=\sum_g w_g \ket{g}$, where $\ket{g}$ is a convenient encoding of the group elements $g\in G$ into ancilla registers.
The group Fourier transformation $F_G$ is a map from an arbitrary quantum state of the group elements $\sum_g w_g \ket{g}$ to a new basis $\{\ket{\Gamma, ij}\}$, which is defined as
\begin{align}
    \hat{F}_G =  %\sum_k^{N_{\rm Conj}} 
    \sum _{\Gamma}
    \sum_{ij} \sum_{g\in G} \sqrt{\frac{d_{\Gamma}}{|G|}} D_{ij}^{\Gamma}(g) \ket{\Gamma, ij} \bra{g},
    \label{MainTextgroup-fourier}
\end{align}
where for a given group element $g \in G$, $D_{ij}^{\Gamma}(g)$ is the $(i,j)$-th element of the irreducible representation $\Gamma$ for $g$
\begin{equation}
    \boldsymbol{D}^{\Gamma}(g) =\begin{bmatrix} D_{11}^\Gamma(g)&\cdots & D_{1d_{\Gamma}}^\Gamma(g) \\
    D_{d_{\Gamma}1}^\Gamma(g) & \cdots & D_{d_{\Gamma}d_{\Gamma}}^\Gamma(g) \end{bmatrix} .
\end{equation}
The dimension of the irrep $\Gamma$ is denoted by $d_{\Gamma}$. Importantly, $\ket{\Gamma, ij}$ and $\ket{g}$ are encodings of the triple $(\Gamma, i, j)$ and the group elements $g \in G$ under the computational basis. In general, the indices $i, j$ depend on the irrep label $\Gamma$ as each irrep $\Gamma$ can have different dimension $d_{\Gamma}$. To simplify the notation, this explicit dependence has been removed.

From the irreducible representation $\boldsymbol{D}^\Gamma$, the character of $g$ in irrep $\Gamma$ is defined as
\begin{equation}
    \chi_{\Gamma}(g)=\sum_iD_{ii}^\Gamma(g) ,
\label{eq:character-def}
\end{equation}
which satisfies the orthogonality relation
%The character $\chi_{\Gamma}(g)$ of a group element $g$ in the $\Gamma$-th irrep is the trace $\chi_{\Gamma}(g)=\sum_iD_{ii}^\Gamma(g)$ of the representation matrix $\boldsymbol{D}^{\Gamma}(g)$. The orthogonality relations for characters are
\begin{equation}
\frac{1}{|G|} \sum_{g \in G} \chi_{\Gamma}(g)\chi^*_{\Gamma'}(g) = \delta_{\Gamma,\Gamma'}
\ ,
\label{eqn:char_orthog}
\end{equation}
where $\delta_{\Gamma,\Gamma'}$ is the Kronecker delta, and $*$ denotes complex conjugation.

It has been shown that $\hat{F}_G$ can serve as a building block to construct a more interesting sub-routine, the generalized quantum phase estimation (GQPE), which has been widely used in various quantum information applications \cite{harrow2005applications}. The circuit is shown in Fig.~\ref{GQPE}. Denote GQPE as $\hat{U}_{\rm GQPE}$ and for an initial state $\ket{0}\otimes \ket{\psi}$ with $\ket{0}$ the qubit encoding of the trivial irrep and $\ket{\psi}$ an arbitrary quantum state with support on the representation, it can be shown that $\hat{U}_{\rm GQPE}$ can transform an initial state in the following way
\begin{align}
    \hat{U}_{\rm GQPE}\ket{0}\otimes\ket{\psi} = \sum_\Gamma \frac{1}{\sqrt{d_\Gamma}} \sum_{ij} \ket{\Gamma, ij} \otimes\ket{\psi^{\Gamma}_{i,j}} := \ket{\Psi}
    \label{GQPEAction}
\end{align}
where $\ket{\psi^{\Gamma}_{i,j}}=\hat{P}^\Gamma_{ij}\ket{\psi}$, for
\begin{align}
    \hat{P}^\Gamma_{ij} = \frac{d_\Gamma}{|G|} \sum_{g \in G} D_{ij}^\Gamma(g) \hat{\rho}(g)
    \label{projector-trivial-irrep}
\end{align}
defined as the projector into a particular state in the irrep $\Gamma$. Note that $\hat{\rho}(g)$ is a unitary matrix that represents the action of the group element $g$ to the system. Formally, $ \rho: G \to \text{GL}(V) $ is a mapping from a group $G$ to a general linear group on vector space $V$. For the abelian group $G=\mathbb{Z}_M$ with $M=2^m$ for an integer $m$, Eq.~\eqref{GQPEAction} can be recognized as the well-known quantum phase estimation. This can be seen by defining the unitary representation of $G=\mathbb{Z}_M$ as $\hat{\rho}(x)=\hat{V}^x$ for some unitary operator $\hat{V}$, where $x=x_1 2^{m-1}+x_2 2^{m-2} +x_m 2^0$ and by assuming that the state $\ket{\psi}$ has support on the Hilbert space that $\hat{V}$ acts on.

\begin{figure}[!htpb]
\centering
\begin{quantikz}
\lstick{$|0\rangle$}& \qwbundle{n} &\gate{\hat{F}_G^\dagger} &   \ctrl{1}  & \gate{\hat{F}_G}  & \qw  \rstick{$\ket{\Gamma,i,j}$} \\
\lstick{$|\psi\rangle$} & \qw & &\gate[1]{\hat{\rho}(g)}  &  &\qw  & \qw  \rstick{$|\psi^{\Gamma}_{i,j} \rangle$} \\
\end{quantikz}
    \caption{Quantum circuit for generalized quantum phase estimation.}
\label{GQPE}
\end{figure}

%%%%%%%%%%%%%%%%%%%%%%%%%%%%%%%%%%%%%%%%%%%%%%%%%%%%%%%%%%%%%%%%%%%%%%%%
\subsection{From Group Fourier Transform to Quantum Character Transform}
\label{ssec:gft2qct}

Despite efficient implementations of the group Fourier transform $F_G$ \cite{harrow2005applications}, three qubit registers are needed to encode the triple $\Gamma,i,j$. In some situations, interest lies in projecting to a given irrep $\Gamma$ while ignore the finer structures with the irrep by ignoring the indices $i,j$. In this section, we will show that such a coarse-grained group Fourier transform on only $\Gamma$ can be constructed by performing a measurement under an entangled basis on the $\ket{i,j}$ registers.

\subsubsection{Group Fourier Transform and its Action Under Entangled Basis Measurement}

Given a group $ G $ and an element $ g \in G $, the conjugacy class of $ g $ in $ G $ is the set $\mathcal{C}(g) = \{ hgh^{-1} \mid h \in G \}$. From now on, the conjugacy classes will be labeled by $C$ to simplify the notation. In the case of abelian groups, all the irreps are one-dimensional, and the number of conjugacy classes equals the number of group elements because all elements commute according to $hgh^{-1} = g$. 

On a quantum computer, we encode the conjugacy class $C$ as $\ket{C}$, therefore, a group element $\ket{g}$ can instead be labeled by three registers as $\ket{C, i, j}$ by leveraging the conjugacy class, where the two registers $\ket{i}$ and $\ket{j}$ have a total dimension equal to the number of group elements in $C$, i.e., the size of the conjugacy class $|C|$. In addition, register $\ket{C}$ will contain $n_{\rm conj} = \log_2(N_{\rm conj})$ qubits where $N_{\rm conj}$ is the total number of conjugacy classes of the, register $\ket{i}$ and $\ket{j}$ each will have $n_{\rm dim} = \log_2(\sqrt{|C|})$ qubits. From group theory, we know that there exists a $\Gamma$ such that for any $C$, $|C| = d_{\Gamma}^2$ is satisfied where $d_\Gamma$ is the dimensionality of irrep $\Gamma$. This guarantees that $\sqrt{|C|}$ can always be an integer.

Consider the modified group Fourier transform circuit in Fig. \ref{fig:ft2qct}, where a Hadamard transform is performed on the lower two registers, and $\hat{V}_{ent}$ is an entangling unitary such that it prepares an entangled state $\ket{\Omega_\Gamma} = \frac{1}{\sqrt{d_\Gamma}} \sum_{i=0}^{d_\Gamma -1}  \ket{i, i}$ from the all zero state
\begin{align}
    \ket{\Omega_\Gamma} = \hat{V}_{ent}^\dagger \ket{0,0} .
\end{align}
After post-selecting the measurement result at the lower two registers to be both $\ket{0}$ in Fig. \ref{fig:ft2qct}, the resulting transformation on the first register is (see Appendix \ref{app:proof-ft2qct} for a proof)
\begin{align}
    &\left( \bra{0} \bra{0} \hat{V}_{ent} \right) \hat{F}_G \left(H \ket{0} \otimes H\ket{0} \right) \nonumber \\
    =& \sum _{\Gamma}
     \sum_{C} \sqrt{\frac{|C|}{|G|}} \chi_{\Gamma}(C) \ket{\Gamma}\bra{C}
     \label{eq:ft2qct} 
\end{align}
where $\chi_\Gamma(C)$ is the character for conjugacy class $C$. 

Formally, we define the final transformation as the \emph{Quantum Character Transform} (QCT)
\begin{equation}
    QCT_G = \sum_{C}^{N_{\rm conj}} \sum_{\Gamma}^{N_{\rm conj}} \frac{\sqrt{|C|}\cdot\chi_{\Gamma}(C)}{\sqrt{|G|}}\ket{\Gamma}\bra{C},
\label{eq:qctBracketNotation}
\end{equation}
which is essentially a quantum encoding of the character table for a given group. It is known that the column index of the character table runs through conjugacy classes $C$, and the row index represents the irreps $\Gamma$. The number of conjugacy classes is equal to the number of distinct irreps \cite{Fulton2004}, which means the character table is a square matrix. The characters $\chi_\Gamma(C)$ of a given conjugacy class $C$ in different irrep subspaces $\Gamma$ are encoded in each column of the character table. It is guaranteed by group theoretical results that any two columns of the character table are orthogonal. Similarly, any two rows of the character table are also orthogonal if the order of each conjugacy class is properly included
\begin{equation}
    \sum_{C}^{N_{\rm conj}} \frac{|C|}{|G|} \chi_{\Gamma}(C)\chi^*_{\Gamma'}(C) = \delta_{\Gamma,\Gamma'}
    \ ,
    \label{eqn:Conjugacychar_orthog}
\end{equation}
which can be derived from Eq.~\eqref{eqn:char_orthog} by using conjugacy classes~\cite{Fulton2004}. These properties guarantee $QCT_{G}$ in Eq. \eqref{eq:ft2qct} is a unitary.

\begin{figure}[!htpb]
    \centering
    \begin{quantikz}
        \lstick{$\ket{C}$}& \qwbundle{n_{conj}} & \qw  & \gate[3]{\hat{F}_G} & \qw & \qw  \rstick{$\ket{\Gamma}$} \\
        \lstick{$\ket{0}$}& \qwbundle{n_{dim}}   & \gate{H}  &  & \gate[2]{\hat{V}_{ent}} & \meter{} \rstick{$\ket{0}$} \\
        \lstick{$\ket{0}$}& \qwbundle{n_{dim}}   & \gate{H}  &  &  &  \meter{}  \rstick{$\ket{0}$}
    \end{quantikz}
    \caption{Group Fourier transform with entangled basis measurement .}
    \label{fig:ft2qct}
\end{figure}

\subsubsection{GQPE and the Effect of Projective Measurement in an Entangled Basis}

Motivated by the results of the previous section, a projector $\hat{\Pi}_{\Gamma,\Omega_{\Gamma}}=\ket{\Gamma,\Omega_{\Gamma}}\bra{\Gamma,\Omega_{\Gamma}}$ can be defined. The measurement of this observable in the state $\ket{\Phi}$ in Eq.~\eqref{GQPEAction} defines classical variables $\Gamma$ and $\Omega_{\Gamma}$ with a joint probability
\begin{align}
p_{\Gamma,\Omega_{\Gamma}}=\bra{\Psi}\hat{\Pi}_{\Gamma,\Omega_{\Gamma}}\ket{\Psi}=\frac{\bra{\psi^{\Gamma}}\psi^{\Gamma}\rangle}{d_{\Gamma}^2}
,
  \label{JointProb}
\end{align}
where $\ket{\psi^{\Gamma}}=\sum_i\ket{\psi^{\Gamma}_{i,i}}=\sum_i \hat{\mathbf{P}}^\Gamma_{ii} \ket{\psi}$
% is obtained by applying the projector $\hat{P}^{\Gamma}=\sum_i\hat{\mathbf{P}}^\Gamma_{ii}$ to $\ket{\psi}$, 
with $\hat{\mathbf{P}}^\Gamma_{ii} $ defined in Eq. \eqref{projector-trivial-irrep}. 
% with $\hat{P}^{\Gamma}=\frac{d_{\Gamma}}{|G|} \sum_{g \in G} \chi^*_{\Gamma}(g) \rho(g)$. 
It follows that the state after this projective measurement is given by
\begin{align}
    \ket{\Psi_{\Gamma,\Omega_{\Gamma}}}=\frac{\hat{\Pi}_{\Gamma,\Omega_{\Gamma}}\ket{\Psi}}{\sqrt{p_{\Gamma,\Omega_{\Gamma}}}}=\frac{1}{d_{\Gamma}\sqrt{p_{\Gamma,\Omega_{\Gamma}}}}\ket{\Gamma,\Omega_{\Gamma}}\otimes\ket{\psi^{\Gamma}}.
  \label{JointProb-2}
\end{align}
Eq. \eqref{JointProb-2} means that after projective measurement under the entangled basis $\ket{\Omega_\Gamma}$, it is sufficient to label the system state $\ket{\psi^{\Gamma}}$ with only one register $\ket{\Gamma}$ that stores the irrep label $\Gamma$. The additional state $\ket{\Omega_\Gamma}$ is redundant in this case. This motivates us to study directly the transform between the irrep $\Gamma$ and its conjugate variable, the character $C$, without resorting to obtaining them from Fig. \ref{fig:ft2qct}.

% In the next sections, we will presented a detailed discussion of the characters, the conjugacy classes and the projectors, which will allow to define the Quantum Character Transform directly and symmetry-adapted quantum routines that exploit it.

%\subsection{Quantum Character Transform and Group Symmetry Transformations}
%\label{ssec:general}
%%%%%%%%%%%%%%%%%%%%%%%%%%%%%%%%%%%%%%%%%%%%%%%%%%%%%%%%%%%%%%%%%%%%%%%%
\subsection{Group Symmetry Projection from Quantum Character Transform}
\label{ssec:symm-projection-from-qct}

In this section,we derive quantum circuits for projection operators onto the subspace spanned by irrep $\Gamma$, by explointing the orthogonality relations of the characters $\chi_\Gamma$ directly without resorting to the irrep $\boldsymbol{D}^\Gamma$ itself in Eq. \eqref{projector-trivial-irrep}.

% Let $g$ be a group element with the representation matrix for irrep $\Gamma$ as 
% \begin{equation}
%     \boldsymbol{D}^{\Gamma}(g) =\begin{bmatrix} D_{11}^\Gamma(g)&\cdots & D_{1d_{\Gamma}}^\Gamma(g) \\
%     D_{d_{\Gamma}1}^\Gamma(g) & \cdots & D_{d_{\Gamma}d_{\Gamma}}^\Gamma(g) \end{bmatrix} .
% \end{equation}
% Then, its character is defined as the trace of this representation matrix
% \begin{equation}
%     \chi_{\Gamma}(g)=\sum_iD_{ii}^\Gamma(g) ,
% \label{eqn:char_orthog}
% \end{equation}
% which satisfies the orthogonality relation
% %The character $\chi_{\Gamma}(g)$ of a group element $g$ in the $\Gamma$-th irrep is the trace $\chi_{\Gamma}(g)=\sum_iD_{ii}^\Gamma(g)$ of the representation matrix $\boldsymbol{D}^{\Gamma}(g)$. The orthogonality relations for characters are
% \begin{equation}
% \frac{1}{|G|} \sum_{g \in G} \chi_{\Gamma}(g)\chi^*_{\Gamma'}(g) = \delta_{\Gamma,\Gamma'}
% \ ,
% \label{eqn:char_orthog}
% \end{equation}
% where $|G|$ is the order of the group, $\delta_{\Gamma,\Gamma'}$ is the Kronecker delta, and $*$ denotes complex conjugation. 

\subsubsection{Constructing Symmetry Projectors from Group Characters} 

Consider the following definition for the projection operator $\hat{P}^{\Gamma}$ into the $\Gamma$-th irrep 
%derived using the characters of the group elements. The goal is to construct an operator that projects any quantum state $\ket{\psi}$ into the subspace corresponding to the $\Gamma$-th irrep. The projection operator $\hat{P}^{\Gamma}$ for the $\Gamma$-th irrep can be defined as
\begin{equation}
    \hat{P}^{\Gamma} = \frac{d_{\Gamma}}{|G|} \sum_{g \in G} \chi_{\Gamma}(g) \hat{\rho}(g)
 ,
\label{eqn:projector_bavk}
\end{equation}
% where $d_{\Gamma}$ is the dimension of the $\Gamma$-th irrep and 
% $\rho(g)$ is a unitary operator representation acting on vector space $V$ corresponding to the group element $g$. Formally, $ \rho: G \to \text{GL}(V) $ is a mapping from a group $G$ to a general linear group on vector space $V$. 
% To see how this works, 
its action on an arbitrary quantum state $\ket{\psi}$ is
$\hat{P}^{\Gamma}\ket{\psi} = \frac{d_{\Gamma}}{|G|} \sum_{g \in G} \chi_{\Gamma}(g) \hat{\rho}(g)\ket{\psi} 
= a_\Gamma |\psi^\Gamma\rangle $.
Here, $a_\Gamma$ is the amplitude of the projection of $|\psi_\Gamma\rangle$ on $|\psi\rangle$ and $\hat{\rho}(g)$ acts on the Hilbert space of $\ket{\psi}$. 

% % Given a group $ G $ and an element $ g \in G $, the conjugacy class of $ g $ in $ G $ is the set $\mathcal{C}(g) = \{ hgh^{-1} \mid h \in G \}$. From now on, the conjugacy classes will be labeled by $C$ to simplify the notation. In the case of abelian groups, all the irreps are one-dimensional, and the number of conjugacy classes equals the number of group elements because all elements commute according to $hgh^{-1} = g$. 

For groups with structure where the conjugacy class has more than one element, Eq. \eqref{eqn:projector_bavk} can be factorized on conjugacy classes because the characters of all group elements in the same conjugacy class are equal
\begin{equation}
    \hat{P}^{\Gamma} = \frac{d_{\Gamma}}{|G|} \sum^{N_{\rm conj}}_C \chi_{\Gamma}(C) \left( \sum^{|C| }_{g \in C} \hat{\rho}(g) \right) ,
    \label{eqn:Conjugacy_factorised}
\end{equation} 
where $N_{\rm conj}$ is the number of conjugacy classes and $|C|$ is the number of group elements in a given conjugacy class $C$.
% Here $\chi_{\Gamma}(C)$ denotes the character for a given conjugacy class $C$. 
Despite the similarity of Eq. \eqref{eqn:Conjugacy_factorised} with \eqref{eqn:projector_bavk}, Eq. \eqref{eqn:Conjugacy_factorised} has the advantage that only the irrep and conjugacy class label need to be dealt with, without explicitly handling each dimension of an irrep (or each individual group element in the same conjugacy class). When a linear combination of projectors in Eq. \eqref{eqn:Conjugacy_factorised} is applied to a state $\ket{\psi}$, a superposition of states spanning each irrep $\Gamma$ is obtained.
\begin{equation}
    \sum_\Gamma \hat{P}^{\Gamma}\ket{\psi} = \sum_\Gamma a_{\Gamma} \ket{\psi^{\Gamma}}
    \ ,
\label{eqn:TGA_back}
\end{equation}
which allows coherent manipulation of all the irreps of a symmetry group simultaneously. Here, $a_\Gamma$ is again the amplitude of the projection of $|\psi\rangle$ onto $|\psi_\Gamma\rangle$. 

% In the next section, it will be shown how to achieve this on quantum computers by defining a new unitary transformation, the \emph{Quantum Character Transform}.

\subsubsection{Constructing Group Symmetry Transformations from QCT}
\label{sec:symmetry-QuantumChar-transform}

QCT as defined in Eq. \eqref{eq:qctBracketNotation} is a unitary matrix as guaranteed by the orthogonal relationship in Eq. \eqref{eqn:Conjugacychar_orthog}. This immediately means that QCT can be realized directly on a quantum computer without resorting to the measurement protocol in Fig. \ref{fig:ft2qct}! 
% In more detail, the \emph{Quantum Character Transform} (QCT) for a group $G$, which we denote $QCT_G$, can be written explicitly as
% \begin{equation}
%     \widehat{QCT}_G = \sum_{C}^{N_{\rm Conj}}\sum_{\Gamma}\frac{\sqrt{|C|}\cdot\chi_{\Gamma}(C)}{\sqrt{|G|}}\ket{C}\bra{\Gamma}
%     \ .
% \label{eq:qctBracketNotation}
% \end{equation}
% As a result, the matrix elements of the QCT are given by the expression $QCT_{(C,\Gamma)}=\bra{C} \widehat{QCT}_G\ket{\Gamma} = \frac{\sqrt{|C|}\cdot\chi_{\Gamma}(C)}{\sqrt{|G|}}$.
As a familiar example, the quantum Fourier transform is nothing but the QCT for the cyclic group $\mathbb{Z}_N$ of $N$ elements $\{e,g,g^2,\cdots,g^{N-1}\}$. As $\mathbb{Z}_N$ is an Abelian group, the number of group elements equals the number of conjugacy classes. 

The QCT can be combined with the following controlled group action to realize a quantum coherent superposition of all irrep sectors. Specifically, the controlled group action can be thought of as a $\select$ operator indexed on the conjugacy classes, with all the group elements of the conjugacy classes controlled on the conjugacy class index $\ket{C}\bra{C}$
\begin{equation}
\begin{split}
    \select[\tilde{\rho}_G]=&\sum_{C}^{N_{\rm conj}} |C\rangle \langle C | \otimes \frac{1}{|C|} \sum^{|C|}_{g \in C} \hat{\rho}(g) \\
    =& \sum_{C}^{N_{\rm conj}} |C\rangle \langle C | \otimes \widehat{\tilde{\rho}}(C)
    \ ,
    \\
\end{split}
\label{eqn:select_conj}
\end{equation}
where $N_{\rm conj}$ is the number of conjugacy classes for the group of interest, and the conjugacy classes are iterated over in the summation. To simplify the notation, from now on the upper limit $N_{\rm conj}$ of the sum over conjugacy classes will be dropped. For convenience, $\widehat{\tilde{\rho}}(C)=\frac{1}{|C|} \sum^{|C|}_{g \in C} \hat{\rho}(g)$ has been defined, which is an equally-weighted Linear Combination of Unitaries (LCU), with each unitary being a unitary representation of the group element $g$. If $|C| > 1$, this can be implemented with a single additional $\prepare$ register when mid-circuit measurement and reset are used. This is shown in Sec. \ref{sec:select}, with resource estimates for the general case. 

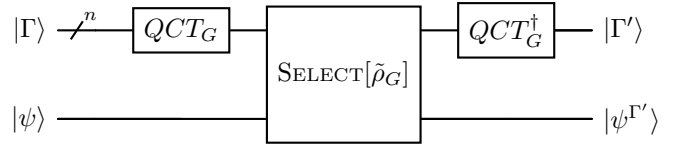
\begin{figure}[!htpb]
\centering
\begin{quantikz}
\lstick{$|\Gamma\rangle$}& \qwbundle{n} &\gate{QCT_G^\dagger} & \gate[2]{\select[\tilde{\rho}_G]}  & \gate{QCT_G}  & \qw  \rstick{$\ket{\Gamma'}$} \\
\lstick{$|\psi\rangle$} & \qw & \qw  &  &\qw  & \qw  \rstick{$|\psi^{\Gamma'} \rangle$} \\
\end{quantikz}
    \caption{Quantum circuit for $\hat{T}_{\rm GSA}$ showing symmetry transformation circuits of $|\psi\rangle$ under irrep $\Gamma$, where $QCT_G$ is the quantum character transform as defined in Eq. \eqref{eq:qctBracketNotation}. The square box is a multi-control that runs across all indexes on the $\Gamma$ register.}
\label{tgsa_circ}
\end{figure}

The Generalized Symmetry-Adapted Transform, $\hat{T}_{\rm GSA}$, is obtained by combining the QCT and the $\select$. The abstract quantum circuit that implements $\hat{T}_{\rm GSA}$ is shown in Fig. \ref{tgsa_circ}.
The $\hat{T}_{GSA}$ circuit acts on a quantum state $\ket{\Gamma}\otimes\ket{\psi}$ as follows
\begin{align}
&\hat{T}_{\rm GSA}\ket{\Gamma}\otimes\ket{\psi} = \sum_{\Gamma'}\sum^{N_{\rm conj}}_C \frac{|C|}{|G|} \chi^*_{\Gamma}(C)\chi_{\Gamma'}(C) \ket{\Gamma'}\otimes \widehat{\tilde{\rho}}(C) \ket{\psi} \nonumber \\
&= \sum_{\Gamma'} \frac{1}{|G|} \sum^{N_{\rm conj}}_C \chi^*_{\Gamma}(C)\chi_{\Gamma'}(C)  \cdot  \sum_{g \in C}^{|C|} \ket{\Gamma'}\otimes \hat{\rho}(g) \ket{\psi} . 
\label{eqn:TGA__back_cg}
\end{align}

This circuit formulation slightly modifies Eq. \eqref{eqn:TGA_back}. The character product can be expanded further using the Clebsch-Gordan relation $\chi^*_{\Gamma}(C)\chi_{\Gamma'}(C) = \sum_{\lambda} c^\lambda_{\Gamma \Gamma'} \chi_{\lambda}(C),\quad \forall C \in G$, where $c^\lambda_{\Gamma \Gamma'}$ are the Clebsch Gordan coefficients~\cite{gilmore1992clebsch}. When $\Gamma = 0$ (the trivial representation), $\chi_{0}(C) = 1, \forall C \in G$. Eq. \eqref{eqn:TGA__back_cg} simplifies to a rescaled a projector:
\begin{align}
\hat{T}_{\rm GSA}\ket{0}\otimes\ket{\psi} &= \sum_{\Gamma'} \frac{1}{|G|} \sum^{N_{\rm conj}}_C \chi_{\Gamma'}(C) \cdot  \sum^{|C|}_{g \in C} \ket{\Gamma'}\otimes \hat{\rho}(g) \ket{\psi} \nonumber \\
& = \sum_{\Gamma'} \frac{1}{d_{\Gamma'}}\ket{\Gamma'}\otimes\hat{P}^{\Gamma'}\ket{\psi} \nonumber \\
&= \sum_{\Gamma'}\frac{a_{\Gamma'}}{d_{\Gamma'}}\ket{\Gamma'}\otimes \ket{\psi^{\Gamma'}} ,
\label{eqn:TGA__back_dash}
\end{align}
where the post-selection probability to obtain a state in irrep $\Gamma'$ will be $|a_{\Gamma'} / d_{\Gamma'}|^2$, when the irrep register is measured. A derivation of the $\hat{T}_{\rm GSA}$ from the character projection is presented in Appendix~\ref{AppendixC}.  Therefore, the $\hat{T}_{\rm GSA}$ circuit rescales the projection by the dimension $d_\Gamma$ of the irrep $\Gamma$. If the input state on the ancilla register is a superposition of different irrep states, one should use $QCT^\dagger_G$. This can allow coherent control and simulation of states within individual irreps, where the number of required ancilla qubits to label a given irrep only scales logarithmically as the number of irreps increases. Such individual control on different irreps is desirable for applications such as quantum chemistry where a triplet electronic state will respond differently to a magnetic field than a singlet one; therefore, it may require different quantum subroutines to simulate. We leave further exploration of the advantages of $\hat{T}_{GSA}$ in various applications to future works.

However, if the ancilla starts from $\ket{0}^{\otimes n}$ such as when using the $\hat{T}_{\rm GSA}$ as a projector to a specific irrep $\Gamma$, a simpler $\prepare$ can be used as shown in Appendix \ref{app:prepare}, which can reduce the gate count significantly and also remove the $d_\Gamma$ by introducing a uniform renormalization factor. A similar circuit construction has been proposed in \cite{heredge2024nonunitaryquantummachinelearning}, with application to quantum machine learning.

To find the relevant symmetry transformation circuits, two fundamental operations are required: $\select[\tilde{\rho}_{G}]$ and $QCT$ for the relevant group. In particular, the controlled group actions and the $QCT$ are bespoke to each group arising from their structure. Some examples are presented in the next section.

%%%%%%%%%%%%%%%%%%%%%%%%%%%%%%%%%%%%%%%%%%%%%
\subsection{Single Finite Groups}
\label{ssec:single-group}

In this section, examples of the QCT for the cyclic group, symmetric group, and general finite groups will be discussed.

\subsubsection{Cyclic Group $\mathbb{Z}_{M}$}
\label{ssec:cyclic}

The QCT simplifies for the cyclic group $\mathbb{Z}_{M}$ of $M$ elements because the group is abelian; hence, the number of group elements equals the number of conjugacy classes and $|C| = 1$. In this case, each irrep $\Gamma = k$ is labeled by an integer $k = 0, 1, 2, \dots, M-1$ (mod $M$) and is one-dimensional, i.e. $d_k = 1$, owing to the abelian structure. Furthermore, each conjugacy class is labeled by $v$, where $v = 0, 1, 2, \dots, M-1$, corresponding directly to the group elements. The $k$th character of $\mathbb{Z}_M$, evaluated for $v$, is given by:

\begin{equation}
    \chi_{k}(v)=e^{2\pi\mathrm{i}kv/M} \ .
\end{equation}

As irreps of $\mathbb{Z}_M$ are one-dimensional, therefore the irreducible representations can be defined in terms of the characters.  

With all these elements at hand, from Eq.~\eqref{eq:qctBracketNotation} the expression for the QCT for $G=\mathbb{Z}_M$ is obtained.

\begin{align}
    QCT_{\mathbb{Z}_M}=\hat{U}_{\text{QFT}} =  
   \frac{1}{\sqrt{M}} \sum _{k\in \mathbb{Z}_M}
     \sum_{v\in \mathbb{Z}_M}  e^{2\pi\mathrm{i}kv/M} \ket{\bm{k}}\bra{\bm{v}}
    \label{Cyclicgroup-fourier}
    \ .
\end{align}
This expression is recognized as the quantum Fourier transform for an $m$-qubit system, provided that $M=2^m$ and by representing the group elements using the binary representation $\bm{v}=v_{1}v_{2}\cdots v_m$. More formally, the integer $v$ can be written as $v=v_1 2^{m-1}+v_2 2^{m-2} +v_m 2^0$. From now on, $v$ will denote integers (mod $M$), while $\bm{v}$ will denote their binary representation.

An interesting example is the case $M=2$, where the group Fourier transform acts on a single qubit and is given by a Hadamard gate $H$, as follows:
\begin{align}
    QCT_{\mathbb{Z}_2}=H=  
   \frac{1}{\sqrt{2}} \sum^1 _{a=0}
     \sum^1 _{b=0}  (-1)^{ab} \ket{a}\bra{b}
    \label{Cyclicgroup-fourier_1}
    \ .
\end{align}

\begin{figure}
\begin{tikzpicture}
\node[scale=0.8] {
\centering
    \begin{quantikz}
\lstick{$\ket{\bm{j}}$} & \qwbundle{} & \gate{\hat{T}} & \qw  \rstick{$\ket{\bm{j+1}}$}\\
\end{quantikz} \hspace{0.15cm}
$\Rightarrow$ \hspace{0.15cm}
\begin{quantikz}[wire types={q,n,q,q,q}]
 & \qw& \qw & \qw  & \qw & \targ{} & \qw \\
 & & \vdots &  &  \vdots &  & &  \\
\qw& \qw& \qw & \targ{}  & \qw & \ctrl{-2} & \qw \\
\qw& \qw & \targ{}  & \ctrl{-1} & \qw & \ctrl{-1} & \qw  \\
\qw& \targ{}  & \ctrl{-1} & \ctrl{-1} & \qw & \ctrl{-1} & \qw  \\
\end{quantikz}
};
\end{tikzpicture}\\
    \caption{Quantum circuit for incrementing an integer $\bm{j}=j_1j_2\cdots j_m$ by 1 to $\bm{j+1}$, where each qubit in the right diagram represents a binary digit $j_l$ of $\bm{j}$. The binary decomposition $j=j_1 2^{m-1}+j_2 2^{m-2} +j_m 2^0$ of the integer $j$ is used.}
        \label{fig:increment_circ}
\end{figure}

\begin{figure}[!htbp]   
\label{fig:uid}
\begin{quantikz}[row sep=0.2cm, column sep=0.2cm]
         & \octrl{1} & \octrl{1} & \ctrl{1} & \ctrl{1} & \qw\\
         & \octrl{1} & \ctrl{1} & \octrl{1} & \ctrl{1} & \qw \\
         & \gate[3]{\hat{T}^0} & \gate[3]{\hat{T}^1} & \gate[3]{\hat{T}^2} & \gate[3]{\hat{T}^3} & \qw\\
         & & & & & \qw\\
         & & & & & \qw
\end{quantikz}
= 
\begin{quantikz}[row sep=0.2cm, column sep=0.2cm]
         & \qw & \ctrl{2} & \qw\\
         & \ctrl{1} &  \qw & \qw \\
         & \gate[3]{\hat{T}^{2^0}} & \gate[3]{\hat{T}^{2^1}} & \qw\\
         & & & \qw\\
         & & & \qw
\end{quantikz}
\caption{Circuit identity for $
    \sum_{v=0}^{M-1} |\bm{v}\rangle\langle \bm{v}| \otimes \hat{T}^v$ showing how a linear combination of sequentially increasing powers of unitary $\hat{T}$ \cite{rosenkranz2024quantumstatepreparationmultivariate} can be efficiently factorized.}
\label{fig:factor_identity}
\end{figure}
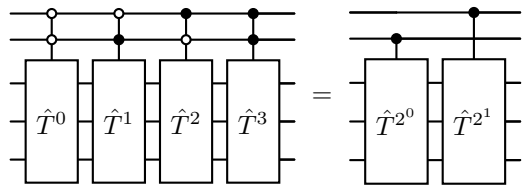

The next step is to construct the controlled group action $\select[\tilde{\rho}_{\mathbb{Z}_M}]$. In the case of the cyclic group, the group elements $\ket{\bm{v}}=\hat{T}^v\ket{\bm{0}}=\bigotimes^m_{l=1}\hat{T}^{(v_l 2^{m-l})}\ket{0_l}$ appearing in Eq.~\eqref{Cyclicgroup-fourier} are generated by powers of the cyclic shift operator $\hat{T}$ such that $\hat{T}\ket{\bm{j}}=\ket{\bm{j+1}}$ and $\hat{T}^M=\hat{1}$ with $\hat{1}$ being the identity operation. When acting on the state representation spanned by $\ket{\psi}$, this is the binary modular $+1$ increment operator. This is shown in Fig. \ref{fig:increment_circ}. The controlled group action is a controlled indexed power of the same operator. The powers of $\hat{T}$ are implemented by sequential applications of the controlled circuit box. The factorization identity of Fig. \ref{fig:factor_identity} can, therefore, be used to reduce the number of multi-controlled operators  
\begin{align}
    \select[\tilde{\rho}_{\mathbb{Z}_m}] &= \sum_{v=0}^{M-1} |\bm{v}\rangle\langle \bm{v}| \otimes \hat{T}^v \\
    &= \prod_{l=1}^{m} \left[ |0\rangle\langle 0|_{l} \otimes I^{\otimes m} + |1\rangle\langle 1|_{l} \otimes \hat{T}^{2^{(m-l)}}  \right]
    \label{Eq:MultiControlled}
    \ ,
\end{align}
where $v$ is the integer represented by the $m$-bit string $\bm{v}$ and $l$ labels the $l$th bit.

It is well known that the number of two-qubit gates in QFT scales as $O(m^2)$. For the cyclic group of $\mathbb{Z}_{M}$, there are $M$ irreps. Therefore, the irrep register, which the QFT acts on, will have $m=\log_2(M)$ qubits. This cost is assumed to be negligible, with the dominant cost being the powers of controlled increment boxes. This is displayed in Table \ref{tab:cyclic_resources}, where two compilation schemes are defined for the individually controlled group operation and the $\select$. 
%\textcolor{blue}{which for the cycle group is formed of the sequentially controlled powers of incrementers.}

\begin{table*}[hbpt!]
    \centering
    \begin{ruledtabular}
\begin{tabular}{r|rr|rrr|rrr}
  & & &  \multicolumn{3}{c|}{Controlled Increment$(+1)$}    & \multicolumn{3}{c}{$\select$} \\
\hline
Primitive  & Qubits & Ancilla & T & Toffoli  & Depth  & T & Toffoli  & Depth  \\
\hline
Incremeter\cite{khattar2024riseconditionallycleanancillae}  & $m+1$ & $\log_2(m)$ & $12(m+1)$ & $3(m+1)$  &$O(m+1)$ & $O(\exp(m))$ & $O(\exp(m))$  &$O(\exp(m))$      \\
Adder \cite{Gidney2018halvingcostof}  & $m$ & $m$ &$8m + O(1)$ & $4m$ & $O(m)$ & $8m^2 + O(1)$ & $4m^2$ & $O(m^2)$    \\
\end{tabular}
\end{ruledtabular}
\caption{Resource estimates for two different implementations of the $\select$ operation in the cyclic group projector and the controlled group action of a cyclic shift by binary increment $+1$ operation, for a cyclic group $\mathbb{Z}_M$ where $m=\log_2(M)$ is the number of ancilla qubits.}
\label{tab:cyclic_resources}
\end{table*}

\begin{figure}
    \centering
    \begin{tikzpicture}
    \node[scale=0.85] {
    \begin{quantikz}
    \lstick{$\ket{\Gamma}_0$} & \gate[3]{\hat{U}_{\text{QFT}}^\dagger}  & \qw & \qw & \ctrl{3} &\gate[3]{\hat{U}_{\text{QFT}}} & \qw \rstick{$\ket{\Gamma'}_0$}\\
    \lstick{$\ket{\Gamma}_1$}  &  & \qw & \ctrl{2} & \qw & &\qw \rstick{$\ket{\Gamma'}_1$}\\
    \lstick{$\ket{\Gamma}_2$} &  & \ctrl{1} & \qw & \qw  & & \qw \rstick{$\ket{\Gamma'}_2$} \\
    \lstick{$\ket{\psi}$} & \qwbundle{}  & \gate{\hat{T}^{2^0}}  & \gate{\hat{T}^{2^1}}  & \gate{\hat{T}^{2^2}} & \qw  & \qw \rstick{$\ket{\psi^{\Gamma'}}$}  \\
    \end{quantikz}
    };
    \end{tikzpicture}\\
    \caption{Symmetry group transform quantum circuit for the cyclic group $\mathbb{Z}_8$.}
    \label{fig:stm_proj_circ}
    \end{figure}
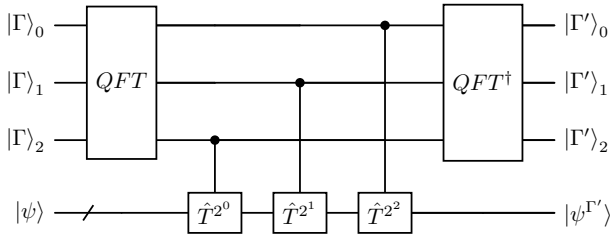

The naive compilation using the incrementer uses the compilation presented in Ref. \cite{khattar2024riseconditionallycleanancillae}, using $\log_2(m)$ conditionally clean qubits. The cost of the individual incrementation step is efficient; however, to apply the $\select$ operator, one must apply the controlled incrementer box $2^{m-1}$ times, as each $2^j$ controlled power requires $2^j$ individual sequential incrementation steps, which scales exponentially and is not suitable for large $m$. This can be improved using controlled addition circuits \cite{Gidney2018halvingcostof}, at the cost of an extra $m$ qubits. This is because modular adders can be used to implement the cyclic shift, where each controlled power of the cyclic shift can be implemented by adding mod-$2^j$ via the initialization of the extra addition register at a cost independent from the size of the shift. Putting these all together, Fig. \ref{fig:stm_proj_circ} displays an overall circuit for $T_{GSA}$ for the cyclic group $\mathbb{Z}_8$.

\subsubsection{Symmetric Group $S_N$}
\label{ssec:symmetric}

The Symmetric group is the group of $N!$ permutations of $N$ elements. The elements of the group can be represented in cycle notation. Each cycle can then be decomposed into a maximum of $N-1$ \emph{transpositions}, where a \emph{transposition} is a permutation of only two elements. There are $\frac{1}{2}[N(N-1)]$ unique transpositions which can generate the set of $N!$ permutations. Consider the permutation $g$ on the set $\{1, 2, 3, 4\}$ defined by $g = \begin{pmatrix} 1 & 2 & 3 & 4 \\ 2 & 3 & 4 & 1 \end{pmatrix}$. This permutation can be written in cycle notation as $
g = (1\ 2\ 3\ 4)$, which can be decomposed into a sequence of three transpositions $g= (1\ 4)(1\ 3)(1\ 2)$. Hence, the controlled group actions in the $\select$ can be formed as products of SWAP operations. 

The conjugacy classes $\{ C_j \}$ of the symmetric group are labelled by $ \lambda $, a partition of $ N $, $ \lambda = (\lambda_1, \lambda_2, \dots, \lambda_k) $ with $ \lambda_1 + \lambda_2 + \dots + \lambda_k = N $. For example, a permutation $g'$ of the symmetric group of 5 elements $S_5$ can be expressed efficiently with the notation $g' \equiv (1,3,5)(2,4)$. Such a permutation of a three-cycle and a two-cycle is said to have a $(3,2)$ cycle structure, which is a partition of $5$ elements. It is then readily seen that the conjugate of $g'$ by $g$ (for any $ g \in S_5$) is the element $
g g' g^{-1} = (g_1, g_3, g_5)(g_2, g_4)$, which has the same cycle structure as $ g' $ and is therefore in the same conjugacy class. Hence, the conjugacy classes can be indexed by all possible cycle structures, which themselves can be labelled by partitions. 

In addition to the conjugacy classes, the partitions $\lambda$ also label the irreps $\Gamma$, which can be shown through the use of Young symmetrisers and Young diagrams \cite{Fulton2004}. For small symmetric groups, the characters are tabulated, but one general formula for the corresponding irreducible character $ \chi_\Gamma $ of $ S_N $ can be expressed using the Frobenius character formula given in Appendix \ref{app:frob_char}.

As a simple example, the symmetry transformation circuits for $S_2$ can be constructed as follows. The group elements of $S_2$ are $\{e, (1,2)\}$, where $e$ is the identity element. Thus giving the group order of $|G|=2$. In the circuit implementation, the controlled group actions are simply an identity and a controlled swap gate on qubits $1$ and $2$. Here, the qubits are the objects being permuted. The characters of $S_2$ are well known as there are two irreps $\Gamma_1=(2,0)={\tiny\yng(2)}$ (totally symmetric) and $\Gamma_2=(1,1)={\tiny\yng(1,1)}$ (totally anti-symmetric). The characters are given by Table \ref{tab:S2_character_table}

\begin{table}[!htb]
\centering
\begin{tabular}{c|cc}
\hline
 & $ e $ & $ (12) $ \\
\hline
$ \chi_{(2,0)} $ & 1 & 1 \\
$ \chi_{(1,1)} $ & 1 & -1 \\ \hline
\end{tabular}
\caption{Character table for the symmetric group $ S_2 $.}
\label{tab:S2_character_table}
\end{table}

\begin{figure}[!htbp]
\centering
\begin{quantikz}
\lstick{$\ket{\Gamma}_0$}  & \gate{H} & \ctrl{1} &  \gate{H}  & \qw \rstick{$\ket{\Gamma'}_0$} \\
\lstick{$\ket{\psi}_0$}  & \qwbundle{m}  & \swap{1}  & \qw & \qw \rstick{$\ket{\psi^{\Gamma'}}_0$}  \\
\lstick{$\ket{\psi}_1$}  & \qwbundle{m}  &  \swap{0} & \qw & \qw \rstick{$\ket{\psi^{\Gamma'}}_1$}  \\
\end{quantikz}
\caption{$S_2$ symmetry projection circuit, acting on two system registers $\ket{\Psi}_0$ and $\ket{\Psi}_1$ of $m$ qubits each and one ancilla qubit $\ket{\Gamma}_0$. Comparing with Fig. \ref{tgsa_circ}, the $QCT$ is reduced to a Hadamard gate, and the $\select$ operation is implemented by an identity (no operation) and a controlled SWAP in the above. In this case, $T_{\rm GSA}$ of the group $S_2$ reduces to the familiar SWAP test on the two system registers. This circuit can be used to project an arbitrary initial state on two system registers to their symmetric and anti-symmetric sectors.}
\label{fig:s2_proj_circ}
\end{figure}
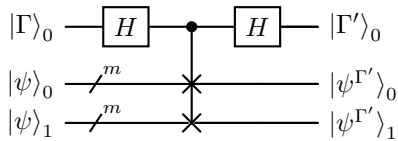

The Quantum Character Transform is therefore given by
\begin{equation}
    QCT_{S_2} = \frac{1}{\sqrt{2}}\begin{bmatrix} 1& 1 \\
    1 & -1 \end{bmatrix} = H,
    \label{eq:QCTS2}
\end{equation}
which is a simple Hadamard gate formed from the normalized rows Table~\ref{tab:S2_character_table}. Putting this all together in circuit form, we obtain the symmetry-adapted transformation as depicted in Fig. \ref{fig:s2_proj_circ}.

The first example of a non-abelian group is given by $G=S_3$ which has $|S_3|=3!=6$ elements. It is straightforward to calculate its conjugacy classes labeled by $C=1,2,3$. The first conjugacy class $[1]=\{e\}$ with size $|1|=1$ corresponds to the identity. The second $[2]=\{c_{(1,2)},c_{(1,3)},c_{(2,3)}\}$ with $|2|=3$ is composed by three transpositions, while $[3]=\{c_{(1,2,3)},c_{(1,3,2)}\}$ with size $|3|=2$ contains two 3-cycles.
The group $G=S_3$ has two one-dimensional irreps $\Gamma_1=(3,0)={\tiny\yng(3)}$ and  $\Gamma_2=(1,1,1)={\tiny\yng(1,1,1)}$, while the irrep $\Gamma_3=(2,1)={\tiny\yng(2,1)}$ is two-dimensional. These irreps are in direct correspondence to the conjugacy classes mentioned before.
The character table associated is given in Table~\ref{tab:S3_character_table}.
\begin{table}[!htb]
\centering
\begin{tabular}{c|ccc}
\hline
 & $e$ & $\rm Trans.$ & $ $3-\rm Cyc$. $\\
\hline
$\chi_{(3,0)} $ & 1 & 1& 1 \\
 $ \chi_{(1,1,1)} $ & 1 & -1& 1
 \\
 $ \chi_{(2,1)} $ & 2 & 0& -1\\ \hline
\end{tabular}
\caption{Character table for the symmetric group $ S_3 $.}
\label{tab:S3_character_table}
\end{table}
The Quantum Character Transform is then given by
\begin{equation}
    QCT_{S_3} = \frac{1}{\sqrt{6}}\begin{bmatrix} 1& \sqrt{3} & \sqrt{2} \\
    1 & -\sqrt{3} & \sqrt{2} \\
    2& 0 & -\sqrt{2}
    \end{bmatrix} .
    \label{eq:QCTS2}
\end{equation}
As a next step, the group action $\select[\tilde{\rho}_{S_3}]$ will be constructed. One basic building block of this operation is the LCU $\widehat{\tilde{\rho}}(C)=\frac{1}{|C|} \sum^{|C|}_{g \in C} \hat{\rho}(g)$. It is instructive to build this LCU for each conjugacy class labelled by $C=1,2,3$. In this case
\begin{align}
    \widehat{\tilde{\rho}}(1)&=\hat{1} \\
    \widehat{\tilde{\rho}}(2)&= \frac{1}{3}\left(\text{SWAP}_{1,2}+\text{SWAP}_{1,3}+\text{SWAP}_{2,3}\right)\\
    \widehat{\tilde{\rho}}(3)&= \frac{1}{2}\left(\text{SWAP}_{1,2}\text{SWAP}_{2,3}+\text{SWAP}_{1,3}\text{SWAP}_{3,2}\right)
    \label{Eq:S3GroupAction}
    \ .
\end{align}
In the last expression, the 3-cycles $c_{(1,2,3)}=c_{(1,2)}c_{(2,3)}$ and $c_{(1,3,2)}=c_{(1,3)}c_{(3,2)}$ are decomposed in terms of transpositions. On a quantum computer, the action of the transpositions is represented by a SWAP gate. The LCU defined above can be implemented using the quantum circuit in Fig.~\ref{fgr:select_c}. Later on in Sec.~\ref{ssec:Symmetrization} it will be shown how to use the $\hat{T}_{\rm GSA}$ transformation for $G=S_3$ to symmetrize 3-particle wave functions.

In the general case of $S_N$ for $N>2$, the symmetric group circuit for the QCT is more complicated due to the branching structure in the Bratelli diagram arising from the group subduction chain $S_N \supset S_{N-1} \supset \cdots \supset S_1$. For an efficient method to generate the QCT, a recursive method for the generation of characters based on this Bratelli diagram \cite{bratteli1972inductive,kawano2016quantum} will need to be used, as the Bratelli diagram encodes a factorization of the internal subspaces. Possible candidates could be the Murnaghan–Nakayama rule or the more recent method proposed by Holmes \cite{holmes2017recursionformulairreduciblecharacters}. However, an efficient circuit construction for the full Symmetric group Fourier transform is known as shown in Appendix \ref{app:sym_fourier}, from which it is known that the characters could be obtained from by tracing over the irreps and factorizing on the conjugacy class. Efficient quantum algorithms to calculate the characters have also been presented\cite{bravyi2025classicalquantumalgorithmscharacters}, which could be adapted to build a QCT for the symmetric group.

A possible method for the general construction of the $\select$ primitive could use the fact that the symmetric group for $N!$ group elements can be factorized into $\frac{1}{2}N(N-1)$ transpositions. A quantum circuit has been presented to generate the indexed set of $N!$ group elements from the transpositions using controlled SWAP gates by Laborde et al. \cite{laborde2024quantumalgorithmsrealizingsymmetric}. However, it is unclear how this strategy could be combined with the methodology presented here, which needs to group the conjugacy classes containing $|C|$ group elements under a common index governed by the common cycle structure of the permutations. To form an efficient control structure indexed on the conjugacy classes generated from the group generators, one must know how the conjugacy classes are produced from the generators and encode that information as efficient quantum circuits.

\subsubsection{General Case for Finite Groups}
\label{sec:select}

As long as the group is finite, the group action on the space of $|\psi\rangle$ is known, and the quantum character transform can be formed, the approach presented in Fig. \ref{tgsa_circ} can be used in general. In this section, some general strategies for constructing the QCT and the controlled group action will be presented.

\textbf{Quantum Character Transform.}
The QCT is a square matrix containing the normalized character table of the group and 1 on the rest of the diagonal for the indices greater than the number of conjugacy classes $N_{\rm Conj}$. Often the character tables are known or can be calculated efficiently classically, scaling sub-exponentially with the size of the group. Even for the $O(N!)$ scaling symmetric group of $N$ elements, the number of conjugacy classes scales approximately:  $\frac{1}{4N \sqrt{3}} \exp\left( \pi \sqrt{\frac{2N}{3}} \right)$ as $N \to \infty
$ \cite{Fulton2004}. Therefore, it is not unquestionable that generating a QCT circuit could be achieved in a brute force manner by combining classically-generated character tables and recent algorithms such as the fast Hadamard transform\cite{HadmardTransform}. However, it is likely that the optimal strategy is a recursive structure encoded into the circuit, such as generating the characters recursively from the branching of the sub-group chain of $S_N$ via the method of \cite{holmes2017recursionformulairreduciblecharacters}.

\textbf{SELECT Circuits.}
\begin{figure*}[htbp!]
\centering
\begin{tikzpicture}
\node[scale=1] {
\begin{quantikz}
\lstick{$|C\rangle$} &  \gate[2]{\select[\tilde{\rho}_G]} & \qw \\
\lstick{$|\psi\rangle$}  &   &\qw \\ 
\end{quantikz}
\hspace{0.25cm}
=
\begin{quantikz}[wire types={q,n,q,q,q}]%
\lstick{$|c_n\rangle$} &  \octrl{1} & \octrl{1} & \octrl{1} & \octrl{1} & \ldots\\
\lstick{$\vdots$} & \vdots & \vdots  & \vdots  & \vdots  &   \\ 
\lstick{$|c_2\rangle$} & \octrl{1}\wire[u]{q} & \octrl{1}\wire[u]{q} & \ctrl{1}\wire[u]{q} & \ctrl{1}\wire[u]{q} &  \ldots \\
\lstick{$|c_1\rangle$} & \octrl{1} & \ctrl{1} & \octrl{1} & \ctrl{1} & \ \ldots\\ 
\lstick{$|\psi\rangle$} & \gate{\widehat{\tilde{\rho}}\left({C} \right)}  &  \gate{\widehat{\tilde{\rho}}\left(C' \right)}  &  \gate{\widehat{\tilde{\rho}}\left(C''\right)}  &  \gate{\widehat{\tilde{\rho}}\left(C''' \right)}  &  \ \ldots\ \\ 
\end{quantikz}
};
\end{tikzpicture}
\caption{$\select$ circuit implementing $\sum^{N_{\rm conj}}_{C} |C\rangle \langle C | \cdot \widehat{\tilde{\rho}}(C)$ showing the first 4 terms of the controlled conjugacy classes of group actions acting on the space of $|\psi\rangle$ indexed on the conjugacy class $C$. %\YL{@Victor: Please check index} 
The ancilla qubits are labeled as $\ket{c_1}, \ket{c_2}, \cdots, \ket{c_{n_{conj}}}$, where $n_{conj} = \lceil\log_2(N_{\rm conj})\rceil$ for a bit representation $c_1 c_2 \cdots c_n$ of the integer $C$ labeling the conjugacy classes. $n_{conj}$ is the number of qubits on the conjugacy class index register. $\widehat{\tilde{\rho}}(C)$ stands for the LCU acting on the system wave function $\ket{\psi}$ corresponding to the conjugacy class $C$.}
\label{fgr:select_rho}
\end{figure*}
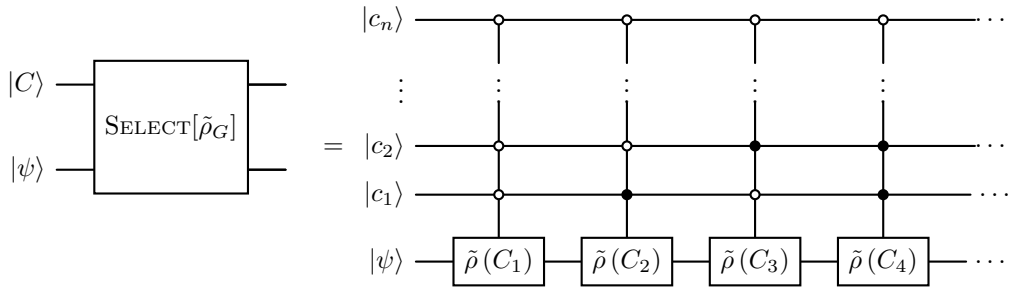
The quantum circuits for implementing the $\select$ oracle applying the group actions grouped into their conjugacy classes acting on a state $|\psi\rangle$ indexed on a control register are presented in this section. The circuits implementing Eq.\eqref{eqn:select_conj} are shown in Figs.~\ref{fgr:select_c} and~\ref{fgr:select_rho}. The sum over the conjugacy classes  $\sum_{C}^{N_{conj}} |C\rangle \langle C | \cdot \widehat{\tilde{\rho}}(C)$ is implemented by a series of multi-controlled operations which index the $N_{\rm Conj}$ conjugacy classes on the register $|C\rangle$. This can be seen in Fig.~\ref{fgr:select_rho} with a general formulation. Each $\widehat{\tilde{\rho}}(C) = \frac{1}{|C|} \sum^{|C|}_{g \in C} \hat{\rho}(g)$ needs to be implemented as a nested Linear Combination of Unitaries (LCU) operation \cite{childs2012hamiltonian} to account for the action of all group elements in $C$. This is an equal superposition of group elements in the conjugacy class $C$. This is implemented with a unitary state preparation acting on the group element register, where the state is encoded in the first column of the following unitary: 
%It is implemented with a state preparation unitary acting on the group element register where the state is encoded in the first column of the unitary

\begin{equation}
%\begin{split}
  \prep\left(\frac{1}{\sqrt{|C|}}\right) = 
  \sum_{g \in C}^{|C|} \frac{1}{\sqrt{|C|}} |g\rangle\langle 0 | + \hat{\Pi}_\perp
  = \begin{bmatrix}
    \frac{1}{\sqrt{|C|}} & \cdot & \hdots \\ 
    \frac{1}{\sqrt{|C|}} & \cdot & \hdots  \\ 
    \vdots &  \ddots & \hdots   \\ 
    \frac{1}{\sqrt{|C|}}  & \cdot  &  \hdots
    \end{bmatrix}
%  \end{split}
\label{equation:prepare}
\end{equation}
Here, $\hat{\Pi}_\perp$ is a projector to the orthogonal space spanned by the column vectors other than the first column. Additionally, the subscript $j$ in $C_j$ has been dropped, and $C$ is used instead, since \eqref{equation:prepare} works for any conjugacy class. The same notation will be used in the following. This equal superposition can be implemented efficiently using the approach presented in Ref. \cite{Lineartgate}. Mid-circuit measurement and reset can be used, which will only ever result in the use of $\lceil\log_2(\max(C))\rceil$ extra ancilla qubits as shown in Fig. \ref{fgr:select_c}.

%where $\Pi_\perp$ is a projector to the orthogonal space spanned by the column vectors other than the first column, and we have dropped in the subscript $j$ in $C_j$ and use $C$, since \eqref{equation:prepare} works for any conjugacy class. 

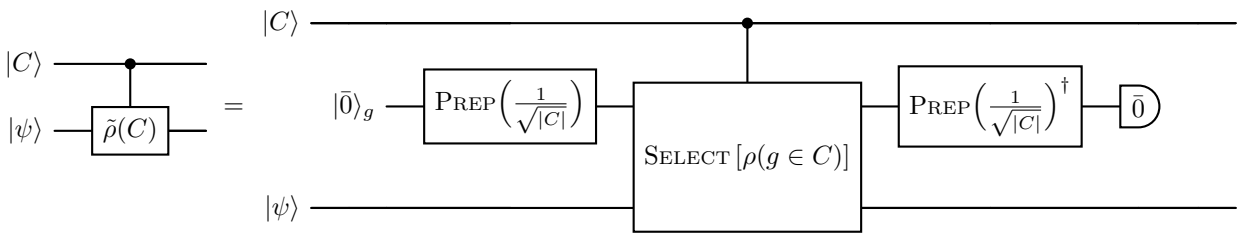
\begin{figure*}[!htpb]
\centering
\begin{tikzpicture}
\node[scale=1] {
\begin{quantikz}
\lstick{$|C\rangle$}& \ctrl{1}&   \qw \\
\lstick{$|\psi\rangle$}  & \gate{\widehat{\tilde{\rho}}(C)} & \qw
\end{quantikz}
=
\begin{quantikz}[wire types={q,n,q}]
\lstick{$|C\rangle$}&\qw& \qw& \qw & \ctrl{1} & \qw  & \qw &\qw&\qw  \\
& & \lstick{$|\bar{0}\rangle_{g}$}  & \gate{\prep\Big(\frac{1}{\sqrt{|C|}}\Big)} \setwiretype{q} &\gate[2]{\select\left[\rho(g \in C)\right]}   & \gate{\prep\Big(\frac{1}{\sqrt{|C|}}\Big)^\dagger} &\meterD{\bar{0}}\\
\lstick{$|\psi\rangle$} & \qw& \qw & \qw & & \qw & \qw & \qw& \qw\\
\end{quantikz}
};
\end{tikzpicture}
\caption{Circuit showing the controlled LCU implementing $\widehat{\tilde{\rho}}(C) = \frac{1}{|C|} \sum^{|C|}_{g \in C} \hat{\rho}(g)$ which is an equal superposition of representations of group elements within a conjugacy class $g\in C$. The equal superposition over $\frac{1}{|C|}$ is implemented via the two $\prepare$ oracles acting on the group element register $g$. The summation $\sum^{|C|}_{g \in C} \hat{\rho}(g)$ is implemented by the $\select$ oracle. Mid-circuit measurement, selecting the $\bar{0}$ results and reset of the group element register is used for efficient qubit count.}
\label{fgr:select_c}
\end{figure*}

 The nested $\select$ operations can be implemented with unary iteration \cite{Lineartgate}. The upper bound of indexed operations is $D = N_{\rm Conj} \cdot \max(|C|)$ with  $O(4D-4)$ $T$ gates, $O(D-1)$ Toffolis, at the addition of $O(\log(N_{\rm Conj}) + \log(\max(|C|) - 1)$ ancilla qubits. $N_{\rm Conj}$ is the number of conjugacy classes, and $\max(|C|)$ is the size of the largest conjugacy class. The $\select$ operation, combined with the controlled group action $\hat{\rho}(g)$ itself, is expected to be the dominant cost of the algorithm. This assumption can be made because the $QCT$ acts on $\lceil\log_2(\max(C))\rceil$ qubits with the size of the conjugacy class known to scale sub exponentially even for the $N!$ scaling of the symmetric group $S_N$ \cite{Fulton2004}. To prepare the equal superposition across all conjugacy class elements, we require a linear number of qubits. The cost of the controlled group action will vary depending on the relevant group.

\subsection{Direct Product Groups}
\label{ssec:product-groups}

\subsubsection{General Discussion}
\begin{figure*}[!htpb]
\centering
\begin{tikzpicture}
\node[scale=1] {
\begin{quantikz}
    \lstick{$|\Gamma_{G'}\rangle$}& \qwbundle{m_{G'}} &\gate{QCT^\dagger_{G'}} & \qw & \gate[3]{\select[\tilde{\rho}_{G'}]}  & \gate{QCT_{G'}}  & \qw  \textbf{}\rstick{$\ket{\Gamma'_{G'}}$} \\
\lstick{$|\Gamma_G\rangle$}& \qwbundle{m_G} &\gate{QCT^\dagger_{G}} & \gate[2]{\select[\tilde{\rho}_G]}  &  & \gate{QCT_{G}}  & \qw  \rstick{$\ket{\Gamma'_G}$} \\
\lstick{$|\psi\rangle$} & \qw & \qw  & & & \qw  & \qw \rstick{$|\psi^{\Gamma_G\Gamma_{G'}} \rangle$} \\
\end{quantikz}
};
\end{tikzpicture}
\caption{Quantum circuit for $\hat{T}_{GSA}$ showing symmetry projection circuits of $|\psi\rangle$ onto the dual irrep $(\Gamma_G,\Gamma_{G'})$, where QCT is the quantum character transform as defined in Eq. \eqref{eq:qctBracketNotation}. The $\select[\tilde{\rho}_{G'}]$ acts over the first and third wire only.}
\label{fig:tgsa_circ_prod}
\end{figure*}
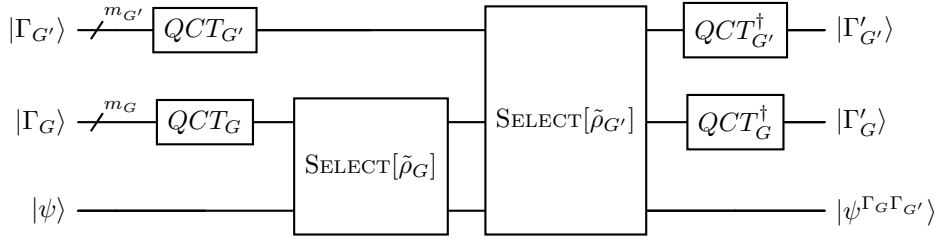

Direct product groups serve as a mathematical tool for investigating physical systems containing multiple symmetries. In this section, the direct product of groups of the form $G \times G' = \{ (g, g') \mid g \in G, g' \in G' \}$ will be implemented as a $\hat{T}_{GSA}$. Each element of $G \times G'$ is a pair $(g, g')$, and the group operation on $G \times G'$ is defined component-wise. This means that if $(g_1, g'_1)$ and $(g_2, g'_2)$ are elements of $G \times G'$, their product is given by $(g_1, g'_1) \cdot (g_2, g'_2) = (g_1 \cdot g_2, g'_1 \cdot g'_2)$, where $\cdot$ denotes the respective group operations in $G$ and $G'$. As representations of these groups acting on finite Hilbert spaces are considered, their respective representations can be defined as $\rho_G: G \to \text{GL}(V)$ and $\rho_{G'}: G' \to \text{GL}(W)$. The group $G \times G'$ therefore acts on the tensor product space $V \otimes W$ by $[\rho_{G}(g) \otimes \rho_{G'}(g')](v \otimes w) = \rho_G(g)(v) \otimes \rho_{G'}(g')(w)$, where $v \in V$, and $w \in W$.
As a first step to define the transformation $\hat{T}_{\text{GSA}}$ for a direct product of groups, $m=m_g+m_{g'}$ qubits need to be considered in the quantum circuit of Fig. ~\ref{fig:tgsa_circ_prod} to encode the group elements $g\in G\times G'$. Due to the direct product structure of the group, the quantum character transform is naturally factorized as a tensor product of the individual character transforms
\begin{equation}
   QCT_{G \times G'}=QCT_{G}\otimes QCT_{G'} .
\end{equation}
The direct product of controlled group actions can be defined by
\begin{align}
    \select[&\tilde{\rho}_G  \cdot \tilde{\rho}_{G'}]:= \left(\sum_{C_G} |C_G\rangle \langle C_G | \otimes \widehat{\tilde{\rho}}(C_G) \right) \nonumber \\
    \otimes& \left(\sum_{C_{G'}} |C_{G'}\rangle \langle C_{G'} | \otimes \widehat{\tilde{\rho}}(C_{G'}) \right)
    \ .
\end{align}
As both the Fourier and controlled group action factorize, the symmetry-adapted transformation takes the form $\hat{T}_{GSA}=\hat{T}^{(G)}_{GSA}\otimes \hat{T}^{(G')}_{GSA}$. The notation $\widehat{\tilde{\rho}}(C_G)=\frac{1}{|C_G|} \sum_{g \in C_G} \hat{\rho}(g)$ and $ \widehat{\tilde{\rho}}(C_{G'})=\frac{1}{|C_{G'}|} \sum_{g' \in C_{G'}} \hat{\rho}(g')$ is used as they are representations acting on the same space. Note that the notation $\tilde{\rho}_G  \cdot \tilde{\rho}_{G'}$ is used in the argument of the select operation because $\hat{\rho}(g)$ and $\hat{\rho}(g')$ act on the same vector space containing $|\psi\rangle$ and cannot be represented as an explicit tensor product in a circuit implementation. Its action on a quantum state is given by
\begin{equation}
    \begin{split}
    \hat{T}_{GSA}\ket{0,0}\otimes\ket{\psi} &= \sum_{\Gamma_G \Gamma_{G'}}  \ket{\Gamma_G,\Gamma_{G'}}\otimes\frac{\hat{P}_{G}^{\Gamma_G}\hat{P}_{G'}^{\Gamma_{G'}}}{d_{\Gamma_G}d_{\Gamma_{G'}}}\ket{\psi} \\
    &= \sum_{\Gamma_G \Gamma_{G'} }\frac{a_{\Gamma_G}}{d_{\Gamma_G}} \frac{a_{\Gamma_{G'}}}{d_{\Gamma_{G'}}} \ket{\Gamma_G,\Gamma_{G'}}\otimes\ket{\psi^{\Gamma_G \Gamma_{G'}}}.
    \end{split}
    \label{eqn:TGA__back_dash_1}
\end{equation}
The abstract quantum circuit implementing the $T_{GSA}$ circuit is shown in Fig. \ref{fig:tgsa_circ_prod}.

When working with multiple groups acting on the same vector space, their representations may not commute with each other $ \rho_{G}(g) \rho_{G'}(g') \neq \rho_{G'}(g')  \rho_{G}(g) $. Despite this non-commutativity, the products of their respective projectors still correctly project onto the subspaces corresponding to their irreps, $\hat{P}_{G'}^{\Gamma_{G'}} \hat{P}_{G}^{\Gamma_G} \ket{\psi} = \hat{P}_{G}^{\Gamma_G} \hat{P}_{G'}^{\Gamma_{G'}}\ket{\psi} =  |\psi^{\Gamma_G} \rangle \cap |\psi^{\Gamma_{G'}} \rangle$, because this projects onto the intersection of two symmetric sup spaces spanning the irreps $\Gamma_G$ and $\Gamma_{G'}$

\subsubsection{Example on $\mathbb{Z}_N \times \mathbb{Z}_2$}
In the previous section, the general framework of the symmetry-adapted transformation for direct products of groups was discussed. In this section, a concrete example for the direct product group $\mathbb{Z}_N \times \mathbb{Z}_2$ will be provided.

For convenience, a representation of the Hilbert state basis using binary strings of length $N$ will be used. From now on, $\ket{\bm{x}}$ will be used such that $Z_l\ket{\bm{x}}=x_l\ket{\bm{x}}$ with $x=0,1,2,\dots 2^{N}-1$ to denote the basis and its binary representation $\bm{x}=x_{1}x_{2}\cdots x_N$.
Next, the group elements of the commuting pair of groups $G=\mathbb{Z}_N$ and $G'=\mathbb{Z}_2$ will be defined. The representation $\rho(G)=\{ \hat{1},\hat{T},\hat{T}^2,\dots, \hat{T}^{N-1}\}$ is generated by powers of the operator $\hat{T}$ such that $\hat{T}^N=\hat{1}$. A representation will be used such that powers of the $\hat{T}$ act as binary circular increment shifts. For example, $\hat{T}^n\ket{\bm{x}}=\ket{\bm{y}}$ is a shift of the digit $x$ to the left giving $y=2^n x$.
Similarly, the representation $\hat{\rho}(G')=\{ \hat{1},\hat{Q}\}$ is generated by powers of the operator $\hat{Q}$ satisfying $\hat{Q}^2=\hat{1}$. Its action on the quantum state is defined as $\hat{Q}\ket{\bm{x}}=\ket{\bar{\bm{x}}}$, which is nothing but a binary NOT operation on each bit of $\bm{x}$.   

As the group $\mathbb{Z}_N \times \mathbb{Z}_2$ has a direct product structure, the corresponding group quantum character transform $QCT_{G\times G'} = \hat{U}_{\text{QFT}} \otimes H$ is nothing but a product of the quantum Fourier transformation $\hat{U}_{\text{QFT}}$ associated with the group $\mathbb{Z}_N$ and a Hadamard gate $H$, being the group Fourier transform of $\mathbb{Z}_2$. In this case, the symmetry-adapted transformation reads
\begin{align}
    \hat{T}_{\text{GSA}} =  H\hat{U}^{-1}_{\text{QFT}} \otimes \select[\tilde{\rho}_{\mathbb{Z}_N \times \mathbb{Z}_2}] \otimes \hat{U}_{\text{QFT}}H 
    \label{eq:ExplicitTGSA}
    \ ,
\end{align}
with the controlled group action defined as $\select[\tilde{\rho}_{\mathbb{Z}_N} \otimes \tilde{\rho}_{\mathbb{Z}_2)}]$. In this example, $\hat{\rho}(gg')=\hat{\rho}_1(g)\otimes \hat{\rho}_2(g')$ is a one-dimensional representation of the group element $n\otimes \mu\in \mathbb{Z}_N\otimes \mathbb{Z}_2$ acting as $\hat{\rho}(gg')\ket{\bm{x}}=\ket{\bm{y}'}$ where $\bm{x}=x_{1}x_{2}\cdots x_N$, $\bar{\bm{x}}=\bar{x}_{1}\bar{x}_{2}\cdots \bar{x}_N$ and $\bar{x}_{l}=1-x_{l}$. We use $\bm{y}'$ to denote the binary representation of $y'=2^n(\bar{x}_{1} 2^{m-1}+\bar{x}_{2} 2^{m-2} +\bar{x}_{N} 2^0)$. 

With these elements at hand, the explicit action of the symmetry-adapted transformation in Eq.\eqref{eq:ExplicitTGSA} can be calculated as follows:
\begin{align}
    \hat{T}_{\text{GSA}}\ket{0} \ket{\psi} =  \sum _{k\in \mathbb{Z}_N} \sum _{\sigma\in \mathbb{Z}_2}\ket{\bm{k},\sigma} \ket{\psi^{k,\sigma}}
    \label{schur-Z_N-irrep}
    \ ,
\end{align}
where $k$ and $\sigma$ label the irreps of $\mathbb{Z}_N$ and $\mathbb{Z}_2$, respectively. The projection operator
\begin{align}
    \hat{P}^{k,\sigma}=\frac{1}{2N} \sum_{\mu\in \mathbb{Z}_2}\sum_{n\in \mathbb{Z}_N}  (-1)^{\mu\sigma}e^{\frac{2\pi\mathrm{i}kn}{N}} \hat{T}^n\hat{Q}^{\mu}
    \label{IsingWignerOperator}
\end{align}
The symmetry-adapted state $\ket{\psi^{k,\sigma}} = \hat{P}^{k,\sigma}\ket{\psi}$ is defined to project the state into the irreps labeled by $k$ and $\sigma$. Importantly, as discussed above in Eq~\eqref{eqn:TGA__back_dash_1}, the projector itself factorizes as $\hat{P}^{k,\sigma} =\hat{R}_{k}\hat{Q}_{\sigma}$, where $\hat{Q}_{\sigma}=(\hat{1}+(-1)^{\sigma}\hat{Q})/2$ and $\hat{R}_{k}=1/N\sum_{n\in \mathbb{Z}_N} e^{\frac{2\pi\mathrm{i}kn}{N}} \hat{T}^{n}$ are projection operators into the symmetric subspaces labeled by $\sigma$ and $k$. In Sec. \ref{ssec:3d-h2}, an example will be provided on how to use this representation of the transformation $T_{\text{GSA}}$ to select from fermionic and bosonic, singlet and triplet on an $H_2$ system.
%%%%%%%%%%%%%%%%%%%%%%%%%%%%%%
%
\section{Examples and Applications}
\label{sec:example-application} 
So far, a framework for symmetry-adapted transformation for general groups has been presented. In this section, concrete examples are provided to illustrate the power of these unified symmetry-adapted transformations for quantum simulations. Sec. \ref{sec:symmetry-adapted-algorithms} starts with an example of how to use the symmetry transformation with many standard quantum simulation routines such as the quantum phase estimation algorithm, to determine the spectrum of a many-body system when it is restricted to a given irrep of interest. Sec. \ref{ssec:1d-ising} discusses the cyclic group with applications to a one-dimensional Ising model. Sec. \ref{ssec:2d-hh} expands consideration to a two-dimensional Harper-Hofstadter model with a pair of commuting cyclic groups. Sec. \ref{ssec:3d-h2} further expands the scope to three-dimensional \emph{ab initio} quantum chemistry problems with permutation groups where H$_2$ is used as a specific example. An additional example of para-statistics with a non-Abelian finite group is given in Sec. \ref{ssec:Symmetrization}.

\subsection{Symmetry-Adapted Quantum Subroutines}
\label{sec:symmetry-adapted-algorithms}

\begin{figure}
\centering
\begin{quantikz}
\lstick{$\ket{0}$} & \qwbundle{m} & \gate[2]{\hat{T}_{\rm GSA}} & \qw & \meter{}\\
\lstick{$\ket{\psi}$} & \qwbundle{} & \qw & \gate[2]{\rm \hat{U}_{\text{QPE}}} & \qw  \\
\lstick{$\ket{0}$} & \qwbundle{n} & \qw & \qw & \meter{}\\
\end{quantikz}\\
\caption{Circuit for symmetry-adapted quantum phase estimation, where the first two registers carry information of the irreps and the quantum state $\psi$ of our many-body system, while the last register with $n$ qubits is the ancilla used for QPE. The input state $\psi$ is first decomposed into a coherent superposition of all irreps by $\hat{T}_{\rm GSA}$, and then a QPE or any other quantum subroutine can be applied to transform all irrep sectors of the quantum state simultaneously. The eigen energies and the irreps of quantum states will be stored in the first and the last register, where measurement (or amplitude amplification, not shown) can be performed to estimate the energies of states within a desired irrep.}
\label{Fig2:tgsa+qpe}
\end{figure}
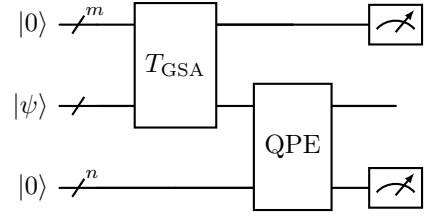

To demonstrate how $T_{GSA}$ can be combined with quantum phase estimation (QPE), a quantum circuit with three registers is defined in Fig. \ref{Fig2:tgsa+qpe}. The first two registers carry information of the irreps and the quantum state of the many-body system, while the last register with $n$ qubits is the ancilla used for QPE. The symmetry-adapted quantum phase estimation is defined as $\hat{V}_{\rm SQPE}=\hat{U}_{\text{QPE}}\hat{T}_{\text{GSA}}$. This unitary performs a symmetry-controlled quantum phase estimation for a desired irrep, as follows:
\begin{align}
       &\hat{V}_{\rm SQPE}|0 \rangle^m \ket{\psi}|0 \rangle^n 
       \nonumber\\&= %\sum_{u,y=0}^{2^n - 1}
       \sum^{2^{n}-1}_{u,v}\sum_{\Gamma} \frac{e^{-i2\pi v (E_{\Gamma} - u/2^n)}}{2^n\sqrt{d_\Gamma}}  \ket{\Gamma}\ket{\psi^\Gamma}  \ket{u},
    \label{eq:qpe}
\end{align}
for $d_\Gamma$ being the dimension of the irrep $\Gamma$.
This circuit allows the determination of the energy $E_{\Gamma}$ when the system is restricted to a desired irrep $\Gamma$. After the first and last registers are measured, the quantum state $\ket{\psi}$ is automatically projected to a given irrep, with the projection probability to each irrep $\Gamma$ determined by the overlap between $\ket{\psi}$ and the state within an irrep $\Gamma$. This approach has many potential applications in diverse fields, ranging from quantum simulation to quantum chemistry. In the next sections, three examples will be used to demonstrate how this approach works for abelian and non-abelian groups.

%\begin{figure}[htbp]
%            \begin{center}
%            \mbox{
%                \Qcircuit @C=1.0em @R=1.5em {
%                |0 \rangle^m \quad\quad    & \qw    /  & \multigate{1}{T_{\rm GSA}} & \qw    
                %\ctrl{2} & \qw & \multigate{1}{T_{\rm GSA}^\dagger}
%                & \meter    \\
                %|\psi \rangle \quad\quad     & \qw /    & \ghost{T_{\rm %GSA}} & \multigate{1}{\rm QPE} & \qw   \\
                %|0 \rangle^n \quad\quad & \qw / & \qw &\ghost{\rm QPE} & \meter 
                %}
           % }
            %\end{center}
            %\caption{Circuit for symmetry-adapted quantum phase estimation.}
            %\label{Fig2:tgsa+qpe}
%\end{figure}

\subsection{One-dimensional Ising model in random transverse and longitudinal fields}
\label{ssec:1d-ising}
% \subsection{Generalized quantum phase estimation: Application to the Ising model in a random transverse and longitudinal fields}
In this section, a simple example of a problem in condensed matter physics related to the irreps of the direct product group $G=\mathbb{Z}_N \times \mathbb{Z}_2$ will be considered, where $\mathbb{Z}_N$ is a cyclic group of order $N=2^n$. 
%

%
%%%%
\subsubsection{The one-dimensional quantum Ising model in random transversal and longitudinal fields} %%%%
% In the previous section we discussed in general the group Fourier transform for the group $G=\mathbb{Z}_N \times \mathbb{Z}_2$. In this section, we will investigate a particular example to show the versatility of generalised quantum phase estimation and its relation to symmetries of a physical system. With this goal in mind, we 

Consider the one-dimensional quantum Ising model in a random transverse and longitudinal fields~\cite{petHo2022random,Berkovits2022}
%%%
\begin{align}
          \label{eq:IsingModel}
\hat{ \mathcal{H}}&= -\sum^{N}_{j=1}a_j X_j- J\sum^{N}_{j=1}Z_j Z_{j+1}-\sum^{N}_{j=1}w_j Z_j
\ ,
\end{align}
%%%
where $X_j, Y_j$ and $Z_j$ are Pauli matrices at the $i$-th site of the chain while $J$ denotes the coupling strength. The transverse $a_j\in [-W_T,W_T]$ and longitudinal $w_j\in [-W_L,W_L]$ fields are chosen from a random distribution of disorder with strengths smaller than $W_T$ and $W_L$, respectively. This model has a rich structure. For example, in the case of uniform transverse $a_j=a'$ and longitudinal $w_j=w'$ fields, this model is closely related to integrability breaking~\cite{Noh2021}, level crossings~\cite{Vionnet2017}, quantum many-body scars~\cite{Peng2022}, and emergent exceptional group symmetry $E_8$~\cite{coldea2010quantum}.
For the purposes of this section, we consider a chain with periodic boundary conditions $O_j=O_{j+N}$ with $O\in\{X,Y,Z\}$. 

\begin{figure}
    \centering \includegraphics[width=0.48 \textwidth]{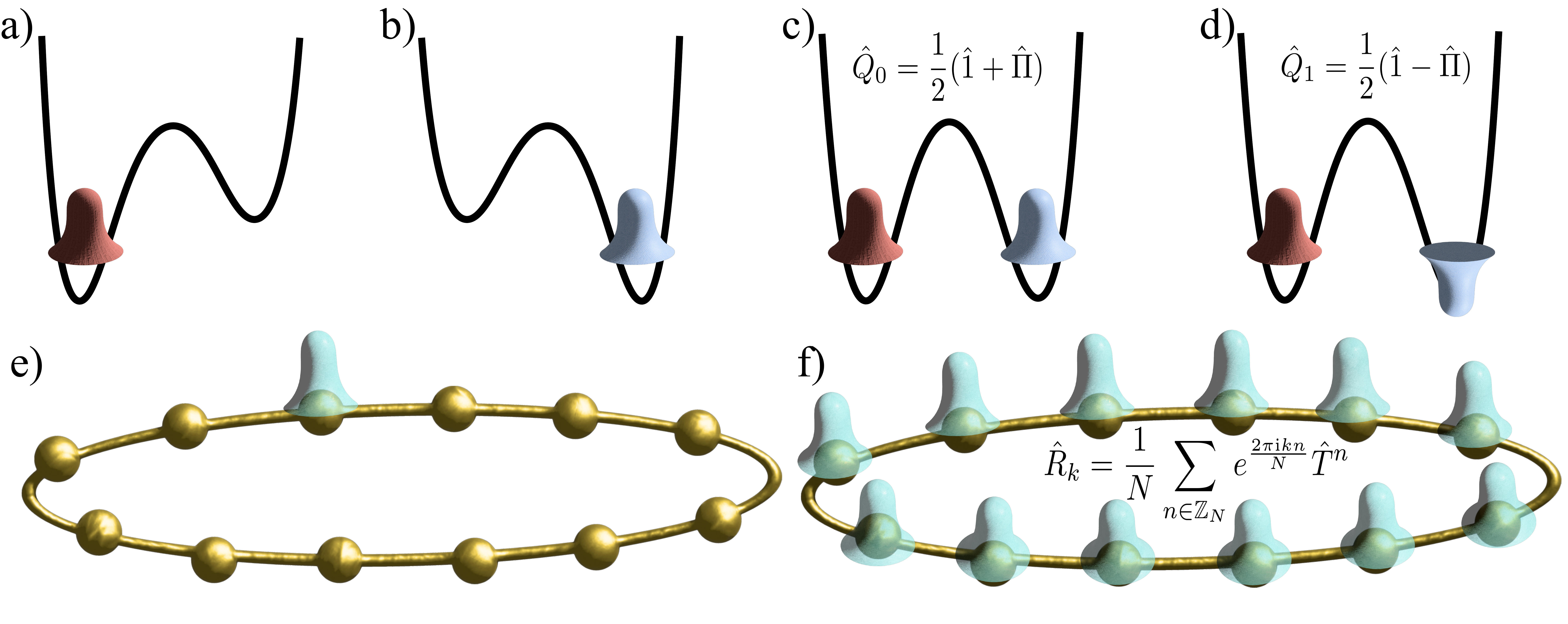}
    \caption{Effect of the symmetry adapted transformation $T_{\rm GSA}$ on the random Ising chain. a) Illustrate initial spin state breaking the parity symmetry of the Ising interaction term while b) depicts the effect of the parity operator on this initial state. c) and d) Illustrate the effect of the projectors $\hat{Q}_0$ and  $\hat{Q}_1$. Similarly, e) shows a defect in the chain breaking translational symmetry. After applying the projector $\hat{R}_k$ into a desired momenta $k$, the state becomes translationally invariant as in f).}
    \label{Fig3N}
\end{figure}

Next, the symmetries of Hamiltonian Eq.~\eqref{eq:IsingModel} in the absence of disorder $(W_T=W_L=0)$ will be discussed. In this case, as the system has periodic boundary conditions and the interaction strengths are uniform along the chain, the system is invariant under spatial translations $\hat{T}^{\dagger}O_j\hat{T}=O_{j+1}$ generated by the operator $\hat{T}$ and $[\hat{\mathcal{H}},\hat{T}]=0$. Further, in the absence of the longitudinal field $(W_L=0)$, the model exhibits an Ising symmetry under the transformation $\hat{Q}^{\dagger}Z_j\hat{Q}= -Z_j$ with the parity operator $\hat{Q}=\bigotimes^N_{j=1} X_j$. In terms of groups, the translational invariance of the chain is related to the group $\mathbb{Z}_N$, while the Ising symmetry is related to $\mathbb{Z}_2$. If these two symmetries are considered, the combined symmetry group is given by $G=\mathbb{Z}_N \times \mathbb{Z}_2$. This symmetry group defines two good quantum numbers $k$ and $\sigma$, which are related to momentum and parity, respectively.

\subsubsection{Projection to symmetric subspaces without second quantization}
% After discussing the basic aspects of the Ising model and its symmetries in the absence of disorder, in this section we discuss the symmetry adapted transformation into symmetric subspaces. 
The goal in this section is to use the symmetry-adapted transformation to project a state with no symmetries to a desired symmetry subspace.

As discussed in Appendix~\ref{AppendixB}, a similar procedure can be attempted by exploiting tools from condensed matter physics, such as fermionization via the Jordan-Wigner transformation. These tools are available in second quantization and for the fermionic operators in real and momentum space. For example, the parity (Ising symmetry) subspaces can be identified by defining the right boundary conditions for the fermions, which also restricts the values of the momenta. In general, the second quantization approach does not simplify the problem, and the treatment of the symmetries remains obscure, as the Hamiltonian maps to a system of interacting fermions with highly nonlocal interactions.

In our approach, Pauli matrices in real space will be used directly without employing second quantization. Consider a concrete example of a quantum state $\ket{\psi}=\ket{110\dots0}$ with a localized domain wall that breaks both translational and Ising symmetries. To see this, the symmetry operations can be applied to this state, obtaining $\hat{T}\ket{\psi}=\ket{10\dots01}$ and $\hat{\Pi}\ket{\psi}=\ket{001\dots1}$. Note that in the last expression, the string $001\dots1$ is the bit-wise negation of $110\dots0$, defining the state $\ket{\psi}$. After the symmetry-adapted transformation is applied, the state $\ket{\psi^{k,\sigma}} = \hat{P}^{k,\sigma}\ket{\psi}$ is effectively obtained, where $\hat{P}^{k,\sigma}$ is the projector defined in Eq.~\eqref{IsingWignerOperator}. This state is explicitly given by  
\begin{align}
   \ket{\psi^{k,\sigma}}&=\frac{1}{2N}(\ket{110\dots0}+e^{\frac{2\pi\mathrm{i}k}{N}}\ket{100,\dots,1}+\ldots)
   \\
   &+\frac{(-1)^{\sigma}}{2N}(\ket{001,\dots,1}+e^{\frac{2\pi\mathrm{i}k}{N}}\ket{011\dots0}+\ldots
   \nonumber
\end{align}
and has well-defined momentum $k$ and parity $\sigma$.
Another simple example is a state $\ket{\psi}=\ket{000\ldots 0}$ with translational symmetry but breaking the Ising symmetry. After the projection, a Greenberger–Horne–Zeilinger state (GHZ state) is obtained.
%%%%
\begin{align}
   \ket{\psi^{0,\sigma}}&=\frac{1}{2}(\ket{000\dots0}+(-1)^{\sigma}\ket{111,\dots,1})  
    \ .
\end{align}
%%%%
For this example, the only projector that survives is for the irrep labelled by $k=0$. This state is very close to an exact energy eigenstate of the Ising chain with weak transverse field $a_j \leq J$ for a uniform finite chain in the absence of the longitudinal field $w_j=0$.

Next, the consequences of the symmetry-adapted transformation when calculating expectation values will be explored. Assume that the expectation value of the Hamiltonian Eq.~\eqref{eq:IsingModel} with disorder is calculated in the projected state $\ket{\psi^{k,\sigma}} = \hat{P}^{k,\sigma}\ket{\psi}$. The expectation value $\langle \hat{ \mathcal{H}}\rangle =\langle \psi|\hat{\mathcal{H}}_{k,\sigma}|\psi\rangle$ can be explicitly written.
%%%%
\begin{align}
      \label{eq:GeneralProjection}
   \hat{ \mathcal{H}}_{k,\sigma} =[\hat{P}^{k,\sigma}]^{\dagger}\hat{\mathcal{H}}\hat{P}^{k,\sigma}=\hat{Q}^{\dagger}_{\sigma}\hat{R}^{\dagger}_{k}\hat{ \mathcal{H}}\hat{R}_{k}\hat{Q}_{\sigma}
    \ ,
\end{align}
%%%%
where $\hat{R}_{k}=1/N\sum_{n\in \mathbb{Z}_N} e^{\frac{2\pi\mathrm{i}kn}{N}} \hat{T}^{n}$ and $\hat{Q}_{\sigma}=(\hat{1}+(-1)^{\sigma}\hat{Q})/2$ are projection operators into the symmetric subspaces labeled by $k$ and $\sigma$. Next, the effect of the projection $\hat{Q}_{\sigma}$ on the Hamiltonian will be investigated.
%%%
\begin{align}         
      \label{eq:ProjectionParity}
\hat{Q}^{\dagger}_{\sigma}\hat{\mathcal{H}}\hat{Q}_{\sigma}=\hat{Q}_{\sigma}\left(-\sum^{N}_{j=1} a_j X_j- J\sum^{N}_{j=1}Z_j Z_{j+1}\right)
    \ .
\end{align}
%%%
It can be observed that the longitudinal field term $\sum^{N}_{j=1} w_j Z_j$ is odd under parity, and thus, the symmetry-adapted transformation eliminates this term.

After this, the projection into a given $k$ irrep can be calculated. As the Ising interaction term is invariant under translations, the action of the projection on that term is trivial. The most interesting term is the random transverse field $\hat{\mathcal{H}}_T=-\sum^{N}_{j=1} a_j X_j$ because it explicitly breaks the translational invariance of the lattice and the projection acts in a nontrivial fashion on it, as follows
%%%
\begin{align}         
      \label{eq:ProjectionTranslationTransverse}
\hat{R}^{\dagger}_{k}\hat{\mathcal{H}}_T\hat{R}_{k} 
&=
\frac{1}{N^2}\sum_{n,m\in \mathbb{Z}_N} e^{\frac{2\pi\mathrm{i}k(n-m)}{N}} \hat{T}^{n-m}\left(-\sum^{N}_{j=1} a_{j+n}X_{j}\right)
\nonumber\\
     &=\frac{\hat{R}^{\dagger}_{k}}{N}\sum_{n\in \mathbb{Z}_M} e^{\frac{2\pi\mathrm{i}kn}{N}} \hat{T}^{n}\left(-\sum^{N}_{j=1} a_{j+n}X_{j}\right)
    \ .
\end{align}
%%%
After the projection, the Hamiltonian has translational invariance. The projection chooses an irrep with a well defined momentum $k$ that here acts as a label for the irreps of the translation group.
The full projected Hamiltonian in Eq.~\eqref{eq:GeneralProjection} reads
%%%%
%\begin{align}
%      \label{eq:FinalProjection}
%   \hat{H}_{k,\sigma} =\hat{R}^{\dagger}_{k}\hat{Q}_{\sigma}\sum_{n\in \mathbb{Z}_N} \frac{e^{\frac{2\pi\mathrm{i}kn}{N}} \hat{T}^{n}}{N}\left(-\sum^{N}_{i=1}G_{i+n}X_{i}- J\sum^{N}_{i=1}Z_i Z_{i+1}\right)
%    \ .
%\end{align}

%\begin{align}
%    \label{eq:FinalProjection}
%    \hat{\mathcal{H}}_{k,\sigma} &= -\hat{R}^{\dagger}_{k}\hat{Q}_{\sigma} \sum_{n\in \mathbb{Z}_N} \frac{e^{-\frac{2\pi\mathrm{i}kn}{N}} \hat{T}^{n}}{N} 
%     \sum^{N}_{j=1}A_{j+n}X_{j}  \nonumber \\
%    &-J \hat{R}^{\dagger}_{k}\hat{Q}_{\sigma} \sum_{n\in \mathbb{Z}_N} \frac{e^{-\frac{2\pi\mathrm{i}kn}{N}} \hat{T}^{n}}{N} \sum^{N}_{j=1}Z_i Z_{j+1} \ \nonumber .
%\end{align}

\begin{align}
    \label{eq:FinalProjection}
    \hat{\mathcal{H}}_{k,\sigma} &= -\hat{R}^{\dagger}_{k}\hat{Q}_{\sigma} \sum_{n\in \mathbb{Z}_N} \frac{e^{\frac{2\pi\mathrm{i}kn}{N}} \hat{T}^{n}}{N} 
     \sum^{N}_{j=1} a_{j+n}X_{j}  \nonumber \\
    &-J \hat{R}^{\dagger}_{k}\hat{Q}_{\sigma} \sum_{n\in \mathbb{Z}_N} \frac{e^{\frac{2\pi\mathrm{i}kn}{N}} \hat{T}^{n}}{N} \sum^{N}_{j=1}Z_i Z_{j+1} \ .
\end{align}
%%%%
It is important to note that the full Hamiltonian Eq.~\eqref{eq:IsingModel} couples all the irreps of the symmetry groups due to the disorder and the longitudinal field. What the projection is doing is to extract a diagonal block corresponding to irreps labeled by $k$ and $\sigma$.

% \subsection{Symmetry-adapted quantum phase estimation and Irreps of abelian cyclic groups: Application to the Harper-Hofstadter model}
\subsection{Two-dimensional Harper-Hofstadter model}
%%%%%
\label{ssec:2d-hh}

In the previous section, it was shown how to use the symmetry-adapted transformation to project a given quantum state into a desired combination of irreps of relevant symmetry groups of the system. In this section, it will be demonstrated that quantum phase estimation can be effectively performed while restricting the system to a given irrep.

\subsubsection{The Harper Hofstadter model }
\begin{figure}
    \centering
    \includegraphics[width=0.47 \textwidth]{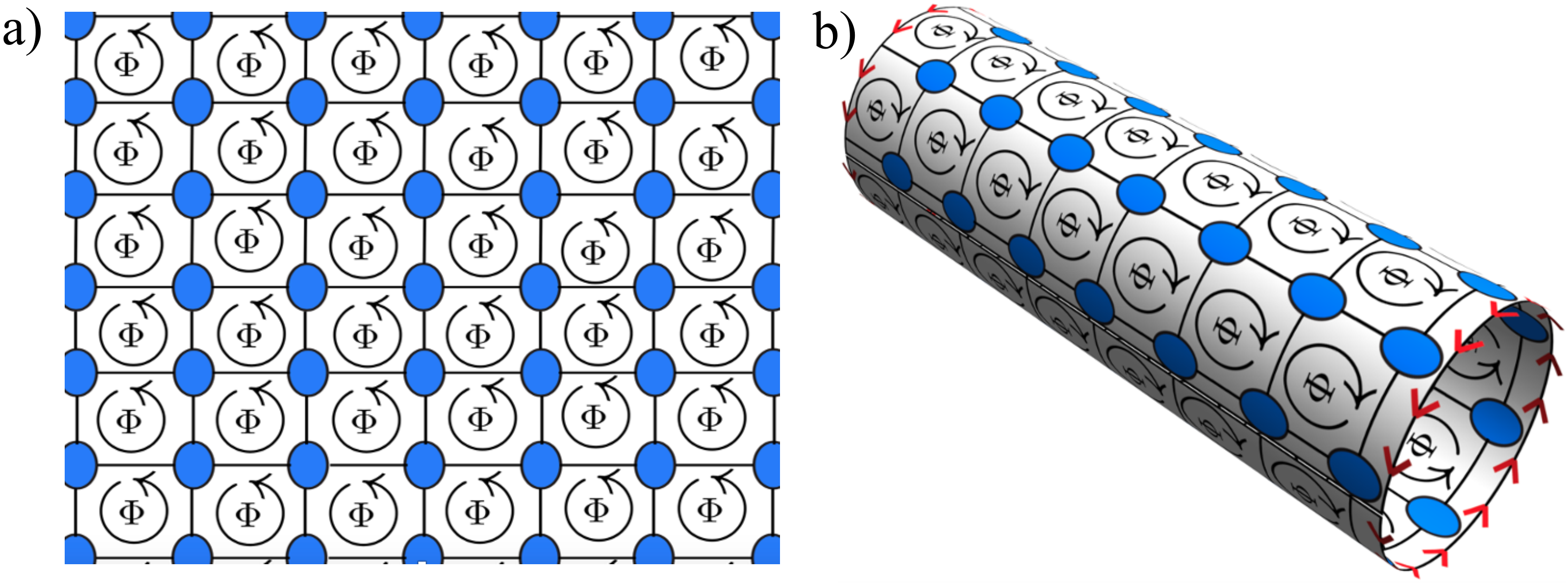}
    \caption{The Harper-Hofstadter model. a) A two dimensional lattice of electrons under the effect of a magnetic field and b) the lattice with periodic boundary conditions defining the dynamics on a cylinder. The stationary modes (red arrows) along the compact dimension have momenta $k_y$. The parameter $\Phi$ denotes the magnetic flux in a given plaquette of the 2D lattice. }
    \label{Fig3}
\end{figure}
After the mathematical details of the group Fourier transform for $G=\mathbb{Z}_M$, also known as the quantum Fourier transform, have been discussed, it is instructive to investigate how the quantum circuit in Fig.~\ref{fig:stm_proj_circ} can be applied to a specific example. It is worth reminding the reader that in this work $\bm{x}=x_{1}x_{2}\cdots x_m$ denotes the binary representation of an integer defined by $x=x_1 2^{m-1}+x_2 2^{m-2} +x_m 2^0$.

The circuit for symmetry-adapted quantum phase estimation will be applied to resolve the energy spectrum of the Harper-Hofstadter model, a paradigmatic model of condensed matter physics~\cite{Hofstadter1976}. In its original form, the Harper-Hofstadter model describes a two-dimensional system of electrons on a lattice under the effect of a magnetic field (see Fig. ~\ref{Fig3}) governed by the Hamiltonian
%%%
%\begin{align}
%         \label{eq:HamSinglePartRest}
% \hat{\mathcal{H}}
% = J_x(\hat{\mathcal{U}}_b +\hat{\mathcal{U}}^{\dagger}_b)   + %J_y (\hat{\mathcal{V}}_b +\hat{\mathcal{V}}^{\dagger}_b)  
% \ .
%\end{align}

\begin{align}
         \label{eq:HamSinglePartRest}
 \hat{\mathcal{H}}
 = J_x(\hat{\mathcal{U}}_{b} + \hat{\mathcal{U}}^{\dagger}_{b}) + J_y (\hat{\mathcal{V}}_{b} + \hat{\mathcal{V}}^{\dagger}_{b})  
 \ .
\end{align}
%%%%

This Hamiltonian is written as a linear combination of two unitaries $\hat{\mathcal{U}}_b=\sum_{x,y}\ket{\bm{x},\bm{y}}\bra{\bm{x+1},\bm{y}}$ and $\hat{\mathcal{V}}_b=\sum_{x,y}e^{2i\pi x b}\ket{\bm{x},\bm{y}}\bra{\bm{x},\bm{y+1}}$ known as magnetic translations. The first term is responsible for the motion along the $x$ direction. The second term is responsible for the motion along the $y$ direction and carries the effect of the magnetic field, as depicted in Fig. ~\ref{Fig3}. In fact, the ratio $b=\Phi/\Phi_0$ between the magnetic flux per plaquette $\Phi$ and the magnetic flux quanta $\Phi_0=h/e$ ($h$ and $e$ being the Planck constant and electron charge, respectively) defines a competition between two characteristic length scales of the model~\cite{Hofstadter1976}.

On a quantum computer, the evolution under this Hamiltonian can be represented as a unitary $\hat{U}=e^{-i\hat{\mathcal{H}} t}$ acting on two sets of registers $\ket{\bm{x}}$ and $\ket{\bm{y}}$ encoding the element $(x,y)$ of the group $\mathbb{Z}_M\times \mathbb{Z}_M$. The integer shift $\ket{\bm{x}}\mapsto \ket{\bm{x+1}}$ can be implemented on a quantum computer using the circuit shown in Fig.~\ref{fig:increment_circ}.

% \begin{widetext}
%    \begin{align}
%       \ket{\Phi}
%       & = \frac{1}{N^{2}} \sum_{x,y, v, w=0}^{N - 1} \sum_{\tilde{k},k}    \ket{\tilde{k}} \ket{\mathbf{Q}^{\tilde{k},k} \Psi} e^{-i2\pi y (E_{k} - \frac{x}{N})}e^{i2\pi x  \delta_{k,k'} \varphi}  e^{i2\pi v( E_{k}-\frac{x}{N}+\frac{w}{N})} \ket{w}  
 %       \ ,
%        \label{eq:schur-inverse-hsb}
%    \end{align}
%    \end{widetext}
   
%
The magnetic translations exhibit interesting algebraic properties. For example, the action of the operators acting in different orders can be calculated as follows:
%%%
\begin{align}
         \label{eq:MagTransComm}
\hat{\mathcal{V}}_b \hat{\mathcal{U}}_b\ket{\bm{x},\bm{y}}&= e^{2i\pi (x -1)b}\ket{\bm{x-1},\bm{y-1}}
 \nonumber\\
 \hat{\mathcal{U}}_b \hat{\mathcal{V}}_b\ket{\bm{x},\bm{y}}&=e^{2i\pi x b} \ket{\bm{x-1},\bm{y-1}}
  \ ,
\end{align}
%%%%
the relation $\hat{\mathcal{U}}_b \hat{\mathcal{V}}_b=e^{2i\pi b }\hat{\mathcal{V}}_b \hat{\mathcal{U}}_b$ is obtained. For irrational $b$, this algebra is known as the quantum torus in non-commutative geometry~\cite{connes1994noncommutative}.

The relation between the lattice constant of the lattice and the magnetic length leads to interesting effects~\cite{Hofstadter1976}. For example, the spectrum is a fractal when it is plotted as a function of the magnetic field. Such a fractal spectrum is known in the literature as the Hofstadter butterfly~\cite{Hofstadter1976}, and it has been realized on quantum simulators~\cite{Geim_spectroscopic}.
Beyond condensed matter physics, the Harper-Hoftstadter model plays a very important role in the study of Mathieu operators~\cite{puig2004cantor}, magnetic quantum walks~\cite{Sajid2019} and non-commutative geometry in mathematics~\cite{connes1994noncommutative}.

Due to the lattice symmetries, as shown in Appendix~\ref{AppendixA}, it can be demonstrated that Eq.~\eqref{eq:HamSinglePartRest} transforms under $\hat{U}_{\text{QFT}}$ for the vertical direction [labeled by $y$ in Eq.~\eqref{eq:MagTransComm}] as
$\hat{U}_{\text{QFT}}\hat{\mathcal{H}}\hat{U}^{\dagger}_{\text{QFT}}=\sum_{k} \hat{\mathcal{H}}_{k_y}\otimes\ket{\bm{k_y}}\bra{\bm{k_y}}$ with $\hat{\mathcal{H}}_{k_y}$ defined as
\begin{align}
         \label{eq:SymmAdapt}
\hat{\mathcal{H}}_{k_y}&=\sum_{x}J_xe^{2i\pi( x b-k_y/N)}\ket{\bm{x}}\bra{\bm{x}}
\nonumber \\
&+\sum_{x}J_y\ket{\bm{x}}\bra{\bm{x+1}}+\text{H.c}
  \ .
\end{align}
%%%%

The particular example $b=1/2$ defining a rational magnetic translation will be considered. For this value, the magnetic translations anticommute $\hat{\mathcal{U}}_{1/2} \hat{\mathcal{V}}_{1/2}=-\hat{\mathcal{V}}_{1/2} \hat{\mathcal{U}}_{1/2}$. In terms of lattice symmetries, an effective dimmerization of the lattice in the horizontal axes is originated from the relation $\hat{\mathcal{U}}_{1/2} \hat{\mathcal{V}}^{2}_{1/2}=\hat{\mathcal{V}}^2_{1/2}\hat{\mathcal{U}}_{1/2}=\hat{\mathcal{U}}^{2}_{1/2} \hat{\mathcal{V}}_{1/2}=\hat{\mathcal{V}}_{1/2}\hat{\mathcal{U}}^2_{1/2}$. That is, a magnetic translation by two lattice constants leaves the system invariant.
In other words, by choosing $b=1/2$, an effective pseudo-spin that tremendously simplifies the problem is defined. The pseudospin structure emerges by considering even $x=2n$ and odd $x=2n-1$ sites with integer $n$. The even sites are affected by an onsite potential $J^{\text{Even}}_x=J_xe^{-2i\pi k_y/N)}$, while the potential for odd sites is $J^{\text{Odd}}_x=-J_xe^{-2i\pi k_y/N)}$. Now, to fully describe the system, a ket $\ket{\bm{n},\Lambda}$ can be defined, where $n=0,1,\ldots M/2$ labels the space and $\Lambda=\uparrow,\downarrow$ the pseudospin degree of freedom representing even and odd sites. By using this notation, the Hamiltonian Eq.~\eqref{eq:SymmAdapt} can be written as spin-$1/2$ particle moving in a lattice under the effect of a magnetic field and spin-orbit coupling, as follows
\begin{align}
         \label{eq:SpinEffective}
\hat{\mathcal{H}}_{k_y}&=\sum_{n}\left(J^{\text{Even}}_x\ket{\bm{n},\uparrow}\bra{\bm{n},\uparrow}+J^{\text{Odd}}_x\ket{n,\downarrow}\bra{\bm{n},\downarrow}+\text{H.c}\right)
\nonumber \\
&+\sum_{n}\left(J_y\ket{\bm{n},\uparrow}\bra{n,\downarrow}+J_y\ket{n,\downarrow}\bra{\bm{n+1},\uparrow}+\text{H.c}\right)
  \ .
\end{align}
%%%%}
In terms of symmetries, the value $b=1/2$ defines the symmetry group $\tilde{G}=\mathbb{Z}_{M/2}\otimes \mathbb{Z}_{M}$ that is a subgroup of $G=\mathbb{Z}_M\otimes \mathbb{Z}_{M}$. This structure can be generalized for any rational number of the form $b=p/q$, where $p$ and $q$ are coprimes on such a way that the system has a symmetry group $\tilde{G}=\mathbb{Z}_{M/q}\otimes \mathbb{Z}_{M}$. When $b$ is irrational, the symmetry group is solely determined by $G=\mathbb{Z}_M$ as the system has translational invariance only in the $y$ axis.

Geometrically, the irrep labels $k_y$ are used to label the foliation of the cylinder by lines. As depicted in Fig.~\ref{Fig3}~b), this corresponds to a family of decoupled lattices along a cylinder. The irreps $k_y$ denote the modes of stationary waves in the $y$ direction.

\subsubsection{Symmetry-adapted quantum phase estimation for the Harper model}
\begin{figure}
    \centering
    \includegraphics[width=0.5 \textwidth]{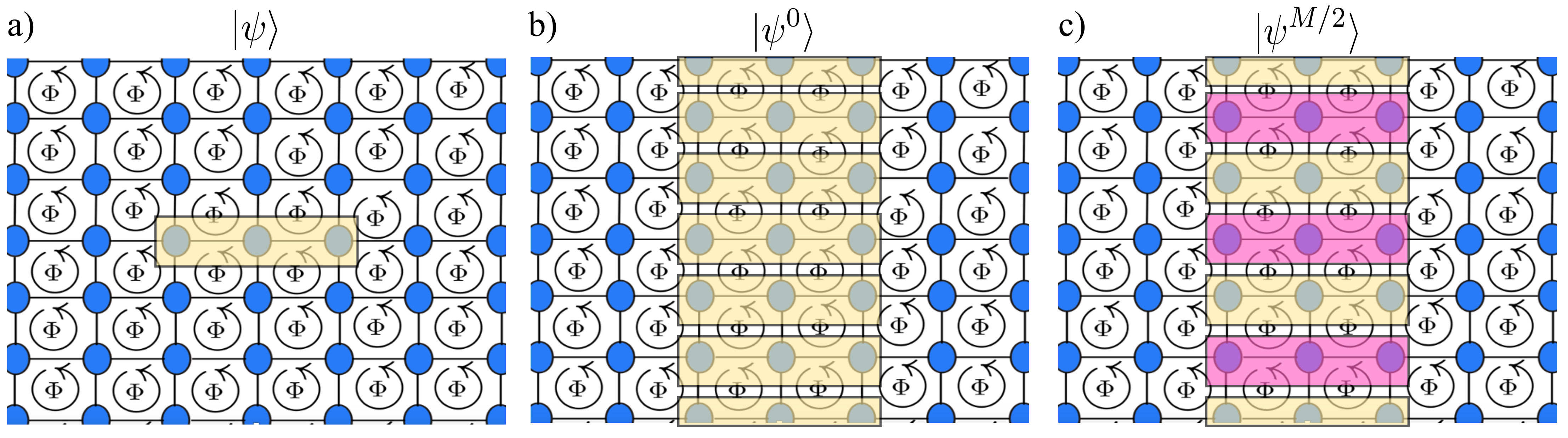}
    \caption{Symmetry adapted projection of a quantum state. a) An initial state $\ket{\psi}=\sum_{x,y\in \mathbb{Z}_M}c_{x,y}\ket{\bm{x,y}}$ that breaks the translational symmetry. Such a state is localized along three lattice sites.  b) and c) The projections $\ket{\psi^{0}}$ and  $\ket{\psi^{M/2}}$ out of the state $\ket{\psi}$ as in a) into the irreps $k=0$ and $k=M/2$, describing a long- and short-wavelength excitations in the lattice, respectively. We assume periodic boundary conditions along the $y$ directions in the lattice, which effectively allows us to represent the system as a lattice on a cylinder as in Fig.~\ref{Fig3} b). The parameter $\Phi$ denotes the magnetic flux in a given plaquette of the 2D lattice.}
    \label{Fig4}
\end{figure}
Next, the action of the quantum circuit of Fig.~\ref{Fig2:tgsa+qpe} for a state $\ket{\psi}=\sum_{x,y\in \mathbb{Z}_M}c_{x,y}\ket{\bm{x,y}}$ with no translational invariance at all, as depicted in Fig.~\ref{Fig4}~a), will be investigated. The symmetry-adapted transformation in this case has the form
\begin{align}
   \hat{T}_{\text{GSA}} =  (\hat{U}_{\text{QFT}}\otimes \hat{1}) \cdot \select[\tilde{\rho}_{\mathbb{Z}_m}] \cdot (\hat{U}^{-1}_{\text{QFT}}\otimes \hat{1})
    \label{eq:ExplicitTGSA-2}
    \ ,
\end{align}
with the controlled group action defined as $\select[\tilde{\rho}_{\mathbb{Z}_m}]=\sum_u\ket{\bm{u}}\bra{\bm{u}}\otimes \hat{\rho}(u)$. Here $\widehat{\tilde{\rho}}(u)=\hat{\rho}(u)$ is a one-dimensional representation of the group element $u\in \mathbb{Z}_M$ acting as a translation $\hat{\rho}(u)\ket{\bm{x,y}}=\ket{\bm{x,y+u}}$. The explicit action of the symmetry adapted transformation reads 
\begin{align}
    T_{\text{GSA}}\ket{\bm{0}} \ket{\psi} =  \sum _{k_y\in \mathbb{Z}_M}\ket{\bm{k_y}} \ket{\psi^{k_y} }
    \label{symm_adap_trans}
    \ ,
\end{align}
where $\ket{\psi^{k_y}} = \frac{1}{M} \sum_{x,y,v\in \mathbb{Z}_M} c_{x,y}  e^{2\pi\mathrm{i}k_y v/M} \ket{\bm{x,y+v}}$. Note that the first register in Eq.~\eqref{symm_adap_trans} keeps the information of a given irrep labelled by $k_y$, while the group element $v\in \mathbb{Z}_M$ acts as a translation of the ket $\ket{\bm{x,y}}\mapsto \ket{\bm{x,y+v}}$. If the circuit is stopped here and the first register is measured in an irrep labelled by $k'_y$, the state is effectively projected into the symmetry-adapted state $\ket{\psi^{k'_y}}$. 

Let us discuss for a moment the physical interpretation of this projector. The integer $k_y$ labels the irreps but it also defines the wavelength of the excitations in the system. Such a wavelength is defined as $\lambda_y=M/k_y$. That is, irreps with $k_y=0$ represent long-wavelenght excitations with $\ket{\psi^{0}} = \frac{1}{M} \sum_{x,y,v\in \mathbb{Z}_M} c_{x,y}  \ket{\bm{x,y+v}}$ while $k_y=M/2$ are rapidly oscillating excitations represented by $\ket{\psi^{M/2}} = \frac{1}{M} \sum_{x,y,v\in \mathbb{Z}_N} (-1)^v c_{x,y}  \ket{\bm{x,y+v}}$ in our system. These states are illustrated in Figures~\ref{Fig4}~b)~and~c). What the projector does is to select a given wavelength in our system for a desired $k_y$ so that such irrep is fixed for quantum phase estimation in our circuit
\begin{align}
       &V_{\rm SQPE}|\bm{0} \rangle^m \ket{\psi}|0 \rangle^n 
       \nonumber\\&= %\sum_{u,v=0}^{2^n - 1}
       \sum^{2^{n}-1}_{u,v}\sum_{k} \frac{e^{-i2\pi v (E_{k_y} - u/2^n)}}{2^n}  \ket{\bm{k_y}}\ket{\psi^{k_y} }  \ket{\bm{u}},
        \label{eq:sqpe}
    \end{align}
where $E_{k_y}$ are the eigenvalues of $\hat{\mathcal{H}}_{k_y}$ in Eq~\eqref{eq:SymmAdapt}. 

In this section, the versatility of our approach was demonstrated by decomposing a system into different irreps labeled by the index $k_y$, denoting the irreps of the group $\mathbb{Z}_{M}$. After this decomposition, it was briefly shown how to apply QPE, a quantum subroutine that can be used to efficiently access the energies of a many-body system using quantum computers. In the case of the Hamiltonian $\hat{\mathcal{H}}_{k_y}$ in Eq~\eqref{eq:SymmAdapt}, QPE enables the energies $E_{k_y}$ to be obtained. After plotting them as a function of the parameter $b$ in Eq~\eqref{eq:SymmAdapt}, the resulting spectrum has fractal properties and is known as the Hofstadter butterfly. Some features of this spectrum have been observed on a quantum simulation by using a superconducting qubit array with $9$ qubits~\cite{roushan2017spectroscopic}. The fractal nature of the Hofstadter butterfly has been observed in moir{\'e}~\cite{dean2013hofstadter} and graphene~\cite{ponomarenko2013cloning} superlattices. A more recent work uses high-resolution scanning tunnelling microscopy to resolve the fractal nature of the spectrum in twisted bilayer graphene close to the second magic angle~\cite{nuckolls2025spectroscopy}.

% \subsection{Symmetry-adapted quantum phase estimation and Irreps of the permutation group $S_N$: Application to electronic structure}
\subsection{Three-dimensional \textit{ab initio} electronic structure}
\label{ssec:3d-h2}

It is instructive to consider an example of a chemistry problem where both the group Fourier transformation and symmetry-adapted transformation $T_{\text{GSA}}$ can be explicitly carried out. One of the advantages of this example is that the symmetric group is used to project a problem in first quantization to a fermionic sector of interest for quantum simulation of electronic structure in quantum chemistry.

\subsubsection{The H$_2$ molecule in first quantization}
In this section, an example is considered that shows a practical application of the QCT for $G=S_2$ in Eq.~\eqref{eq:QCTS2} to a two-electron problem in electronic structure. The simplest example that can be thought of is the molecule $\text{H}_2$. For convenience, this two-electron system is described in the formalism of first quantization under the Born-Openheimer approximation. In atomic units, the Hamiltonian reads
%%%
\begin{align}
    \label{eq:H2Hamiltonian}
          \hat{ \mathcal{H}}&=\sum_{m=1}^2\left(-\frac{\nabla_m^2}{2}-\sum^2_{\beta=1} \frac{Z_{\beta}}{|\boldsymbol{r}_m - \boldsymbol{R}_\beta|}\right)+\frac{1}{|\boldsymbol{r}_1 - \boldsymbol{r}_{2}|}
          \nonumber\\
          &=\hat{\mathcal{O}}^{(1)}(\boldsymbol{r}_1)+\hat{\mathcal{O}}^{(1)}(\boldsymbol{r}_2)+\hat{\mathcal{O}}^{(2)}(\boldsymbol{r}_1, \boldsymbol{r}_{2}) 
          \ ,
\end{align}
%%%
where $Z_{\beta}$ is the atomic number of the nucleus $\beta=1,2$, and $|\boldsymbol{r}_m - \boldsymbol{R}_\beta|$ is the distance between the mth electron and the nuclei labeled by $\beta$, while $r_{1,2}$ is the distance between the two electrons. The Hamiltonians $\hat{\mathcal{O}}^{(1)}(\boldsymbol{r}_1)$ and $\hat{\mathcal{O}}^{(1)}(\boldsymbol{r}_2)$ are the core Hamiltonians for electrons 1 and 2, respectively. As a two-particle system is being dealt with, a tensor product basis of the individual states of the particles is chosen to build up the Hilbert space.
%In general, the wave function is not only described by the spatial coordinates $\boldsymbol{r}_i$ but also by the spin degrees of freedom.  We can then define states $\ket{\boldsymbol{x}_i}=\ket{\boldsymbol{r}_i,\sigma_i}$ with $i=1,2$ and $\sigma_i\in\{\uparrow,\downarrow\}$ that carry information about position and spin degrees of freedom. Using this notation, given a two-particle state $\ket{\psi}$, the corresponding wave function is given by $\psi(\boldsymbol{x}_1,\boldsymbol{x}_2)=\langle\boldsymbol{x}_1,\boldsymbol{x}_2\ket{\psi}$. 
It is remarked here that no restriction on the symmetry of the wave function has been imposed so far, as first quantization is being used.

Next, let us consider a minimal basis with only two 1s spatial orbitals $\ket{\phi_1}$ and $\ket{\phi_2}$ localized around the two nuclei labeled by $1$ and $2$. Due to the spatial symmetry of the $\text{H}_2$ molecule, it is convenient to define \textit{gerade} (even parity, bonding) and \textit{ungerade} (odd parity, anti-bonding) molecular orbital as a linear combination of molecular orbitals
%%%
\begin{align} 
    \label{eq:MinimalBasisGerUnger}
     \ket{\Phi_1}&=[2(1+S_{1,2})]^{-1/2}( \ket{\phi_1}+ \ket{\phi_2})
     \nonumber\\
     \ket{\Phi_2}&=[2(1+S_{1,2})]^{-1/2}( \ket{\phi_1}- \ket{\phi_2})
     \ ,
\end{align}
%%%
where $S_{1,2}$ is the overlap integral between the 1s orbitals.
By using this basis set, any quantum state can be written as a linear combination
%%%
\begin{align} 
    \label{eq:MinimalBasis}  \ket{\psi}=\sum_{j_1,\alpha_1;j_2,\alpha_2}C_{j_1,\alpha_1,j_2,\alpha_2}\ket{j_1,\alpha_1}\otimes \ket{j_2,\alpha_2}
     \ .
\end{align}
%%%
Here $\ket{j_m}$ denotes the spatial orbitals with $j_m\in\{\Phi_1,\Phi_2\}$ while $\ket{\alpha_m}$ represent the spin degree of freedom with $\alpha_m=0,1$.
The two-particle wave function in Eq.~\eqref{eq:MinimalBasis} is a linear combination of $D=16$ basis states and for this reason is natural to encode the state using $N=4$ qubits. The odd qubits encode the orbital information and the even qubits encode the spin of the electrons.

\subsubsection{Labeling the fermionic versus bosonic sector}

After the basic aspects of H$_2$ are discussed, the action of the symmetry-adapted group transformation $\hat{T}_{\rm GSA}$ [see circuit in Fig.~\ref{fig:s2_proj_circ}] on a quantum state $\ket{0} \ket{\Psi}$, where the first register encodes a trivial irrep of the group and $\ket{\Psi}$ is the state of the molecule in first quantization as in Eq.\eqref{eq:MinimalBasis}, needs to be calculated
\begin{align}
   \hat{T}_{\rm GSA}\ket{0} \ket{\psi} 
    &= \sum_\Gamma \ket{\Gamma} \hat{P}^\Gamma\ket{\psi} \nonumber \\
    &=\ket{0} \hat{P}^0\ket{ \psi}+\ket{1} \hat{P}^1\ket{\psi}
    \ ,
    \label{schur-trivial-irrep}
\end{align}
where the projector above is given by $\hat{P}^\Gamma= \frac{d_\Gamma}{|G|} \sum_{g \in G} \chi_{\Gamma}(g) \rho(g)$. In our case, the group elements of $S_2$ are $\{e, (12)\}$ and the characters are shown in Table~\ref{tab:S2_character_table}. Thus, the projectors for different irreps are given explicitly by 
% \begin{widetext}
\begin{align}
    \hat{P}^z 
    &= \frac{1}{2}\chi_z(e) \text{I}+\frac{1}{2}\chi_z[(12)]\text{SWAP}_{1,3}\text{SWAP}_{2,4} \nonumber\\
    &=\frac{1}{2} \left[\text{I}+ (-1)^z \text{SWAP}_{1,3}\text{SWAP}_{2,4} \right],
    % \nonumber\\
    %  \mathbf{P}^1 
    %  &= \frac{1}{2}D^1(id) \text{I}+\frac{1}{2}D^1[c_{(1,2)}]\text{SWAP}_{1,3}\text{SWAP}_{2,4} \nonumber \\
    %  &=\frac{1}{2}( \text{I}-\text{SWAP}_{1,3}\text{SWAP}_{2,4})
    %  \ .
    \label{Exampleprojector-trivial-irrep}
\end{align}
where $\chi_z(g)$ is the characters for the group element $z$ under the irrep $z$, for $z = 0,1$ used for a qubit encoding of the irreps $(2,0), (1,1)$ of $S_2$ as in the character Table~\ref{tab:S2_character_table}.
% \end{widetext} 
Here the SWAP gate acting on odd qubits allows a permutation of the orbitals to be performed, and the SWAP acting on even qubits allows the spin degrees of freedom to be permuted. The action of the operators on the quantum state in Eq.~\eqref{eq:MinimalBasis} can be examined more carefully:
\begin{align}
    \label{FermionProj}
    \hat{P}^z\ket{ \psi}&=\sum_{j_1,\alpha_1;j_2,\alpha_2} C_{j_1,\alpha_1,j_2,\alpha_2} \nonumber \\
    & \times  \left(\ket{j_1,\alpha_1; j_2,\alpha_2} + (-1)^z \ket{j_2,\alpha_2; j_1,\alpha_1} \right) ,
\end{align}
% \begin{widetext}
% \begin{align*}
%     \label{FermionProj}
%     \mathbf{P}^0\ket{ \Psi}&=\sum_{j_1,\alpha_1;j_2,\alpha_2}C_{j_1,\alpha_1,j_2,\alpha_2}(\ket{j_1,\alpha_1}\otimes \ket{j_2,\alpha_2}+\ket{j_2,\alpha_2}\otimes \ket{j_1,\alpha_1})
%     \nonumber\\
%     \mathbf{P}^1\ket{ \Psi}&=\sum_{j_1,\alpha_1;j_2,\alpha_2}C_{j_1,\alpha_1,j_2,\alpha_2}(\ket{j_1,\alpha_1}\otimes \ket{j_2,\alpha_2}-\ket{j_2,\alpha_2}\otimes \ket{j_1,\alpha_1})
%     \ ,
% \end{align*}
% \end{widetext}
which gives us exactly the bosonic and fermionic representation in terms of permanents and the slater determinants. 

To illustrate how the projector acts on the spin degree of freedom, it is useful to consider a simple example. Let an initial state $\ket{ \psi'}=\ket{\Phi_1,\uparrow}\otimes \ket{\Phi_1,\downarrow}$ be assumed. Then, the projection in the fermionic sector is given by
\begin{align}
    \label{ExplicitFermionProj}
    \hat{P}^1\ket{ \psi'}&=\frac{1}{2}(\ket{\Phi_1,\uparrow}\otimes \ket{\Phi_1,\downarrow}-\ket{\Phi_1,\downarrow}\otimes \ket{\Phi_1,\uparrow})
    \nonumber \\ 
    &=\frac{1}{2} \ket{\Phi_1,\Phi_1} \otimes (\ket{\uparrow\downarrow}-\ket{\downarrow,\uparrow}).
\end{align}
This state describes a two-electron system in the singlet state with the two electrons occupying the same spatial orbital. States such as $\ket{ \psi'}=\ket{\Phi_1,\uparrow}\otimes \ket{\Phi_1,\uparrow}$ are projected out of the fermionic sector due to the Pauli exclusion principle. For pedagogical reasons, in the last line, we separated the spatial and spin degrees of freedom to show explicitly that the two-electron system is in a spin singlet state.

Similarly, the following will produce a superposition of triplet and singlet states. Starting from the state
\begin{align}
    \ket{\psi''} 
    &= (I \otimes ZH \otimes X \otimes H) \ket{0}^{\otimes 4} \nonumber \\
    &= \frac{\ket{\Phi_1, \uparrow} -\ket{\Phi_1, \downarrow}}{\sqrt{2}} \otimes \frac{\ket{\Phi_2, \uparrow}+\ket{\Phi_2, \downarrow}}{\sqrt{2}} 
    \label{eq:fermionic-h2-product-state}
\end{align}
and projecting it into the fermionic sector, we have
\begin{align}
    \ket{ \psi}_{\rm fermionic} &\equiv \hat{P}^1\ket{ \psi''}   \nonumber \\
    &= \frac{1}{4} \left[
    \ket{\Phi_1, \Phi_2} - \ket{\Phi_2, \Phi_1}\right] \left[ \ket{\uparrow, \uparrow} - \ket{\downarrow, \downarrow} \right] \nonumber \\
    &+ \frac{1}{4} \left[
    \ket{\Phi_1, \Phi_2} + \ket{\Phi_2, \Phi_1}\right] \left[ \ket{\uparrow, \downarrow} - \ket{\downarrow, \uparrow} \right], 
    \label{eq:fermionic-h2-mixed-s-tt}
\end{align}
% Using only 0 and 1 state of the qubit to rewrite for clarity:
% \begin{align}
%     \ket{ \Psi}_{\rm fermionic} \equiv \mathbf{P}^1\ket{ \Psi} 
%     &= \frac{1}{4} \left[
%     \ket{0, 1} - \ket{1, 0}\right] \left[ \ket{0, 0} - \ket{1,1} \right] \nonumber \\
%     +& \frac{1}{4} \left[
%     \ket{0,1} + \ket{1,0}\right] \left[ \ket{0, 1} - \ket{1, 0} \right]
% \end{align}
where the first line is a triplet state, while the second line is a singlet state.

\subsubsection{Labeling the singlet versus triplet sector}

\begin{figure}[h]
    \centering
    \includegraphics[width=\linewidth]{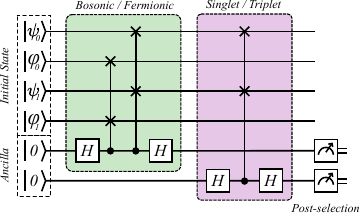}
    \caption{Quantum circuit depicting the implementation of projectors $\hat{P}^z$ to project an initial state of the H$_2$ molecule to different irreps in 1st quantization encoding. Starting from the initial state (first 4 qubits) with two ancilla qubits, the circuit in the green box labels the fermionic and bosonic sector using the first ancilla qubit. The circuit in the purple box then labels the singlet and triplet states using the last ancilla qubit.}
    \label{fig:h2_fb-st}
\end{figure}

Figure \ref{fig:h2_fb-st} is a quantum circuit that achieves the symmetry-adapted transformation for the H$_2$ molecule discussed in the previous section, composing a register representing the initial state in Eq. \eqref{eq:fermionic-h2-product-state} plus the ancilla qubits for manipulating and labeling the irreps. For this initial state, the first part of the circuit allows one to label the bosonic and fermionic sectors, while the second part is the selection of the singlet and triplet sectors. A final measurement is performed on the two ancilla qubits if needed to post-select the desired irrep. Note that the post-selection success probability for a given irrep sector $\Gamma$ is given by $|a_\Gamma|^2$ as defined in Eq. \eqref{eqn:TGA_back}.

\begin{figure}[h]
    \centering
    \includegraphics[width=\linewidth]{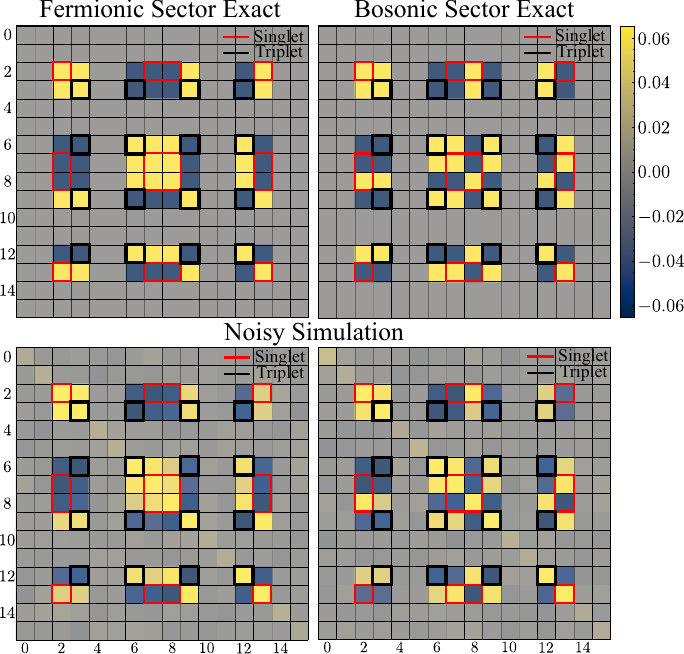}
    \caption{Final density matrix of the H$_2$ molecule, post-selected on the two ancilla qubits (last two qubits in Fig. \ref{fig:h2_fb-st}) that labels the fermionic vs. bosonic or singlet vs. triplet states.}
    \label{fig:h2_simulation}
\end{figure}

Using a standard quantum computing framework such as Qiskit \cite{qiskit2024}, we can simulate the circuit in Fig. \ref{fig:h2_fb-st} and obtain the resulting system density matrix after post-selecting the corresponding sectors. Fig. \ref{fig:h2_simulation} shows the resulting density matrix of the system register from the exact and the noisy simulations.
% The previous figure shows the complete density matrix and the singlet and triplet matrices from the fermionic and bosonic sectors, respectively. 
The noisy simulation was done using the fake provider module in Qiskit, which contains a fake (simulated) backend that mocks the IBM devices, \textit{fake\_7q\_pulse\_v1}.
%The blue and red boxes in Fig.~\ref{fig:h2_simulation}
It can be seen that the singlet and triplet sectors, represented by the red and blue boxes in Fig. \ref{fig:h2_simulation}, respectively, have the same non-zero matrix values location within the complete density matrix representation.

\begin{figure}[h]
    \centering
    \includegraphics[width=\linewidth]
    {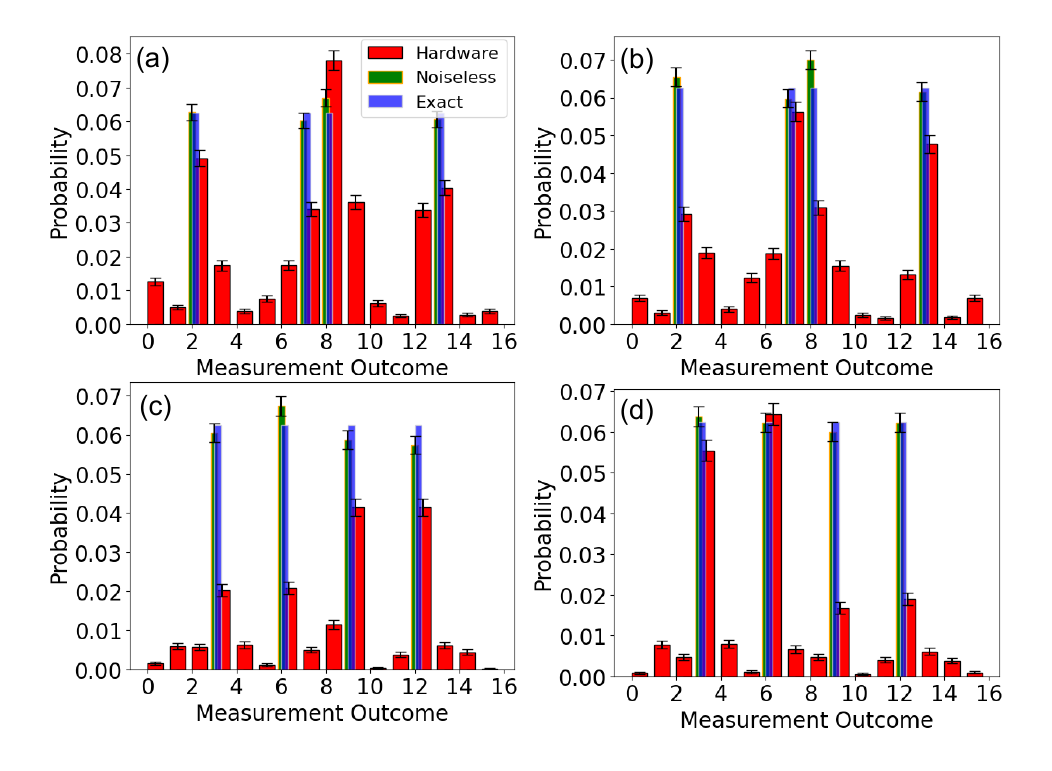}
    \caption{Diagonal elements of the density matrix of the H$_2$ molecule, (a) singlet bosonic state, (b) singlet fermionic state, (c) triplet bosonic state, and (d) triplet fermionic state.}
    \label{fig:h2_tgsa_simulation}
\end{figure}

Figure \ref{fig:h2_tgsa_simulation} presents a comparison of the diagonal elements of the density matrix shown in Fig.~\ref{fig:h2_simulation} for the H$_2$ molecule for different symmetry states. The comparison includes values from a noiseless simulation using the IBM local simulator, results from IBM Nazca (127-qubit) quantum computer, and the theoretical values. Here, in both cases, we run the quantum circuit and obtain 10000 shots to get the output with reduced random error since the output is probabilistic in nature. The noiseless simulation agrees perfectly with the exact results, whereas the hardware results exhibit some deviations, which are attributed to the noise in the quantum hardware.

\subsubsection{Symmetry-adapted quantum phase estimation for H$_2$}
\begin{figure}[h]
    \centering
    \includegraphics[width=\linewidth]{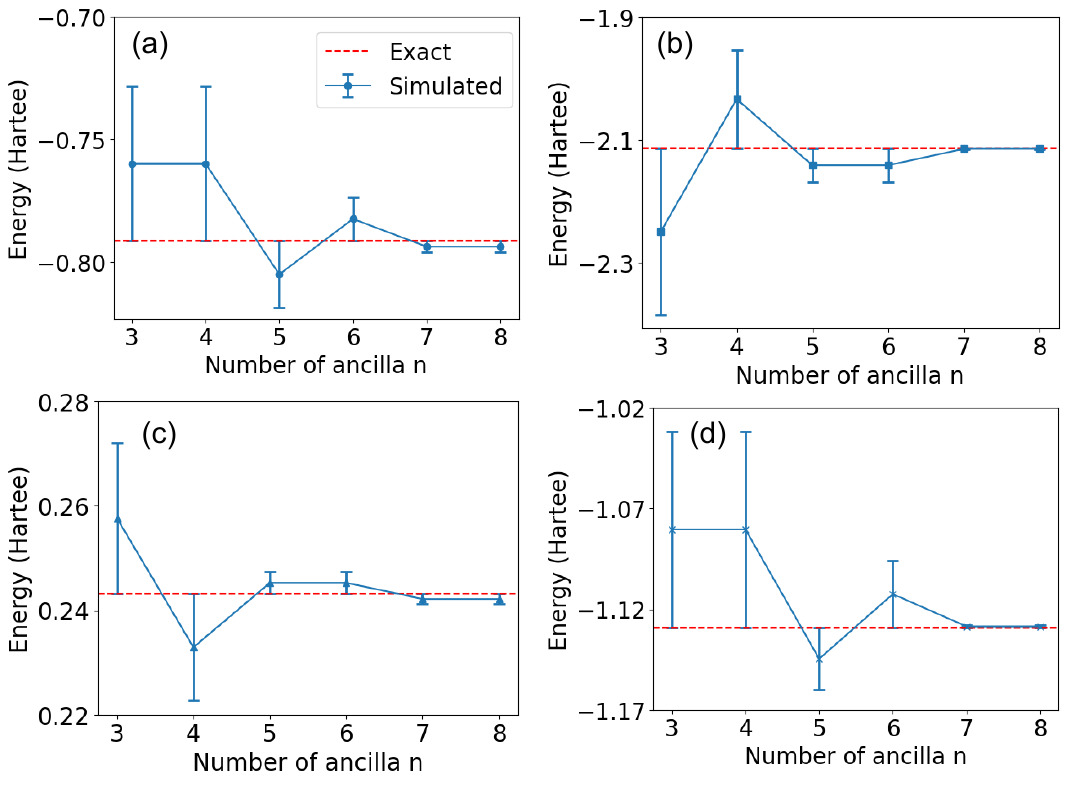}
\caption{Simulated energy by using our symmetry-adapted QPE algorithm for (a) singlet bosonic, (b) singlet fermionic, (c) triplet bosonic, and (d) triplet fermionic states of the H$_2$ molecule. The horizontal axis is the number of ancilla qubits $n$ in the 3rd register of Fig. \ref{Fig2:tgsa+qpe}. 2000 shots are used for the simulation.}
    \label{fig:h2_qpe+tgsa_simulation}
\end{figure}

The symmetry-adapted transformation $\hat{T}_{\rm GSA}$ can be further combined  with quantum phase estimation as shown in Fig. \ref{Fig2:tgsa+qpe} to estimate the energy of singlet and triplet state simultaneously. Fig. \ref{fig:h2_qpe+tgsa_simulation} shows the simulated energies corresponding to the highest count state across different irreps versus the number of ancilla qubits in QPE. The symmetry-adapted QPE circuit was run with 2000 shots in IBM local simulator in a noiseless environment. The dashed red horizontal line in the plot shows the eigenvalues of the corresponding state obtained from the exact diagonalization of the Hamiltonian of the H$_2$ molecule. The results demonstrate that symmetry-adapted QPE can accurately predict the eigenenergies across different irreps. Note that besides the singlet and triplet states for H$_2$ in the case of fermionic particles, the other sectors represent fictitious states of the H$_2$ Hamiltonian with two bosonic particles that can be in singlet and triplet states. 
In this case, if the spin degree of freedom is in the singlet state, the spatial wave function should be antisymmetric to preserve the bosonic statistics of the particles.

% \YL{breaks the Hilbert space into bosonic and fermionic sectors. Thus,
% \begin{align}
%        &V_{\rm SQPE}|0 \rangle^m \ket{\psi}|0 \rangle^n 
%        \nonumber\\&= \frac{e^{-i2\pi v (E_{\Gamma_1} - u/2^n)}}{2^n}  \ket{\Gamma_1}\ket{\mathbf{P}^{\Gamma_1} \psi}  \ket{u}
%        \nonumber\\& + \frac{e^{-i2\pi v (E_{\Gamma_2} - u/2^n)}}{2^n}  \ket{\Gamma_2}\ket{\mathbf{P}^{\Gamma_2} \psi}  \ket{u}
%        .
%         \label{eq:qpe}
%     \end{align}
% For electronic structure, we are interested in the fermionic sector so we need to project into the irrep $\Gamma_2$ to determine the energy of the molecule.}

% \begin{figure}[htbp]
%             \begin{center}
%             \mbox{
%                 \Qcircuit @C=1.0em @R=1.5em {
%                 |0 \rangle^m \quad\quad    & \qw    /  & \multigate{1}{T_{\rm GSA}} & \qw    
%                 %\ctrl{2} & \qw & \multigate{1}{T_{\rm GSA}^\dagger}
%                 & \meter    \\
%                 |\psi \rangle \quad\quad     & \qw /    & \ghost{T_{\rm GSA}} & \multigate{1}{\rm U_H} & \qw   \\
%                 |0 \rangle^n \quad\quad & \qw / & \qw &\ghost{\rm U_H} & \meter 
%                 }
%             }
%             \end{center}
%             \caption{Circuit for symmetry-adapted block encoding.}
%             \label{Fig3}
% \end{figure}

\subsection{Symmetrization of wave functions in first quantization: Fermions, bosons and parastatistics}
\label{ssec:Symmetrization}

In the previous examples only abelian finite groups have been discussed. This section contains an example that uses the simplest non-abelian group $G=S_3$. As a first step the projectors in Eq.~\eqref{eqn:Conjugacy_factorised} will be explictly constructed for the three irreps $\Gamma_1=(3,0)={\tiny\yng(3)}$, $\Gamma_2=(1,1,1)={\tiny\yng(1,1,1)}$, and $\Gamma_3=(2,1)={\tiny\yng(2,1)}$. The projector for the trivial irrep reads
\begin{align}
    \hat{P}^{\Gamma_1} &= \frac{1}{6} \sum^{3}_{C=1} \chi_{\Gamma_1}(C) \left( \sum^{|C| }_{g \in C} \hat{\rho}(g) \right) \nonumber \\
    &= \frac{1}{6}\left[\widehat{\tilde{\rho}}(1)+3\widehat{\tilde{\rho}}(2)+2\widehat{\tilde{\rho}}(3)\right]
    \ ,
    \label{eqn:ProjS31}
\end{align} 
where the LCU's $\widehat{\tilde{\rho}}(1),\widehat{\tilde{\rho}}(2)$ and $\widehat{\tilde{\rho}}(3)$ have been defined in Eq.~\eqref{Eq:S3GroupAction} in terms of SWAP gates.
As a first step, we can consider a $3$-particle quantum state in first-quantization 
%%%
\begin{align} 
    \label{eq:3Particle}  \ket{\psi}
&=\sum_{\boldsymbol{x_1},\boldsymbol{x_2},\boldsymbol{x_3}}C_{\boldsymbol{x_1},\boldsymbol{x_2},\boldsymbol{x_3}}\ket{\boldsymbol{x_1}}\otimes \ket{\boldsymbol{x_2}}\otimes \ket{\boldsymbol{x_3}}
     \ ,
\end{align}
%%%
where $\boldsymbol{x_i}=j_i,\alpha_i$ is used to denote orbital $j_i$ and spin $\alpha_i$ degrees of freedom of the $i$-th electron.
The projector $\hat{P}^{\Gamma_1}$ acts on the  quantum state $\ket{\psi}$ as follows
%\begin{widetext}
\begin{align}
    \label{eq:ActionProjS3}
    \hat{P}^{\Gamma_1}\ket{ \psi}&=\frac{1}{6}\sum_{\boldsymbol{x_1},\boldsymbol{x_2},\boldsymbol{x_3}}C_{\boldsymbol{x_1},\boldsymbol{x_2},\boldsymbol{x_3}} (\ket{\boldsymbol{x_1}}\otimes \ket{\boldsymbol{x_2}}\otimes \ket{\boldsymbol{x_3}} \nonumber \\
    &+\ket{\boldsymbol{x_3}}\otimes \ket{\boldsymbol{x_1}}\otimes \ket{\boldsymbol{x_2}}+\ket{\boldsymbol{x_2}}\otimes \ket{\boldsymbol{x_3}}\otimes \ket{\boldsymbol{x_1}} 
    \nonumber\\ 
    & +
% \sum_{\boldsymbol{x_1},\boldsymbol{x_2},\boldsymbol{x_3}} C_{\boldsymbol{x_1},\boldsymbol{x_2},\boldsymbol{x_3}} 
\ket{\boldsymbol{x_2}}\otimes \ket{\boldsymbol{x_1}}\otimes \ket{\boldsymbol{x_3}} 
+\ket{\boldsymbol{x_1}}\otimes \ket{\boldsymbol{x_3}}\otimes \ket{\boldsymbol{x_2}} \nonumber \\
& +\ket{\boldsymbol{x_3}}\otimes \ket{\boldsymbol{x_2}}\otimes \ket{\boldsymbol{x_1}} )
    \ .
\end{align}
% \end{widetext}
This function is fully symmetric under exchange of two particles and it represent a bosonic state of three particles.

The sign representation $\Gamma_2$ also has an associated projector
\begin{align}
    \hat{P}^{\Gamma_2} &= \frac{1}{6} \sum^{3}_{C=1} \chi_{\Gamma_2}(C) \left( \sum^{|C| }_{g \in C} \hat{\rho}(g) \right) \nonumber \\
    &= \frac{1}{6}\left[\widehat{\tilde{\rho}}(1)-3\widehat{\tilde{\rho}}(2)+2\widehat{\tilde{\rho}}(3)\right]
    \ .
    \label{eqn:ProjS32}
\end{align}
Note that the transpositions are odd and under the sign representation the character associated to the  conjugacy class $C=2$ is $\chi_{\Gamma_2}(2)=-1$,
which allows us to construct a fermionic state
% \begin{widetext}
\begin{align}
    \label{eq:FermionActionProjS3}
    \hat{P}^{\Gamma_2}\ket{ \psi}
    &=
    \frac{1}{6}\sum_{\boldsymbol{x_1},\boldsymbol{x_2},\boldsymbol{x_3}} C_{\boldsymbol{x_1},\boldsymbol{x_2},\boldsymbol{x_3}}(\ket{\boldsymbol{x_1}}\otimes \ket{\boldsymbol{x_2}}\otimes \ket{\boldsymbol{x_3}} \nonumber \\
    &+\ket{\boldsymbol{x_3}}\otimes \ket{\boldsymbol{x_1}}\otimes \ket{\boldsymbol{x_2}}+\ket{\boldsymbol{x_2}}\otimes \ket{\boldsymbol{x_3}}\otimes \ket{\boldsymbol{x_1}}
    \nonumber\\ & -
% \sum_{\boldsymbol{x_1},\boldsymbol{x_2},\boldsymbol{x_3}}C_{\boldsymbol{x_1},\boldsymbol{x_2},\boldsymbol{x_3}}
\ket{\boldsymbol{x_2}}\otimes \ket{\boldsymbol{x_1}}\otimes \ket{\boldsymbol{x_3}}-\ket{\boldsymbol{x_1}}\otimes \ket{\boldsymbol{x_3}}\otimes \ket{\boldsymbol{x_2}} \nonumber \\
&-\ket{\boldsymbol{x_3}}\otimes \ket{\boldsymbol{x_2}}\otimes \ket{\boldsymbol{x_1}} )
    \ .
\end{align}
% \end{widetext}
The two examples provided so far can be used to simulate physical bosonic or fermionic states in first quantization. What is more interesting is that the $\hat{T}_{\rm GSA}$ formalism allows to simulate states with exotic statistics (parastatistics) on a quantum computer. A simple example of a 3-particle state with parastatistical nature can be build by considering the projector into the two-dimensional irrep $\Gamma_3=(2,1)$
\begin{align}
    \hat{P}^{\Gamma_3} &= \frac{2}{6} \sum^{3}_{C=1} \chi_{\Gamma_3}(C) \left( \sum^{|C| }_{g \in C} \hat{\rho}(g) \right) \nonumber \\
    &= \frac{1}{3}\left[2\widehat{\tilde{\rho}}(1)-2\widehat{\tilde{\rho}}(3)\right]
    \ .
    \label{eqn:ProjS33}
\end{align}
The corresponding projected state is
%\begin{widetext}
\begin{align}
    \label{eq:FermionActionProjS3}
    \hat{P}^{\Gamma_3}\ket{ \psi}
    &=
    \frac{1}{3}\sum_{\boldsymbol{x_1},\boldsymbol{x_2},\boldsymbol{x_3}} C_{\boldsymbol{x_1},\boldsymbol{x_2},\boldsymbol{x_3}} (2\ket{\boldsymbol{x_1}}\otimes \ket{\boldsymbol{x_2}}\otimes \ket{\boldsymbol{x_3}} \nonumber \\
    &-\ket{\boldsymbol{x_3}}\otimes \ket{\boldsymbol{x_1}}\otimes \ket{\boldsymbol{x_2}}-\ket{\boldsymbol{x_2}}\otimes \ket{\boldsymbol{x_3}}\otimes \ket{\boldsymbol{x_1}} )
    \ .
\end{align}
%\end{widetext}
Such a state does not correspond to a fermionic or bosonic state. Rather than the conventional statistics, this state satisfies parastatistics, which is unphysical. Nevertheless, it is remarkable that $\hat{T}_{\rm GSA}$ approach allows to coherently simulate the three irreps $\Gamma_1$, $\Gamma_2$ and $\Gamma_3$. After measuring a the irrep register, the state is probabilistically projected into a desired subspace.

\section{Open Problems}
\label{sec:open-problems}

The simulation of many-body quantum systems has been identified as one of the most promising avenues toward quantum advantage \cite{bauer2023quantum,PRXQuantum.5.037001,liu2024hybrid,crane2024hybrid} since its original exposition by Feynman \cite{feynman2018simulating}. As discussed in this paper, the adequate treatment of symmetries in simulating interacting many-body systems on quantum computers opens new possibilities for additional quantum advantage. As the beginning of the endeavor for the systematic treatment of symmetries in simulating many-body problems \emph{on quantum computers with quantum transforms}, a plethora of important problems have been inspired that will be instrumental in quantifying quantum advantage for quantum simulations.

Key questions are summarized in this section, including symmetries related to quantum chemistry in Sec. \ref{ssec:q-chem} and composite fermions and physical laws in Sec. \ref{ssec:c-ferm-phys-law}. Open problems regarding quantum simulation on novel quantum computers that are composed of both continuous (e.g., bosonic modes) and discrete (e.g., qubits) variable components are discussed in Sec. \ref{ssec:symm-q-computer}. Lastly, remarks on practical quantum advantages of symmetries in quantum simulation are provided in Sec. \ref{ssec:remark-qa}.

\subsection{Quantum Chemistry}
\label{ssec:q-chem}
Physical systems often exhibit more symmetries beyond the cyclic shift and permutation group investigated above. For example, a full description of quantum chemistry (including rotation, vibration, and nuclei spins) requires a tensor product of the following five groups \cite{bunker2006molecular}: $E(3) \otimes SO(3) \otimes S^{(e)}_n \otimes G^{\rm CNP} \otimes \mathcal{E}$, where $E(3)$ is the three-dimensional Euclidean translation group, $SO(3)$ is the continuous rigid-body rotation of molecules, $S_n^{(e)}$ is the permutation of $n$ electrons, $G^{\rm CNP}$ is the complete nuclear permutation group for identical nuclei in a molecule, and $\mathcal{E}$ is the inversion of the coordinates of all particles (electrons, nuclei) in the center of mass of the molecule. To describe these symmetries, the following questions immediately arise: 

$\bullet$ A general recipe to construct quantum circuits for QCT of general groups include non-Abelian and continuous groups is needed. The efficiency of such circuits often relies on recursive representation of the branching diagrams of groups~\cite{kawano2016quantum}. Therefore, developing proper ways to encode the branching diagrams on quantum computers is an important subproblem.

$\bullet$ A related mathematical question is that characters for many groups are unknown beyond the symmetric group and unitary group. The characters are important to construct efficient quantum circuits to realize the QCT.

$\bullet$ In addition to QCT, how to best realize generalized phase estimation on digital quantum computers to project into particular irreps of continuous groups is unclear. To achieve this, the construction of the projector in Eq. \eqref{projector-trivial-irrep} has to involve an integral over a continuous set of group elements (instead of the sum). On a digital quantum computer, proper discretization of the continuous group has to be developed, for example, by introducing $t$-design for the continuous group \cite{iosue2024continuous,dankert2009exact}.

$\bullet$ For quantum chemistry applications, an important scenario is that the continuous $SO(3)$ rotational symmetry can be broken into discrete point group symmetries \cite{harris1989symmetry} for highly symmetric molecules. Tackling these spatial symmetries will require local unitary operations on the spatial registers instead of spin registers in the 1st quantization mapping. The design of these local operations will depend on basis set \cite{nagy2017basis} choice in quantum chemistry and careful quantification of the circuit complexity needs further analysis.

$\bullet$ Another major class of molecules in quantum chemistry is transition-metal complexes with spin-orbital coupling \cite{chowdhury2020progress,koseki2019spin}. To investigate this, one needs to implement local unitary operations that mix spin and spatial registers. Moreover, important applications such as calculating the natural lifetime of electronic states will require a relativistic treatment based on the Dirac equation or its reduced version \cite{zhang2024dirac,liu2010ideas}. A full four-component wave function (including positron) in the 1st-quantization may be required to fully leverage symmetries in quantum simulation of these relativistic systems.

Note that most of these questions also apply to material science simulation in general, with slight changes of basis sets and the underlying symmetries.

\subsection{Composite Fermions and Physical Laws}
\label{ssec:c-ferm-phys-law}

Beyond quantum chemistry problems, simulating quantum systems with composite or continuous-variable degrees of freedom (bosonic modes) has become increasingly important \cite{atas2023simulating,martinez2016real}. This inspires the following questions:

$\bullet$ Mathematically, fermions and bosons are associated to irreps of the permutation group acting on $N$ particles in (3+1)-dimensions (three spatial dimensions plus time). However, certain low-dimensional strongly interacting electronic systems can host emergent quasiparticles with exotic statistics such as in the fractional quantum Hall effect in (2+1) dimensions~\cite{Stormer1999}. There, electrons and flux quanta form states known as composite fermions~\cite{Jain1990}. The symmetry group associated to fractional quantum Hall states is the braid group and the quasiparticles above the ground state are referred to as anyons~\cite{Hatsugai1991}. It remains a challenge to simulate these fermionic systems using quantum computers and our approach could be useful to perform this simulation in first quantization.

$\bullet$ Beyond simple Dirac fermions emerging in effective theories for graphene~\cite{Castro2009} and the quantum Ising model~\cite{fradkin2021quantum}, SU$(N)$ fermions~\cite{Halpern1975} are an important class of particles in condensed matter systems exhibiting non-abelian symmetries~\cite{chetcuti2022persistent,cazalilla2009ultracold}. These systems can be experimentally realized in available quantum simulators~\cite{zhang2014spectroscopic,gorshkov2010two}.  %\YL{Victor, can you add citations?}
However, further investigation is required to better leverage symmetry quantum subroutines to treat such SU$(N)$ fermions. This is of utmost importance in condensed matter systems to explore exotic states of matter such as heavy fermions~\cite{Stewart1984} and spin liquids~\cite{Yu2024}.

% $\bullet$ Fundamental interactions have their own symmetries that are intimately related to the physical laws. For example, the grand unification theory suggests $SU(3) \times SU(2) \times U(1)$ symmetries are sufficient to capture the three fundamental interactions \cite{weinberg1980conceptual}. Existing works have explored how to manually encode these symmetries in quantum simulations, but did not use quantum circuit the symmetry transform How to encode these symmetries of physical laws (instead of symmetries of individual quantum systems) into quantum simulation deserves more study.

$\bullet$ By discretizing the quantum fields, lattice gauge theory requires the quantum simulation of boson-fermi-spin mixed systems \cite{martinez2016real}. Mapping them to quantum computers and leveraging symmetry subroutines to speed up the simulation are important problems to investigate.

\subsection{Symmetries on Hybrid Quantum Computers}
\label{ssec:symm-q-computer}

On the flip side, what if the quantum computers themselves are governed by different symmetries? For example, significant progress has been made on hybrid oscillator-qubit quantum processors \cite{liu2024hybrid}. Given such a hybrid continuous-variable (CV) and discrete-variable (DV) quantum computer, how symmetry transformation should be constructed and encoded in a different fashion as compared to qubit-based quantum computers? Can the CV degrees of freedom on the quantum computer help us better perform group Fourier transform for continuous groups?

One important use of hybrid CV-DV quantum processor is to encode logical qubits into the CV bosonic modes to realize bosonic quantum error-correction \cite{cai2021bosonic,albert2018performance}. Preparation of bosonic code words can be viewed as a projection from an initial CV vacuum state to rotational-symmetric (e.g., cat code) \cite{girvin2017schrodinger} or translational-symmetric (e.g., GKP code) \cite{grimsmo2021quantum} CV state. One natural question is whether it is possible to use symmetry transformation to prepare such code word. If so, how to construct these symmetry transformations using native universal gate sets on hybrid CV-DV quantum computers? Moreover, would this way of formulating bosonic code preparation inspire new bosonic codes to be designed based on a wider class of symmetry groups?

\subsection{Remarks on Quantum Advantage}
\label{ssec:remark-qa}

Above all the technical development, the most important question is to identify where exponential speedup could come from in quantum simulations \cite{PhysRevX.13.041041,lee2023evaluating,RevModPhys.86.153,daley2022practical,PhysRevX.8.021010,chan2024spiers}, given proper treatment of symmetries on quantum computers. Obviously, it is important to quantify the gate complexity of the symmetry-adapted quantum simulation routine itself. We emphasize here a few additional considerations:

$\bullet$ One important aspect is the state preparation problem for quantum simulation \cite{lee2023evaluating,ward2009preparation,PhysRevA.95.022332,PhysRevA.110.012455}. Namely, a simple initial product state is usually transformed to a linear combination of symmetry-adapted final states (Eqs. \eqref{eq:fermionic-h2-product-state} and \eqref{eq:fermionic-h2-mixed-s-tt}). It is not clear how to best design the preparation of the initial product state given a desired target final state. For example, how does the use of non-Clifford gates in the product state preparation affect the coverage or weight of the final state over all irreps of the group. The approach presented in the current work may shed new light on the state preparation problem \cite{lee2023evaluating}.

$\bullet$ The answer to quantum advantage also strongly depends on what observables are measured after the quantum simulation, because extraction of any classical information from quantum computers has to pay a non-negligible cost \cite{yen2023deterministic,PRXQuantum.2.040320,huang2020predicting}. In the presence of symmetry transformations, we conjecture that measuring properties related to symmetries of the physical system may better leverage the advantage brought by symmetry-adapted quantum simulation.

$\bullet$ For \emph{ab initio} quantum simulation, an important aspect (despite often ignored) is how to design or pick the best basis set to discretize the original continuous problem \cite{nagy2017basis,chan2023grid,harrison2004multiresolution,piela2006ideas}. Standard basis in quantum chemistry are often used for simulation on quantum computers. However, it is not clear that existing basis sets that work the best on classical computers will necessarily be the choice for quantum computers. 

$\bullet$ In NISQ era, quantum computers often suffer from noise and decoherence. However, the symmetry of noise may be entirely different than the symmetry of the many-body quantum system under investigation. For example, coherent noise can be viewed as a special symmetry, which has been leveraged to construct novel algorithmic-level error-correction strategy \cite{PhysRevA.107.042429}. This suggests that symmetry transformation in the current work might provide a way to design quantum simulation algorithms that are less sensitive to noises, where Ref. \cite{minh2021faster,nguyen2022digital} provide excellent examples on such endeavor.

The challenge of quantifying practical quantum advantage lies in carefully integrating all the above aspects at each layer to derive concrete, end-to-end resource estimation with a clearly defined resource metric.

%%%%%%%%%%%%%%%%%%%%%%%%%%%%%%%%%%
\section{Conclusions and Outlook}
\label{sec:conclusion-outlook}

In summary, a unified framework for treating symmetries in quantum simulation of many-body physics given quantum computers was formulated. Concrete circuit constructions and resource estimations for symmetry-adapted projection for common groups and pairs of groups in many-body physics were provided. The power of these symmetry-adapted routines, i.e., the generalized phase estimation, was further demonstrated by combining them with other quantum simulation subroutines such as quantum phase estimation. Example applications to important condensed matter and quantum chemistry problems ranging from 1D to 3D were discussed. The provable speedups of these symmetry-adapted quantum subroutines suggest their applicability in the fault-tolerant era.
The symmetry-adapted quantum phase estimation circuit for a small molecule under minimal basis was also executed on noisy quantum hardware, demonstrating the practicality of our framework for the NISQ era as well. The paper concludes with a thorough discussion of major open problems regarding symmetries in the simulation of many-body systems on quantum computers. It is hoped that the unified framework and the related open problems presented in this work serve as a solid stepping stone for future endeavors exploring practical and provable quantum advantages of the simulation of many-body quantum systems in the NISQ and fault-tolerant era.

\acknowledgements
Y.L. thanks Bojko Bakalov for valuable discussions. V.M.B thanks K. Azuma, T. Yamazaki and L. Ruks for valuable discussions. V.M.B and K.J.J. thanks NTT Research Inc. for their support in this collaboration. Y.L. and S.I. acknowledge the support by the U.S. Department of Energy, Office of Science, Advanced Scientific Computing Research, under contract number DE-SC0025384. This research used resources of the Oak Ridge Leadership Computing Facility, which is a DOE Office of Science User Facility supported under Contract DE-AC05-00OR22725. N.F thanks Frederic Sauvage and Jedrzej Burkat for reviewing the paper.

% \newpage
\appendix
%%%%%

\section{Proof of Eq. \eqref{eq:ft2qct}}
\label{app:proof-ft2qct}

Consider the action of $\hat{F}_G$ to the left on the entangled state $\ket{\Omega_\Gamma}$, by substituting Eq. \eqref{MainTextgroup-fourier}, we have
\begin{align}
    & \bra{\Omega_\Gamma} \hat{F}_G \nonumber \\
    =& 
    \left(\frac{1}{\sqrt{d_\Gamma}} \sum_{i'=0}^{d_\Gamma -1} \bra{i', i'} \right) \hat{F}_G \nonumber \\
    =& 
    \frac{1}{\sqrt{d_\Gamma}} \sum_{i'=0}^{d_\Gamma -1}  \sum _{\Gamma}
    \sum_{ij} \sum_{g\in G} \sqrt{\frac{d_{\Gamma}}{|G|}} D_{ij}^{\Gamma}(g) \ket{\Gamma} \langle i', i'\ket{ij} \bra{g} \nonumber \\
    =& \sum _{\Gamma}
    \sum_{i} \sum_{g\in G} \sqrt{\frac{1}{|G|}} D_{ii}^{\Gamma}(g) \ket{\Gamma} \bra{g} \nonumber \\
    =& \sum _{\Gamma}
    \sum_{i} \sum_{C} \sum_{g_C \in C} \sqrt{\frac{1}{|G|}} D_{ii}^{\Gamma}(g_C) \ket{\Gamma} \bra{C, g_C},
    \label{app1:proof-1}
\end{align}
where in the last line, we have split the sum over $g$ as sum over the conjugacy class and all elements within the conjugacy class $\sum_{C} \sum_{g_C \in C}$.

Now using the definition of character, we have 
\begin{align}
    \sum_{i} D^\Gamma_{ii}(g_C) = \chi_\Gamma(g_C) = \chi_\Gamma(C)
    \label{app1:proof-2}
\end{align}
for any $g_C \in C$. Substitute this into \eqref{app1:proof-1}, we have
\begin{align}
    \bra{\Omega_\Gamma} \hat{F}_G  
    =& 
    \sum _{\Gamma}
    \sum_{C} \chi_{\Gamma}(C) \sum_{g_C \in C} \sqrt{\frac{1}{|G|}} \ket{\Gamma} \bra{C, g_C} \nonumber \\
    =& \sum _{\Gamma}
    \sum_{C} \chi_{\Gamma}(C) \sqrt{\frac{|C|}{|G|}} \ket{\Gamma} \bra{C} \otimes \bra{\theta},
    \label{app1:proof-3}
\end{align}
for $\ket{\theta} \equiv  \frac{1}{\sqrt{|C|}} \sum_{g_C \in C} \bra{g_C} $ being a uniform superposition of all elements in the conjugacy class $C$. This also means that $\ket{\theta}$ can be produced simply by the Hadamard transform, i.e., $\ket{\theta} = H\ket{0} \otimes H \ket{0}$. This means that acting Eq. \eqref{app1:proof-3} to state $\ket{\theta}$ will give
\begin{align}
    &\bra{\Omega_\Gamma } \hat{F}_G \left( H\ket{0} \otimes H\ket{0} \right) \nonumber \\
    =& \sum _{\Gamma}
    \sum_{C} \sqrt{\frac{|C|}{|G|}} \chi_{\Gamma}(C)  \ket{\Gamma} \bra{C} = QCT_G.
\end{align}
This proves Eq. \eqref{eq:ft2qct}.

\section{Character Derivation of $\hat{T}_{\text{GSA}}$\label{AppendixC}}

We prove Eq.~\eqref{eqn:projector_bavk} is a projector, namely
\begin{equation}
    \hat{P}^{\Gamma} \hat{P}^{\Gamma} = \hat{P}^{\Gamma}
\ .
\label{eq:projector2=proj}
\end{equation}
To show this, let us start by substituting Eq. \eqref{eqn:projector_bavk} into \eqref{eq:projector2=proj}
\begin{align}
    \hat{P}^{\Gamma} \hat{P}^{\Gamma} &= \left( \frac{d_{\Gamma}}{|G|} \sum_{g \in G} \chi_{\Gamma}(g) \hat{\rho}(g) \right) \cdot \left( \frac{d_{\Gamma}}{|G|} \sum_{g' \in G} \chi_{\Gamma}(g') \hat{\rho}(g') \right) \nonumber \\
    &= \frac{d_{\Gamma}^2}{|G|^2} \sum_{g \in G} \sum_{g' \in G} \chi_{\Gamma}(g)\chi_{\Gamma}(g') \hat{\rho}(g) \hat{\rho}(g')
    \ .
\end{align}
% Expanding this:
% \begin{equation}
%  \hat{P}^{\Gamma} \hat{P}^{\Gamma} = \frac{d_{\Gamma}^2}{|G|^2} \sum_{g \in G} \sum_{g' \in G} \chi^*_{\Gamma}(g)\chi^*_{\Gamma}(g') \rho(g) \rho(g')
%  \ .\\
% \end{equation}
Next, the definition of homomorphism $\hat{\rho}(g) \hat{\rho}(g') = \hat{\rho}(gg')$ can be used to obtain the relation
\begin{equation}
\hat{P}^{\Gamma} \hat{P}^{\Gamma} = \frac{d_{\Gamma}^2}{|G|^2} \sum_{g \in G} \sum_{g' \in G} \chi_{\Gamma}(g)\chi_{\Gamma}(g') \rho(gg')
\ .
\end{equation}
The next step is to define a new variable $j=gg'$ such that $g=jg'^{-1}$. After reindexing, obtaining 
\begin{equation}
     \label{eq:ProjectorProduct}
\hat{P}^{\Gamma} \hat{P}^{\Gamma} = \frac{d_{\Gamma}^2}{|G|^2} \sum_{j \in G} \sum_{g' \in G} \chi_{\Gamma}(jg'^{-1})\chi_{\Gamma}(g') \hat{\rho}(j)
\ .
\end{equation}
Next, let us calculate following the sum over the group elements
\begin{align}
\label{eq:Derivation}
&\sum_{g' \in G} \chi^*_{\Gamma}(jg'^{-1})\chi^*_{\Gamma}(g') 
=\sum_{g' \in G}\sum_{m,n} D^{\Gamma}_{m,m}(g'j^{-1})D^{\Gamma}_{n,n}(g'^{-1})
\nonumber \\
&=\sum_{m,r,n}D^{\Gamma}_{r,m}(j^{-1}) \left[\sum_{g' \in G}D^{\Gamma}_{m,r}(g')D^{\Gamma}_{n,n}(g'^{-1})\right]
\nonumber \\
&=\sum_{m,r,n}D^{\Gamma}_{r,m}(j^{-1}) \left[\frac{|G|}{d_{\Gamma}}\delta_{m,n}\delta_{r,n}\right]
\nonumber \\
&=\frac{|G|}{d_{\Gamma}}\sum_{n}D^{\Gamma}_{n,n}(j^{-1}) =\frac{|G|}{d_{\Gamma}}\chi^*_{\Gamma}(j) 
\ ,
\end{align}
where the Schur orthogonality relation for irreducible matrix representations of finite groups has been used
\begin{align}
\label{eq:SchurOrthogonality}
\sum_{g' \in G}D^{\Gamma}_{a,b}(g'^{-1})D^{\Gamma'}_{a',b'}(g')=\frac{|G|}{d_{\Gamma}}\delta_{\Gamma,\Gamma'}\delta_{a,a'}\delta_{b,b'}
\ .
\end{align}
%After using the homomorphism $\chi^*_{\Gamma}(jg'^{-1})=\chi^*_{\Gamma}(j)\chi^*_{\Gamma}(g'^{-1})$ and the orthogonality relation in Eq.~\eqref{eq:ProjectorProduct} we obtain
By replacing the complex conjugate of Eq.~\eqref{eq:Derivation} into Eq.~\eqref{eq:ProjectorProduct} we obtain
\begin{equation}
\begin{split}
\hat{P}^{\Gamma} \hat{P}^{\Gamma} &= \frac{d_{\Gamma}^2}{|G|^2} \sum_{j \in G} \frac{|G|}{d_{\Gamma}}\chi_{\Gamma}(j)  \hat{\rho}(j) \\
&= \frac{d_{\Gamma}}{|G|} \sum_{j \in G} \chi_{\Gamma}(j)  \hat{\rho}(j) \\
&= \hat{P}^{\Gamma}
\ .
\end{split}
\end{equation}
Thus, $\hat{P}^{\Gamma}$ is indeed a projection operator. In the case of abelian groups, all the irreps are one-dimensional and this can be applied in the same fashion as the $\hat{T}_{\text{GSA}}$ by summing over all irreps
\begin{equation}
\begin{split}
\hat{T}_{GSA}\ket{0}\otimes\ket{\psi} &= \sum_{\Gamma} \frac{d_{\Gamma}}{|G|} \sum_{g \in G} \chi_{\Gamma}(g) \ket{\Gamma}\otimes \hat{\rho}(g) \ket{\psi} \\
&= \sum_{\Gamma} a_{\Gamma} \ket{\Gamma}\otimes\ket{\psi^{\Gamma}},
\end{split}
\label{eqn:TGA}
\end{equation}
giving the projection of $\ket{\psi}$ into the irrep subspaces $\psi^{\Gamma}$ weighted by $a_{\Gamma}$.

\section{Projection Circuits} \label{app:prepare}

\begin{figure}[!htpb]
    \centering
    \begin{quantikz}
    \lstick{$|0\rangle$}& \qwbundle{n} &\gate{\prepare} & \gate[2]{\select[\tilde{\rho}_G]} & \gate{QCT_G}  &  \meterD{\Gamma}  \\
    \lstick{$|\psi\rangle$} & \qw & \qw  &   &\qw  & \qw  \rstick{$|\psi^\Gamma \rangle$} \\
    \end{quantikz}
        \caption{Quantum circuit for $T_{GSA}$ showing projection of $|\psi\rangle$ onto irrep $\Gamma$ through post selection, where QCT is the quantum character transform as defined in Eq. \eqref{eq:qctBracketNotation}. }
    \label{fig:tgsa_circ_prep}
    \end{figure}
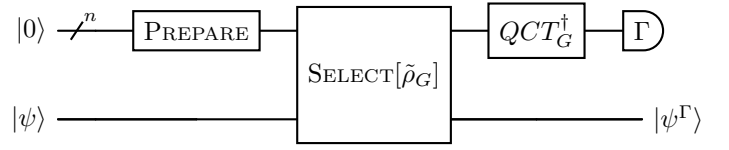

If the projection into a given irrep $\Gamma$ is of interest, the projection equations can be implemented exactly at the cost of a uniform normalization factor.

\begin{equation}
\begin{split}
    \frac{1}{\mathcal{N}}\sum_\Gamma \hat{P}^{\Gamma}\ket{\psi} &=  \frac{1}{\mathcal{N}}\sum_{\Gamma} \frac{d_{\Gamma}}{|G|} \sum_C \chi_{\Gamma}(C) \cdot \sum^{|C|}_{g \in C} \hat{\rho}(g) \ket{\psi} \\
    &= \frac{1}{\mathcal{N}}\sum_{\Gamma} a_{\Gamma} \ket{\psi^{\Gamma}}.
\end{split}
\label{eqn:TGA_back_app}
\end{equation}
The normalization constant is given by $\mathcal{N}= \sqrt{\sum_\Gamma \frac{d^2_\Gamma}{|G|}}$. In this case, a simplified circuit using just a $\prepare$ oracle can be used. The $\prepare$ is a state preparation oracle that stores the normalized coefficients in the first column of a unitary by acting on the $|0 \cdots0 \rangle$ ancilla state 
\begin{equation}
%\begin{split}
  \prepare = 
  \frac{1}{\mathcal{N}}\sum^{N_{\rm Conj}}_{C} \frac{d_{C}}{\sqrt{|G|}}|C\rangle\langle 0 | + \hat{\Pi}_\perp
  = \begin{bmatrix}
    \frac{d_1}{\sqrt{|G|}} & \cdot & \hdots \\ 
    \frac{d_2}{\sqrt{|G|}} & \cdot & \hdots  \\ 
    \vdots &  \ddots & \hdots   \\ 
    \frac{d_{N_{\rm Conj}}}{\sqrt{|G|}}  & \cdot  &  \hdots
    \end{bmatrix}.
%  \end{split}
\label{equation:prepare_1}
\end{equation}
where $\hat{\Pi}_\perp$ is a projector to the orthogonal space in the rest of the matrix except the 1st column.
The weighting coefficient is given by $d_\Gamma = d_C$. The rows of $\prep$ are group conjugacy classes $C$. The abstract circuit to implement the projector is shown in Fig. \ref{fig:tgsa_circ_prep}.

The projection circuits for the cyclic group $\mathbb{Z}_{M}$ can be simplified with $\prepare(\frac{1}{\sqrt{2^n}})$, since the state $\sum_C \frac{1}{\sqrt{2^n}} \ket{C}$ can be prepared by applying $H^{\otimes n}$ to the $\ket{0\cdots0}$ state.
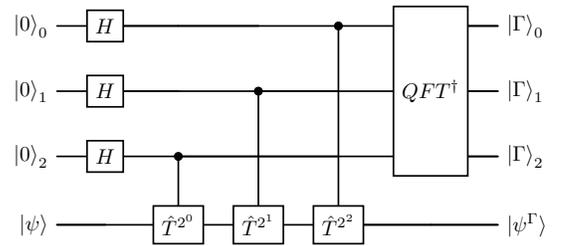
\begin{figure}
\centering
\begin{tikzpicture}
\node[scale=0.8] {
\begin{quantikz}
\lstick{$\ket{0}_0$} & \gate{H} & \qw & \qw & \ctrl{3} &\gate[3]{QFT} & \qw \rstick{$\ket{\Gamma}_0$}\\
\lstick{$\ket{0}_1$}  & \gate{H} & \qw & \ctrl{2} & \qw & &\qw \rstick{$\ket{\Gamma}_1$}\\
\lstick{$\ket{0}_2$} & \gate{H} & \ctrl{1} & \qw & \qw  & & \qw \rstick{$\ket{\Gamma}_2$} \\
\lstick{$\ket{\psi}$} & \qw  & \gate{\hat{T}^{2^0}}  & \gate{\hat{T}^{2^1}}  & \gate{\hat{T}^{2^2}} & \qw  & \qw \rstick{$\ket{\psi^{\Gamma}}$}  \\
\end{quantikz}
};
\end{tikzpicture}\\
\caption{Symmetry group transform for the cyclic group $\mathbb{Z}_8$, using an explicit projection circuit. Since the input state on the ancilla is the trivial irrep $\ket{0}_0 \ket{0}_1 \ket{0}_2$, the QFT in Fig. \ref{fig:stm_proj_circ} can be simply replaced by a Hadamard transform in the above.}
\label{fig:stm_proj_circAppendix}
\end{figure}
The overall circuit is shown in Fig. \ref{fig:stm_proj_circAppendix}, which takes a very similar form as the quantum phase estimation algorithm. However, the ancilla qubits are not read out but explicitly projected via post-selection for a given irrep $\Gamma$.

\section{Frobenius Character Formula}\label{app:frob_char}

The characters of the symmetric group $S_N$ can be calculated by the Frobenius Character formula \cite{Fulton2004}, simply from the coefficients of a multivariate polynomial of $ x =  x_1, x_2, \dots, x_k $, where $k$ is the number of parts of a partition representing an irrep $\Gamma$
\begin{equation}
    \chi_\Gamma(C_\mu) = \left[ \Delta(x) \cdot \prod_q P_q(x)^{j_q} \right]_{(l_1,\dots,l_k)}
    \ ,
\label{eqn:frobenius_character}
\end{equation}
where $ \chi_\Gamma(C_\mu)$ is the character of a conjugacy class $ C_\mu $ represented by the cycle type given by the partition $\mu = (m^{j_m}, \cdots,  2^{j_2}, 1^{j_1}) $, where $ j_q $ is the number of cycles of length $ q $ in the cycle type of the conjugacy class $\mu$. $ [\cdots]_{(l_1,\dots,l_k)} $ means the coefficient of the monomial $ x_1^{l_1} x_2^{l_2} \cdots x_k^{l_k} $ in Eq. \eqref{eqn:frobenius_character}.$\Gamma$ is the irreducible representation corresponding to a partition $\Gamma = ( \lambda_1, \lambda_2,...,\lambda_k )$ of $k$ parts. We introduce an index shift $l_n = \lambda_n + k - n$ for each part $(l_1,\dots,l_k)$. $ \Delta(x) $ is the Vandermonde determinant defined as
\begin{equation}
\Delta(x) = \prod_{1 \leq a < b \leq k} (x_a - x_b) 
\ .
\end{equation}
  
This determinant captures the antisymmetric properties of the conjugacy class, where $\Delta(\dots, x_a, \dots, x_b, \dots) = -\Delta(\dots, x_b, \dots, x_a, \dots)$,  with each variable relating to a part of the partition $\lambda$. $ P_q(x) $ is the power sum symmetric polynomial defined as
\begin{equation}
P_q(x) = x_1^q + x_2^q + \cdots + x_k^q
\ .
\end{equation}
This polynomial is used to encode each part of the cycle structure of the conjugacy class $ C_\mu $. $ \prod_q P_q(x)^{j_q} $ is the product of power sum polynomials raised to the powers corresponding to the cycle structure of the conjugacy class.  Therefore, the coefficients of the polynomial encode the characters for a given cycle type $\mu$ for the various irreps related to the index $(l_1,\dots,l_k)$. Other computational methods exist for calculating the characters of the symmetric group, such as those by Coleman \cite{COLEMAN196883}.

\section{Group Fourier Transform}
\label{app:sym_fourier}
In addition to the definition of group Fourier transform
\begin{align}
    \hat{F}_G =  %\sum_k^{N_{\rm Conj}} 
    \sum _{\Gamma}
    \sum_{ij} \sum_{g\in G} \sqrt{\frac{d_{\Gamma}}{|G|}} D_{ij}^{\Gamma}(g) \ket{\Gamma, ij}\bra{g},
    \label{group-fourier}
\end{align}
the inverse Fourier transform is given by
\begin{align}
    \hat{F}_G^{\dagger} =   \sum _{\Gamma}\sum_{ij} \sum_{g\in G} \sqrt{\frac{d_{\Gamma}}{|G|}}[D_{ij}^{\Gamma}(g)]^* \ket{g} \bra{\Gamma, ij}.
    \label{group-inverse-fourier}
\end{align}
Within a irrep $\Gamma$, a group element $g$ is represented by a $d_{\Gamma}\times d_{\Gamma}$ matrix with elements $D_{ij}^{\Gamma}(g)$. The group Fourier transformation is unitary and it can be efficiently performed on a quantum computer~\cite{moore2006generic}, where a subgroup chain known as a Bratelli diagram is exploited. % The Quantum Character Transform can be obtained by tracing over the $i,j$ indices and factorizing on the conjugacy classes, which is a more compact way of representing the group Fourier transform for high-dimensional irreps.

%\textcolor{red}{[NF: One of the interpretation of the $T_{GSA}$ over a finite group can be viewed as a (unitary) linear combination of orthogonal projection operators onto each subspace associated with the irreducible representations of the group.]}

%\YL{Maybe useful to start from a general discussion of $S_N$ and then continuing with the following special case of $S_2$.}
We briefly summarize the most important aspects of the group Fourier transform for the symmetric group $G=S_n$ for $n$ objectsd~\cite{kawano2016quantum}. This group is finite with an order of $|G|=n!$ and its elements are permutations of $n$ objects. Given a permutation $g\in S_n$, it can be decomposed as a product of elementary transpositions $c_{k,k+1}$. The irreps are then fully characterized in terms of partitions and Young diagrams.In fact, to build the representation $\Gamma_j$, the representation of the elementary transpositions $c_{k,k+1}$ needs to be calculated by looking at the weighted number of steps to reach two numbers in a given standard Young tableaux. For example, for $n=2$, the order is $|S_2|=2$ and there are two irreps $\Gamma_1=(2,0)={\tiny\yng(2)}$ and $\Gamma_2=(1,1)={\tiny\yng(1,1)}$ that are one-dimensional. For $n=3$ the order of the group is $|S_3|=6$ and there are three irreps. $\Gamma_1=(3,0)={\tiny\yng(3)}$ and $\Gamma_2=(1,1,1)={\tiny\yng(1,1,1)}$ and $\Gamma_3=(2,1)={\tiny\yng(2,1)}$. The irreps $\Gamma_1$ and $\Gamma_2$ are one-dimensional, while $\Gamma_3$ is two-dimensional. The irrep $\Gamma_3$ is two-dimensional because there are two standard Young tableux $T^{\Gamma_2}_{A_1}={\scriptsize\young(12,3)}$ and $T^{\Gamma_2}_{A_2}={\scriptsize\young(13,2)}$ that act as basis defining the matrix representation $D^{\Gamma_2}_{i,j}(g)$.

To define the group Fourier transform in Eq.~\eqref{group-fourier} for the symmetric group $G=S_n$~\cite{beals1997quantum}, the irreps and the group elements need to be encoded using ancilla quantum registers. In order to do this, the \emph{canonical coding} of an element $g \in S_n$ is used. This provides a concrete indexing scheme for elements with the implicit map back to permutations as follows:

\begin{align}
    (i_1, i_2, \cdots, i_{n-1}) &\in \mathbb{Z}_2\times\mathbb{Z}_3\times\cdots\times\mathbb{Z}_{n},\\
    g &= c^{i_{n-1}}_{(1,2,\cdots,n)}\cdots c^{i_{2}}_{(1,2,3)}c^{i_{1}}_{(1,2)}. \label{eq:g_index}
\end{align}
An indexing scheme can be generated from group elements following this decomposition as follows:
    \begin{equation} \label{eq:perm_ordering}
        S_n \rightarrow \{g_i\}, \text{ with } i \equiv \sum_{j = 1}^{n - 1} i_j \frac{n!}{(j + 1)!}, 
    \end{equation}
where this index is generated given any $g$ with the form of Eq.~\eqref{eq:g_index} ~\cite{kawano2016quantum}. In general, for non-abelian groups, we also need to encode the label $\Gamma$ and the matrix indices $i,j$ and finally the group elements $g$ associated to the matrix representation $D^{\Gamma}_{i,j}(g)$. We refer the readers to Ref.~\cite{kawano2016quantum} for further details on this encoding. Ref.~\cite{kawano2016quantum} also provides an explicit construction of an efficient quantum circuit implementing the group Fourier transform for the symmetric group based on the group tower structure $S_1\subset S_2\cdots S_{n-1}\subset S_n$ with poly$(n)$ scaling.
 
Next, let us explicitly construct the group Fourier transform in Eq.~\eqref{group-fourier} for the permutation group of two particles $G=S_2$ with group elements $g\in S_2=\{e,c_{(1,2)}\}$. Any group element can be encoded using a single qubit $\ket{i_1}$ as $g=c^{i_1}_{(1,2)}$ where  $i_1\in\mathbb{Z}_2$ as in the canonical encoding. More specifically, the canonical encoding of the group elements is given by $\ket{e}=\ket{0}$ and $\ket{c_{(1,2)}}=\ket{1}$.

As the irreps for $n=2$ are one-dimensional, the labels $i,j$ do not need to be encoded in the group Fourier transform of Eq.~\eqref{group-fourier}. The irrep labels can be encoded using a single qubit, i.e., $\ket{\Gamma_1}=\ket{0}$ and $\ket{\Gamma_2}=\ket{1}$. Using these encodings, the explicit form for the group Fourier transform for $G=S_2$ is obtained.

\begin{align}
    \hat{F}_{S_2} &=  %\sum_k^{N_{\rm Conj}} 
    \sum _{\Gamma}
     \sum_{g\in S_2} \sqrt{\frac{d_{\Gamma}}{|S_2|}} D^{\Gamma}(g) \ket{\Gamma}\bra{g}
     \nonumber \\
     &=\frac{1}{\sqrt{2}}(\ket{0}\bra{0}+\ket{0}\bra{1}+\ket{1}\bra{0}-\ket{1}\bra{1})
     \nonumber \\
     &=\frac{1}{\sqrt{2}} 
    \begin{bmatrix}
        1 & 1 \\
        1 & -1
    \end{bmatrix},
    \label{group-fourierS2}
\end{align}
where $d_{\Gamma}=1$ and $|S_2|=2$. Note that  $D^{\Gamma_1}(e)=D^{\Gamma_2}(e)=D^{\Gamma_1}[c_{(1,2)}]=1$ and that $D^{\Gamma_2}[c_{(1,2)}]=-1$. Thus, in the canonical representation, the group Fourier transform corresponds to a Hadamard gate $F_{S_2}=H$.

%%%%%%%%%%%%%%%%%%%%%
\section{Jordan-Wigner transformation and symmetries of interacting fermionic models \label{AppendixB}}
%%%%%%%%%%%%%%%%%%%%%
In the main text, a non-integrable spin model was discussed in Eq.~\eqref{eq:IsingModel}. In this section, symmetries of this model will be discussed using standard tools of condensed matter physics to show its complexity. It will be shown that this model can be mapped to a system of interacting fermions in momentum space, albeit being highly non-local.

Pauli matrices satisfy an $SU(2)$ algebra and commute at different sites of the spin chain. On the contrary, fermionic operators at different sites anticommute. For this reason, to build a fermionic representation of su$(2)$, one needs to include a highly non-local factor to satisfy the anticommutation relation. The solution for this problem was found by Jordan and Wigner in a seminal paper~\cite{jordan1928}. The Jordan-Wigner transformation is given by
%%%
\begin{align}
          \label{eq:JordanWigner}
 Z_j&=-(\hat{f}^{\dagger}_j+\hat{f}_j)\prod^{j-1}_{m=1}(1-2\hat{f}^{\dagger}_m\hat{f}_m)
\nonumber \\
Y_j&=-\mathrm{i}(\hat{f}^{\dagger}_j-\hat{f}_j)\prod^{j-1}_{m=1}(1-2\hat{f}^{\dagger}_m\hat{f}_m)
\nonumber \\
X_j&=1-2\hat{f}^{\dagger}_j\hat{f}_j
\ .
\end{align}
%%%
Here the operators $\hat{f}^{\dagger}_j$ and $\hat{f}_j$ are the fermionic creation and annihilation operators in real space, satisfying the anticommutation relations $\{\hat{f}_i,\hat{f}^{\dagger}_j\}=\delta_{i,j}$ and $\{\hat{f}_i,\hat{f}_j\}=\{\hat{f}^{\dagger}_i,\hat{f}^{\dagger}_j\}=0$. 

By using this transformation, the Hamiltonian in Eq.~\eqref{eq:IsingModel} becomes
%%%
\begin{align}
          \label{eq:IsingHamitonianFermionSI}
\hat{\mathcal{H}}&=- \sum^{N}_{j=1} a_j(1-2\hat{f}^{\dagger}_j\hat{f}_j)- J\sum^{N-1}_{j=1}(\hat{f}^{\dagger}_j-\hat{f}_j)(\hat{f}^{\dagger}_{j+1}+\hat{f}_{j+1})
\nonumber \\
&-\sum^{N}_{j=1} w_j(\hat{f}^{\dagger}_j+\hat{f}_j)\prod^{j-1}_{m=1}(1-2\hat{f}^{\dagger}_m\hat{f}_m)
\nonumber \\
&- J(\hat{f}^{\dagger}_N-\hat{f}_N)(\hat{f}^{\dagger}_{N+1}+\hat{f}_{N+1})\prod^{N}_{m=1}(1-2\hat{f}^{\dagger}_m\hat{f}_m)
\ .
\end{align}
%%%
From this, it can be seen that the disordered spin chain maps to a highly non-local Hamiltonian of interacting fermions. The last term of the Hamiltonian is the parity operator $\hat{Q}=\bigotimes^N_{j=1} X_j$ discussed in the main text. It can be shown that the boundary conditions of the fermions allow the Hamiltonian to be mapped into two coupled blocks with different parities. This can be achieved by using the projector $\hat{Q}_{\sigma}=[\hat{1}+(-1)^{\sigma}\hat{Q}]/2$ into the subspaces with parity eigenvalue $(-1)^{\sigma}$ for $\sigma = 0,1$, defined in the main text, and the identity $\hat{Q}_{0}+\hat{Q}_{1}=\hat{1}$. The Hamiltonian can be explicitly written as
%%%
\begin{align}
\label{eq:IsingHamitonianParityDecomposition_V1}
(\hat{Q}_{0}+\hat{Q}_{1})\hat{\mathcal{H}}(\hat{Q}_{0}+\hat{Q}_{1})
=
\hat{\mathcal{H}}^{(+)}+\hat{\mathcal{H}}^{(-)}+\hat{\mathcal{H}}^{(+-)}
\ .
\end{align}
%%%
The positive and negative parity sector correspond to fermions with anti-periodic $\hat{f}_{N+1}=-\hat{f}_{1}$ and periodic $\hat{f}_{N+1}=\hat{f}_{1}$ boundary conditions, respectively~\cite{Dziarmaga2005,Vogl2012}. The two diagonal blocks of this decomposition can be written in terms of the integrable part of the Hamiltonians as follows

%%%
\begin{align}
          \label{eq:IsingHamitonianParityDecomposition}
\hat{\mathcal{H}}^{(\pm)}&= \hat{Q}_{\sigma}\left[- \sum^{N}_{j=1} a_j(1-2\hat{f}^{\dagger}_j\hat{f}_j)\right]\hat{Q}_{\sigma}
\nonumber \\
& +\hat{Q}_{\sigma}\left[- J\sum^{N}_{j=1}(\hat{f}^{\dagger}_j-\hat{f}_j)(\hat{f}^{\dagger}_{j+1}+\hat{f}_{j+1})\right]\hat{Q}_{\sigma}
\ .
\end{align}
%%%
The longitudinal field term $\hat{\mathcal{H}}_L=-\sum^{N}_{j=1} w_j(\hat{f}^{\dagger}_j+\hat{f}_j)\prod^{j-1}_{m=1}(1-2\hat{f}^{\dagger}_m\hat{f}_m)$ in Eq.~\eqref{eq:IsingHamitonianFermionSI} breaks the parity of the model and thus couples the different parity sectors giving rise to the term $\hat{\mathcal{H}}^{(\pm )}$. In summary, the different parity sectors can be separated by taking into account the boundary conditions of the fermions, while the longitudinal field is responsible for the coupling between different irreps.

Next, the decomposition of the Hamiltonian in terms of invariant subspaces under the group of spatial translations will be explored. To achieve this, the discrete Fourier transformation 
$\hat{f}_j=\frac{e^{-\mathrm{i}\frac{\pi}{4}}}{\sqrt{N}}\sum_k \hat{F}_{k} e^{\mathrm{i}k j}$ (see Ref.~\cite{Dziarmaga2005}) needs to applied to the Ising model in the decomposition Eq.~\eqref{eq:IsingHamitonianParityDecomposition}, where $\hat{F}_k$ is the fermionic annihilation operator of momentum $k$ in the Fourier space. Of course, from our previous discussion on the relation between parity and boundary conditions, only discrete values of the momenta $k$ are allowed within a given parity sector~\cite{Dziarmaga2005}. We will denote these particular values as $k_{\pm}$. 

Before doing the explicit calculation, it can be seen that the Ising interaction term $\hat{\mathcal{H}}_I=- J\sum^{N}_{j=1}(\hat{f}^{\dagger}_j-\hat{f}_j)(\hat{f}^{\dagger}_{j+1}+\hat{f}_{j+1})$ is invariant under spatial translations and it can be written in diagonal form as $\hat{\mathcal{H}}_I=\sum_k J(\hat{\gamma}^{\dagger}_k\hat{\gamma}_k-1/2)$, where $\hat{F}_k=u_k\hat{\gamma}_{k}+v^*_{-k}\hat{\gamma}^{\dagger}_{-k}$. The total momentum operator $\hat{P}=\sum_k k\hat{\gamma}^{\dagger}_k\hat{\gamma}_k$ is a conserved quantity of the Ising interaction in our Hamiltonian defining the symmetric sectors under translational invariance. 

Both the transverse and longitudinal fields in our model break the translational symmetry and generate coupling between different sectors with momenta $k$. To see how the different sectors are coupled, it is enough to calculate the explicit form of the transverse field term $\hat{\mathcal{H}}_T=- \sum^{N}_{j=1} a_j(1-2\hat{f}^{\dagger}_j\hat{f}_j)$ in the momentum representation, as follows
%%%
\begin{align}
          \label{eq:IsingHamitonianMomentumDecomposition}
\hat{\mathcal{H}}_T&= - \sum^{N}_{j=1} a_j(1-2\hat{f}^{\dagger}_j\hat{f}_j)= - \sum^{N}_{j=1} a_j+2\sum_{k,k'}M_{k,k'}\hat{F}^{\dagger}_{k}\hat{F}_{k'}
\nonumber \\
& =- \sum^{N}_{j=1} a_j+2\sum_{k,k'}\left(B_{k,k'}\hat{\gamma}^{\dagger}_{k}\hat{\gamma}_{k'}+C_{k,k'}\hat{\gamma}^{\dagger}_{k}\hat{\gamma}^{\dagger}_{-k'}\right)
\nonumber \\
& +2\sum_{k,k'}\left(D_{k,k'}\hat{\gamma}_{-k}\hat{\gamma}_{k'}+E_{k,k'}\hat{\gamma}_{-k}\hat{\gamma}^{\dagger}_{-k'}\right)
\ ,
\end{align}
%%%
where $B_{k,k'}=M_{k,k'}u_k^*u_{k'}$, $C_{k,k'}=M_{k,k'}u_k^*v^*_{-k'}$, $D_{k,k'}=M_{k,k'}v_{-k}u_{k'}$, and $E_{k,k'}=M_{k,k'}v_{-k}v^*_{-k'}$.
This expression already exhibits high complexity, as the disorder coupled the different momenta $k$ and while keeping a good parity. Further, the longitudinal term couples all the subspaces with momenta $k$ for different parities. 

In summary, while tools from condensed matter physics can be used, the mapping to fermions makes the problem intractable. For this reason, the novel approach in this work is advantageous, as the spin operators are worked with directly without using fermions.

% \subsection{Questions}

% It would seem like for general groups using the characters is the only way to do it. As using the full representation matrices is not scalable. Furthermore It is not clear to me why you need to use 2 Fourier transforms, as you are just using it to multiply the characters with the group elements. It also seems like it will be very non-trivial to implement the algorithm when the number of group elements does not equal the number of irreps, as the group Fourier transform of characters will be non-square. Also, when the number of group elements is not a power of 2, a uniform superposition would need to be used rather than just a set of hadamards on all ancilla qubits.

\section{Harper model on a cylinder and quantum circuits}
\label{AppendixA}
Our goal in this section is to demonstrate that by considering periodic boundary conditions, the Harper-Hofstadter model~\cite{Hofstadter1976} can be represented using a simple quantum circuit, which simplifies the task of Hamiltonian simulation that is essential for quantum phase estimation.

To start with, a QFT can be directly performed on the register using the digital encoding $\bm{y}$ of the integer $y$ in the Harper model in Eq.~\eqref{eq:HamSinglePartRest}. As the magnetic translations factorize
\begin{align}
         \label{eq:MagTransFact}
\hat{\mathcal{U}}_b&=\left(\sum_{\bm{x}}\ket{\bm{x}}\bra{\bm{x+1}}\right)\otimes\left(\sum_{\bm{y}}\ket{\bm{y}}\bra{\bm{y}}\right)
 \nonumber\\
 \hat{\mathcal{V}}_b&=\left(\sum_{\bm{x}}e^{2i\pi x b}\ket{\bm{x}}\bra{\bm{x}}\right)\otimes\left(\sum_{\bm{y}}\ket{\bm{y}}\bra{\bm{y+1}}\right)
  \ ,
\end{align}
%%%%
the QFT can be performed along the $y$ axis to obtain
\begin{align}
         \label{eq:MagTransQFT}
\hat{U}_{\text{QFT}}\hat{\mathcal{U}}_b \hat{U}^{\dagger}_{\text{QFT}}&=\sum_{\bm{k_y},\bm{x}}\ket{\bm{x}}\bra{\bm{x+1}}\otimes\ket{\bm{k_y}}\bra{\bm{k_y}}
 \nonumber\\
\hat{U}_{\text{QFT}} \hat{\mathcal{V}}_b \hat{U}^{\dagger}_{\text{QFT}}&=\sum_{\bm{k_y},\bm{x}}e^{2i\pi (x b-k_y/N)}\ket{\bm{x}}\bra{\bm{x}}\otimes\ket{\bm{k_y}}\bra{\bm{k_y}}
  \ .
\end{align}
%%%%
By using these identities, it can be shown that  Eq.~\eqref{eq:HamSinglePartRest} transforms under $\hat{U}_{\text{QFT}}$ as
$\hat{U}_{\text{QFT}}\hat{\mathcal{H}}\hat{U}^{\dagger}_{\text{QFT}}=\sum_{k_y} \hat{\mathcal{H}}_{k_y}\otimes\ket{\bm{k_y}}\bra{\bm{k_y}}$ with $\hat{\mathcal{H}}_{k_y}$ defined as
\begin{align}
         \label{eq:SymmAdapt_qft}
\hat{\mathcal{H}}_{k_y}=\sum_{\bm{x}}\left(J_xe^{2i\pi( x b-k_y/N)}\ket{\bm{x}}\bra{\bm{x}}+J_y\ket{\bm{x}}\bra{\bm{x+1}}+\text{H.c}\right)
  \ .
\end{align}
%%%%
By using the expressions discussed above, it can further be shown that the evolution operator for the Harper-Hofstadter Hamiltonian can be written in terms of a control unitary, as follows
\begin{align}
         \label{eq:EvQFT}
e^{-i\hat{\mathcal{H}} t}=\hat{U}^{\dagger}_{\text{QFT}}\left(\sum_{k_y} \hat{V}_{k_y}\otimes\ket{\bm{k_y}}\bra{\bm{k_y}}\right)\hat{U}_{\text{QFT}}
  \ ,
\end{align}
where $\hat{U}_{\text{QFT}}$ is acting on the second register.
Thus, the evolution operator can be simplified using the irreps (momenta) $\ket{\bm{k_y}}$ and the unitary $\hat{V}_{k_y}=e^{-\hat{\mathcal{H}}_{k_y} t}$.

\bibliography{ref.bib}

\end{document}